	\documentclass[preprint2]{aastex}

	\usepackage{epsfig}
	\usepackage{amssymb} %enable lots of maths symbols
	\usepackage{textcomp} %more math symbols
	\usepackage{subfigure}

\def \lsun{\ifmmode{{\rm\ L}_\odot}\else{${\rm\ L}_\odot $}\fi}
\def \msun{\ifmmode{{\rm\ M}_\odot}\else{${\rm\ M}_\odot$}\fi}
\def \rsun{\ifmmode{{\rm\ R}_\odot}\else{${\rm\ R}_\odot$}\fi}
\newcommand{\kms}{km~s$^{-1}$}                         % Kms-1
\def \mdot{\ifmmode{{\rm\dot{M}}}\else{${\rm\dot{M}}$}\fi}

\newcommand{\das}[2]{${#1}''\!\!.{#2}$}                 %decimal arcsec: 0''.5

\newcommand\as{${''}$}

\newcommand{\ha}{H$\alpha${}}
\newcommand{\hb}{H$\beta${}}
\newcommand{\hi}{H\,{\sc i}{}}

\newcommand{\oiii}{[O\,{\sc iii}]}
\newcommand{\oiiil}{[O\,{\sc iii}]\ $\lambda$5007\,\AA}
\newcommand{\none}{[N\,{\sc i}]}
\newcommand{\nii}{[N\,{\sc ii}]}

\newcommand{\sii}{[S\,{\sc ii}]}
\newcommand{\siil}{[S\,{\sc ii}]\ $\lambda$6716\,\AA}
\newcommand{\siib}{[S\,{\sc ii}]\ $\lambda$6731\,\AA}
\newcommand{\feii}{Fe\,{\sc ii{}}}

\def \st{\ifmmode{^{\mathrm{st}}}\else{${^{\mathrm{st}}}$}\fi}
\def \nd{\ifmmode{^{\mathrm{nd}}}\else{${^{\mathrm{nd}}}$}\fi}
\def \rd{\ifmmode{^{\mathrm{rd}}}\else{${^{\mathrm{rd}}}$}\fi}
\def \th{\ifmmode{^{\mathrm{th}}}\else{${^{\mathrm{th}}}$}\fi}

\newcommand{\myemail}{pbw@astro.livjm.ac.uk}

\shorttitle{Dynamical Drivers of Nuclear Activity. I.}
\shortauthors{Westoby et al.}

\begin{document}

%%	LaTeX will automatically break titles if they run longer than one line. However, you may 
%%	use \\ to force a line break if you desire.

\title{A Magellan--IMACS-IFU Search for Dynamical Drivers of Nuclear Activity. \\ 
			I. Reduction Pipeline and Galaxy Catalog}

%%	Use \author, \affil, and the \and command to format author and affiliation information.
%%	Note that \email has replaced the old \authoremail command from AASTeX v4.0. You can use 
%%	\email to mark an email address anywhere in the paper, not just in the front matter. 
%%	As in the title, use \\ to force line breaks.

\author	{P.~B.~Westoby\altaffilmark{1}, 
			C.~G.~Mundell\altaffilmark{1}, 
			N.~M.~Nagar\altaffilmark{2}, 
			W.~Maciejewski\altaffilmark{1}, 
			E.~Emsellem\altaffilmark{3,4}
			M.~M.~Roth\altaffilmark{5}, 
			J.~Gerssen\altaffilmark{5},
			and
			I.~K.~Baldry\altaffilmark{1} \\
			\vspace{4mm}
			\textit{Accepted for publication in ApJS}
			}

	\altaffiltext{1}{Astrophysics Research Institute, Liverpool John Moores University, Twelve Quays House, Egerton Wharf, Birkenhead, CH41 1LD, UK; \myemail}
	\altaffiltext{2}{Astronomy Group, Departamento de F\'{i}sica, Universidad de Concepci\'{o}n, Casilla 160-C, Concepci\'{o}n, Chile}
	\altaffiltext{3}{European Southern Observatory, Karl-Schwarzschild-Str 2, 85748 Garching, Germany}
	\altaffiltext{4}{Centre de Recherche Astronomique de Lyon, 9 avenue Charles Andr{\'e}, Saint-Genis Laval, F-69230, France}
	\altaffiltext{5}{Astrophysikalisches Institut Potsdam, An der Sternwarte 16, D-14482 Potsdam, Germany}

%%	Mark off your abstract in the ``abstract'' environment. In the manuscript style, abstract will 
%%	output a Received/Accepted line after the title and affiliation information. No date will 
%%	appear since the author does not have this information. The dates will be filled in by the
%%	editorial office after submission.

\begin{abstract}
Using the Inamori Magellan Areal Camera and Spectrograph (IMACS) integral-field unit (IFU) on the 6.5\,m Magellan telescope, we have designed the first statistically significant investigation of the two-dimensional distribution and kinematics of ionised gas and stars in the central kiloparsec regions of a well-matched sample of Seyfert and inactive control galaxies selected from the Sloan Digital Sky Survey. The goals of the project are to use the fine spatial sampling (0.2 arcsec~pixel$^{-1}$) and large wavelength coverage (4000--7000\AA) of the IMACS-IFU to search for dynamical triggers of nuclear activity in the central region where active galactic nucleus (AGN) activity and dynamical timescales become comparable, to identify and assess the impact of AGN-driven outflows on the host galaxy and to provide a definitive sample of local galaxy kinematics for comparison with future three-dimensional kinematic studies of high-redshift systems. In this paper, we provide the first detailed description of the procedure to reduce and calibrate data from the IMACS-IFU in `long-mode' to obtain two-dimensional maps of the distribution and kinematics of ionised gas and stars. The sample selection criteria are presented, observing strategy described and resulting maps of the sample galaxies presented along with a description of the observed properties of each galaxy and the overall observed properties of the sample.
\end{abstract}

%%	Keywords should appear after the \end{abstract} command. The uncommented example has been  
%%	keyed in ApJ style. See the instructions to authors for the journal to which you are 
%%	submitting your paper to determine what keyword punctuation is appropriate.

%%	Authors who wish to have the most important objects in their paper linked in the electronic 
%%	edition to a data center may do so in the subject header.  Objects should be in the 
%%	appropriate "individual" headers (e.g., quasars: individual, stars: individual, etc.) with 
%%	the additional provision that the total number of headers, including each individual object, 
%%	not exceed six.  The \objectname{} macro, and its alias \object{}, is used to mark each 
%%	object.  The macro takes the object name as its primary argument.  This name will appear 
%%	in the paper and serve as the link's anchor in the electronic edition if the name is 
%%	recognized by the data centers.  The macro also takes an optional argument in parentheses 
%%	in cases where the data center identification differs from what is to be printed in the paper.

\keywords{galaxies: active --- galaxies: kinematics and dynamics --- galaxies: nuclei --- galaxies: Seyfert --- galaxies: structure.}

%%	From the front matter, we move on to the body of the paper. In the first two sections, notice 
%%	the use of the natbib \citep and \citet commands to identify citations.  The citations are 
%%	tied to the reference list via symbolic KEYs. The KEY corresponds to the KEY in the 
%%	\bibitem in the reference list below. We have chosen the first three characters of the 
%%	first author's name plus the last two numeral of the year of publication as our KEY for 
%%	each reference.

\section{Introduction}

Supermassive black holes are widely accepted to lie at the centres of all bulge-dominated galaxies, with the observed correlation between black hole mass and bulge velocity dispersion evident in both active and inactive galaxies confirming a causal link between previous phases of accretion-driven nuclear activity in the galactic life cycle \citep{geb00,merr01}. Although, such active galactic nuclei (AGN) are recognised to be integral to galaxy formation and evolution, the AGN triggering and fuelling mechanisms remain unknown. Further, the transportation of fuel to the AGN and any corresponding ejection or feedback are key components of cosmological models and are tuned to deliver observed properties of today's galaxies (e.g., \citealt{spring05}). Merger-driven galaxy evolution and nuclear activity is central to these models but is unlikely to be a valid mechanism in the local Universe, where the merger rate has declined and the $\sim$20\% of galaxies currently exhibiting nuclear activity \citep{gould10} are commonly found in early type spiral galaxies in non-cluster environments and do not show signatures of strong interactions \citep{west07,gabor09}. Identifying the triggering and fuelling mechanism and the origin and transportation of fuel in AGN therefore remain an important goal in astronomy.

Optical/IR {\em imaging} studies have remained ambiguous in identifying a single fuelling mechanism \citep{86bus,fuentes88,deRob98,hunt04,tang08,kuo08}. External perturbations, such as tidal interactions or minor mergers, and internal instabilities in non-axisymmetric potentials such as bars or $m=1$ spirals have long been suggested as viable mechanisms for removing angular momentum from host galaxy gas to allow it to move closer to the AGN \citep{wada04,witold97}, and more recently hydrodynamic turbulence in the nuclear interstellar medium has been suggested to contribute to low-level accretion \citep{alig11}, but direct evidence of such physical mechanisms has been hard to obtain (e.g., \citealt{cgm99,martini01}). The availability of large numbers of uniformly derived galaxy properties from the Sloan Digital Sky Survey (SDSS; \citealt{york00}) has allowed statistical comparisons of active and inactive galaxies to be performed and evolutionary trends to be investigated in an attempt to explain the location of active and inactive galaxies across the so-called blue and red sequences in galaxy color-magnitude space (\citealt{bald04,west07,schaw09}). These studies build on earlier work on smaller samples, such as that by \cite{hunt04}, who suggested an evolutionary sequence driven by an underlying, but unidentified, dynamical instability to explain structural differences observed in a sample of 250 active and inactive host galaxies imaged with NICMOS on the \textit{Hubble Space Telescope} (\textit{HST}). Direct determination of the dynamical properties of active and inactive host galaxies is therefore vital.

Radio interferometric spectroscopy provides valuable information on the large-scale gaseous environment and gaseous structure and kinematics of the host galaxy disks to large radii (e.g., \citealt{cgm07,haan08,haan09}) but surface brightness sensitivity constraints of current interferometers limits achievable angular resolutions to $\sim5$\as\ at best (\citealt{mund99,things08}), but more typically $\sim$18\arcsec. Millimetric observations can sample dense gas, such as CO, closer to the nucleus, but this gas may be clumpy and discontinuous, limiting the ease with which the velocity fields can be interpreted (e.g., \citealt{dumas10}). Optical observations of galactic kinematics offer the capability of sampling the velocity field close to the centre of a galaxy; in Seyferts, such kinematic observations were traditionally performed with a combination of narrowband imaging to identify line-emitting regions and long-slit spectroscopy in a preferred position angle (e.g., \citealt{ekers83,tad89a,wilson94}), but disentangling the kinematics of the host galaxy from non-gravitational AGN-driven outflows is difficult and model dependent.

With recent developments in integral-field unit (IFU) technology on large telescopes, which provide 2D imaging spectroscopy, it is now feasible to study the distribution and kinematics of gas {\em and} stars on scales ever closer to the galactic nucleus (e.g., \citealt{fathi06,barb06,riffel08}). Spurious misinterpretation of one-dimensional velocity fields is less likely when full two-dimensional velocity fields are sampled; the comparison of stellar and gaseous velocity fields is vital to identify kinematics associated with the galactic gravitational potential rather than non-circular gaseous streaming motion that may be erroneously interpreted in long-slit data e.g., mistakenly implying the presence of black holes offset from their galactic centres \citep{wilson85,ferruit04}.

The two-dimensional approach was particularly effective in the comparison of stellar and gaseous kinematics in a small matched sample of active and inactive galaxies by \cite{dumas07}, who used the SAURON IFU \citep{bacon01} on the 4.2-m William Herschel Telescope to tentatively identify a kinematic difference between Seyfert and inactive galaxy hosts, after removal of non-gravitational AGN dynamics. The SAURON IFU provides a large field of view (FOV, 33\as $\times$ 41\as) but at the cost of pixel sampling ($\sim 1$\as) and the sample was selected before SDSS data were available. We, therefore, carefully selected a larger, distance-limited and well-matched sample of active and inactive control galaxies from the SDSS (see \citealt{west07}) and observed them with the IMACS-IFU on the 6.5-m Magellan telescope \citep{bigelow98,schmoll04}. The IMACS-IFU is particularly well-suited to the study of galactic nuclei as it has small pixels (\das02) to sample its (4\as $\times$ 5\as) field of view and an unusually large wavelength coverage ($\sim 4000 - 7000$~\AA), which provides access to all major emission lines from \hb\ to \ha\ as well as the underlying stellar continuum and stellar absorption lines such as Mg\,$b$ and \feii. Table~\ref{tab:IFUs} shows a comparison of selected integral-field spectrographs.

IMACS is most commonly used as a multi-object spectrograph (MOS) and published documentation concentrates on this MOS mode. In this paper, we describe the full procedure for obtaining stellar and gaseous distribution and kinematic maps from Magellan IMACS-IFU data. In particular we present, for the first time, the detailed procedure for reducing and calibrating IMACS in IFU `long-mode'. We present the resultant gaseous and stellar maps in catalogue form for our SDSS-selected IMACS-IFU sample of active and matched-inactive galaxies. The detailed dynamical analysis, interpretation and comparison will be presented in Paper II.

\section{Observations and Analysis}

\subsection{Sample Selection}

The purpose of this study was to compare active and inactive galaxies. To do this, it is important to carefully select a sample of Seyfert galaxies with a well-matched control sample of inactive galaxies. Our active galaxy sample was initially selected from the \cite{hao05a} AGN catalogue\footnote{Data publicly available at: \url{http://isc.astro.cornell.edu/$\sim$haol/agn/agncatalogue.txt}}. The Hao catalogue was compiled from the Second Data Release (DR2) of the SDSS, based on the emission-line properties of the galaxies, and their locations in the various line-ratio diagnostic diagrams \citep{BPT}. The catalogue therefore contains 1317 broad-line AGN, 3074 narrow-line AGN (assuming the \cite{kew01a} criteria), and a further 10700 narrow-line AGN that satisfy the \cite{kauf03b} criteria.

A redshift cut of $z < 0.05$, an \textit{r}-band fibre-magnitude cut of $r < 17.5$ and a declination cut of $\delta < 10$\textdegree\ were applied to the Hao catalogue, along with a right ascension cut. The SDSS images and spectra for the remaining galaxies were then examined individually, and a selection made based on the spectra --- i.e. ensuring the presence of the forbidden lines typical of an AGN. 

Once a Seyfert galaxy had been selected (which here can include low-ionization nuclear emission-line region (LINER) galaxies), 
we required a \textit{matched} control galaxy. The matching process used here was similar to that used in a statistical study of the parent Hao catalogue \citep{west07}. Potential control galaxies were obtained from the SDSS DR2 \textit{SkyServer}\footnote{SDSS \textit{SkyServer} currently at \url{http://cas.sdss.org/astro/en/}}, under the same conditions as above ($z < 0.05$, $r < 17.5$, $\delta < 10$\textdegree), and $specClass=2$ (i.e., galaxies). 

Controls were closely matched to the active galaxies through the following photometric properties:
\begin{enumerate}
	\item	Redshift, $z$.
	\item	Absolute $r$-band magnitude, $M_{r}$.
	\item	Aspect ratio, $b/a$, (Isophotal minor/major axis ratio in $r$-band): inclination, in principle.
	\item	Radius, $petroR90$ (in arcsec), containing 90\% of the Petrosian flux, averaged
  over $r$- and $i$-bands.
\end{enumerate}
 
Potential matches were identified by calculating the differences, $\Delta z$, $\Delta [b/a]$, $\Delta M_{r}$ and $\Delta petroR90$, between the AGN and each control. The differences were then weighted, and summed as:
\begin{equation}
	\Delta C = \frac{| \Delta z |}{0.01} + \frac{| \Delta M_{r} |}{0.2} + \frac{| \Delta [b/a]
	|}{0.1} + \frac{| \Delta petroR90 |}{0.2\arcsec}
%	\Delta |C| = 10 \Delta |z| + 0.5 \Delta |M_{r}| + \Delta |b/a| + 0.5 \Delta  |petroR90|
\end{equation}

The SDSS images of the 10 controls with the lowest values of $\Delta C$ were then extracted and visually compared to that of the Seyfert galaxy, with the best visual, morphological match being retained. The SDSS spectrum of the matched control was also checked prior to observation, to ensure that its `inactive' classification was correct. 

\begin{figure*}
		\begin{centering}
		\includegraphics[scale=0.55]{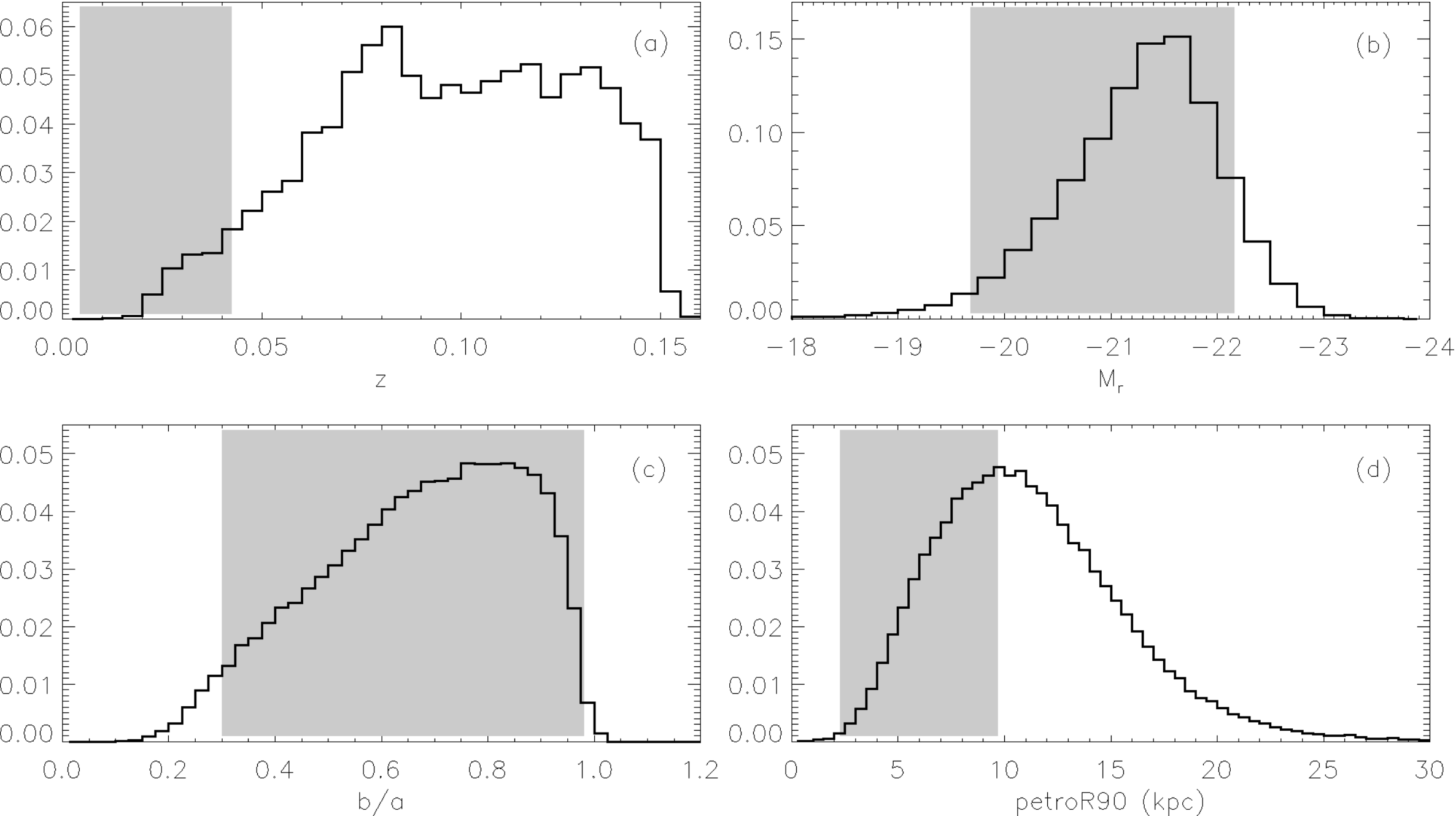}
		\caption{Parameter space probed by the IMACS-IFU survey (\textit{shaded areas}), compared to normalized distributions of the 
			parent SDSS sample. (a) Redshift coverage. (b) Absolute $R$-band magnitude. (c) Galaxy axial ratio ($b/a$). 
			(d) Radius containing 90\% of the Petrosian flux (in kiloparsecs). }
		\label{fig:para-space}
		\end{centering}
	\end{figure*}

\begin{figure*}
		\begin{centering}
		\includegraphics[scale=0.5]{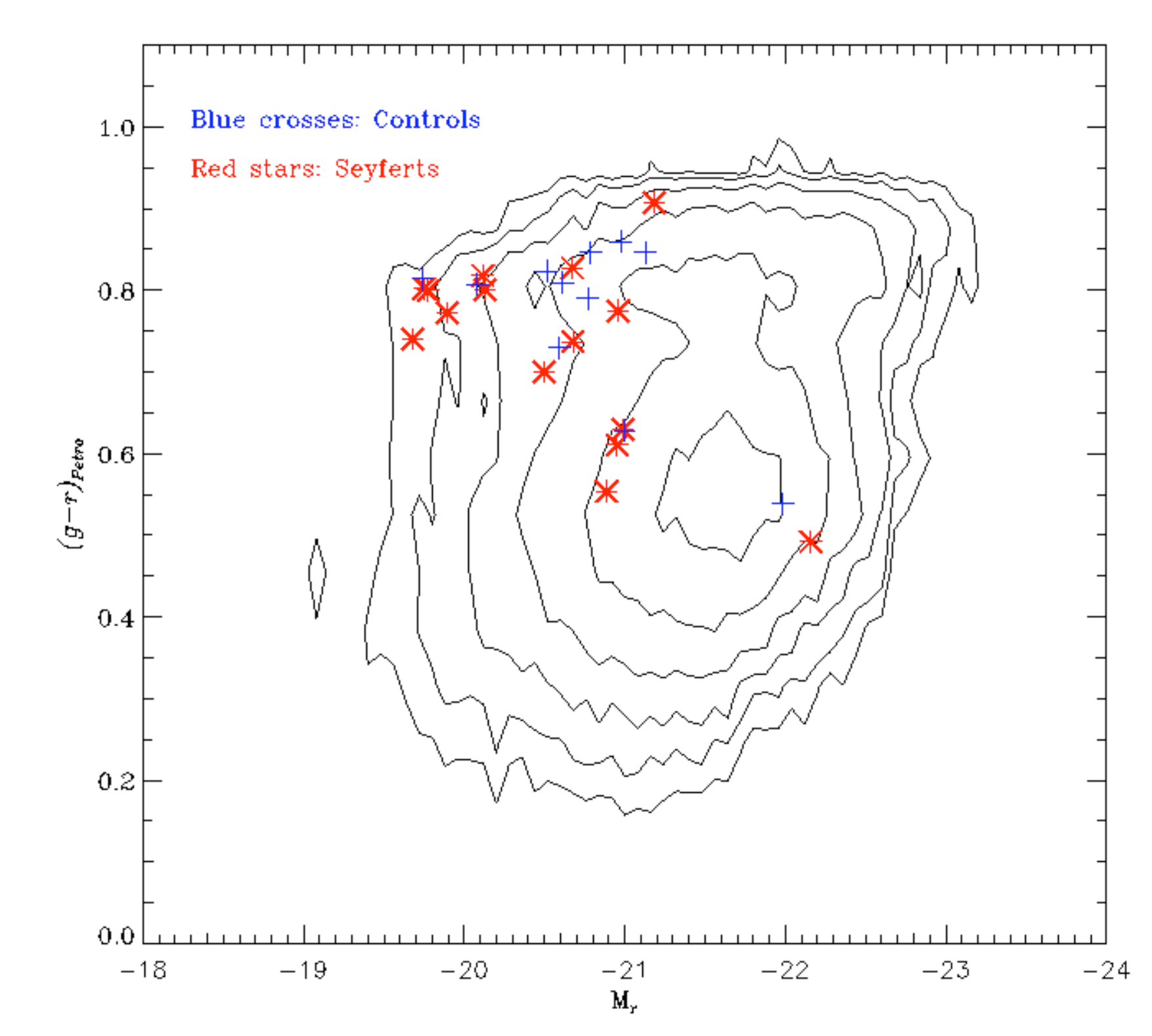}
		\caption{Locations of the IMACS-IFU sample in the colour-magnitude diagram. The contours plotted are those 
				of the `Controls A' sample from \citealt{west07}. }
		\label{fig:cmd-loc}
		\end{centering}
	\end{figure*}

\begin{figure*}[htp]
	\begin{centering}
		\subfigure{\includegraphics[scale= 0.63]{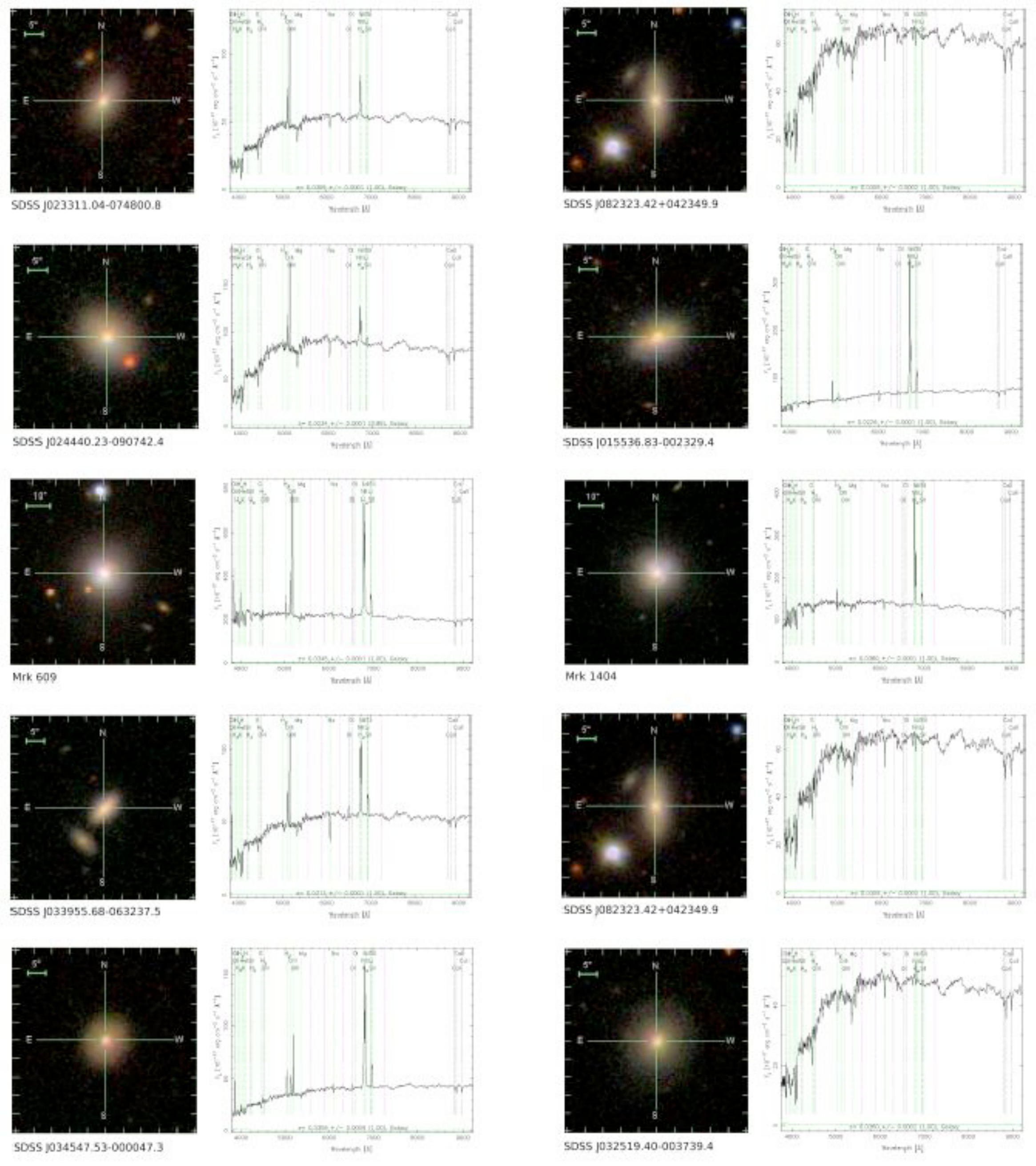}} \\
	\caption{\textit{gri} SDSS colour images of the sample galaxies, along with their SDSS spectrum 
				obtained through a 3\as\ fibre. Each Seyfert galaxy is displayed on the left, with its 
				associated control galaxy on the right. The orientation is such that North is up and East is left. 
				The spatial scale displayed for each image is given in the top-left corner of the image. 
				High-resolution version available at: http://www.astro.ljmu.ac.uk/$\sim$pbw/IMACS-IFU/IMACS-1.pdf.}
	\label{fig:sdss}
	\end{centering}
\end{figure*}

In our final observing run, in August 2007, the MPIA complete galaxy catalogue of SDSS DR4 had become available. We therefore did not restrict ourselves to selecting control galaxies from DR2, thus providing many more galaxies to match to the Seyferts, increasing the probability of obtaining a good match. To date, a total of 28 galaxies have been observed: 17 Seyferts and 11 control galaxies. Table~\ref{tab:properties} summarises the basic properties of the IMACS sample. In addition, the parameter space probed by the IMACS sample is directly compared to that of the parent sample \citep{west07} in Figures \ref{fig:para-space} and \ref{fig:cmd-loc}. In Fig.~\ref{fig:para-space} the distributions of the selection properties for the parent SDSS sample are plotted (histograms), with the range of values covered by the IMACS sample shown by the shaded area. The locations of the IMACS sample in the color-magnitude diagram are shown in Fig.~\ref{fig:cmd-loc}. The contours represent the parent sample and show the bimodal split of late- and early-type galaxies. The locations of the IMACS galaxies are overplotted and show that the majority of the IMACS sample is made up of low-luminosity, early-type galaxies. Despite color not being a matching criteria in selecting controls, Fig.~\ref{fig:cmd-loc} also shows that the Seyferts and controls are still well matched. The SDSS images and spectra for the final sample are shown in Fig.~\ref{fig:sdss}.

Observations were carried out over four observing runs, for a total of nine nights. The first was in December 2005, the second and third in April 2006, and the final run was in August 2007. Table.~\ref{tab:sample} summarises the final sample, and observations thereof. Some of the control galaxies fit the matching criteria for more than one Seyfert, so to maximise telescope usage, some Seyferts share a control galaxy. The total integration time is also given in Table.~\ref{tab:sample}. The original aim was to obtain four times 30 minute exposures for each galaxy to achieve a minimum signal-to-noise ratio (S/N) of 60 \AA$^{-1}$, but due to time constraints resulting from bad weather and technical problems, this was not always possible. For example, the control galaxy MCG~+00-02-006 was excluded from any analysis due to the small integration time resulting in very weak signal-to-noise (S/N $< 20$ \AA$^{-1}$).

\subsection{IMACS-IFU}

The IMACS-IFU is a fibre-fed IFU developed and built for the IMACS at the Magellan-I 6.5m telescope at Las Campanas Observatory, Chile. The IMACS spectrograph itself, located on the Nasmyth platform and mechanically derotated, was designed for imaging, long-slit and multi-slit spectroscopy with a field-of-view (FOV) of 27 arcmin \citep{schmoll04}. The spectrograph operates two different cameras, offering different imaging scales and dispersions. The `short' (\textit{f}/2.5) camera works with grisms, while the `long' (\textit{f}/4) camera utilises reflective gratings. In addition to integral-field spectroscopy (IFS) observations, IMACS can be used for image-slicing multi-slit observations (GISMO), tunable narrow-band imaging (MMTF), and multi-object Echelle observations (MOE).

Our observations were carried out in $f/4$ long-mode, and thus used the IMACS Mosaic1 CCD camera, which consists of eight 2K $\times$ 4K $\times$ 15~$\mu$m SITe detectors, forming an $8192 \times 8192$ mosaic image. 

The IFU is an optional part of IMACS and can be moved into the focal plane by the IMACS slit-mask server in the same manner as an ordinary slit mask. It consists of two identical fields -- one object and one sky field -- separated by 60\as, allowing for classical background subtraction, and beam switching. The basic specification of the IMACS-IFU in its two different modes is summarised in Table~\ref{tab:basics}, along with a summary of our observing setup (in long mode). All observations were run in $1 \times 1$ binning.

\subsection{Observational Strategy}

Integral-field spectroscopy requires a similar observing strategy to that of optical imaging. In traditional fashion, at the start of the night bias frames, domeflat exposures and skyflat exposures were taken in order to undertake basic CCD reduction. In addition, preliminary arc frames were taken to provide an initial wavelength calibration. Arc frames were also taken throughout the night, before and after each science observation. Typically two or three standard stars were also observed throughout the night (again, preceded and followed by arc exposures) to enable flux calibration, and/or to be velocity standards. Table~\ref{tab:observed} summarises the typical observing process for IMACS-IFU.

In fibre-fed IFS, domeflats are required to trace the fibres across the CCD. As such, several domeflat exposures were taken with various exposure times. These domeflats were taken with the grating in place, and Quartz lamps turned on, allowing for all fibres to be illuminated at all wavelengths, and hence traceable across the CCD. 

Skyflats, or twilight flats, can be used to correct for variations in fibre throughput. These are again taken with the grating in place and quartz lamps on, and with the vents of the telescope dome also open, allowing twilight radiation onto the detector. If no skyflats are observed, domeflats can also be used to correct for variations in fibre throughput.

Prior to starting the science observations, an acquisition frame was acquired. A reconstructed image of the acquisition frame could be viewed immediately to ensure that the galaxy was located centrally in the FOV. Science observations were then done in exposures of 1800~s, with the telescope dithered between exposures. Arc frames of 60~s were sufficient for wavelength calibration.

\subsection{Data Reduction}

Prior to our observations, no reduction pipeline had been developed for IMACS in IFU mode (Note: since then a pipeline, {\em kungifu}, has been developed for IMACS-IFU `short-mode'; \citealt{bolton07}). We have therefore produced a full reduction and calibration procedure to ensure reliable science products.

Reduction of IFU data utilises standard CCD imaging reduction techniques, but also requires additional processes. In particular, careful extraction of the multiple spectra is required. For this we used an adaptation of the \texttt{p3d}\footnote{The latest version of \texttt{p3d} \citep{sand10} can be found at: http://p3d.sourceforge.net/} package developed for the Potsdam Multi-Aperture Spectrograph (PMAS; \citealt{pmas}), called \texttt{imacs\_online}. This extraction process is described in \S2.4.4. 

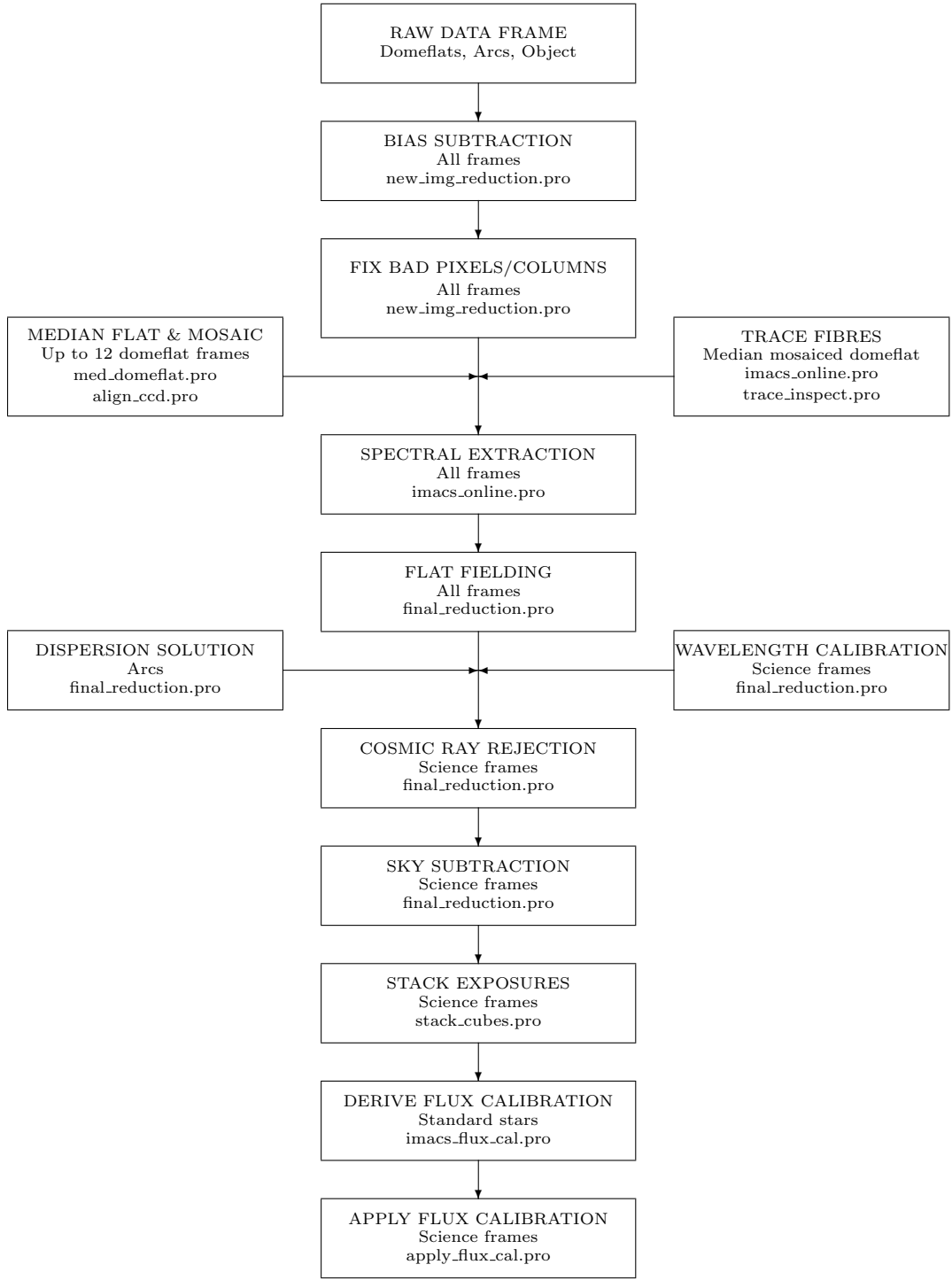
\begin{figure*}
\begin{centering}
\scriptsize
\setlength{\unitlength}{2.1em}
%\vspace{0.5in}
\begin{picture}(24.000000,35.000000)(-8.000000,-35.000000)
	\put(0.0000,0.0000){\framebox(8.0000,2.0000)[c]{\shortstack[c]{
		RAW DATA FRAME \\
		Domeflats, Arcs, Object
	}}}
	\put(4.0000,0.0000){\vector(0,-1){1.0000}}
%	\put(-1.0000,-0.5000){\vector(1,0){5.0000}}
%	\put(-8.0000,-1.5000){\framebox(7.0000,2.0000)[c]{\shortstack[c]{
%		MEDIAN BIAS FRAME \\
%		Up to 8 bias frames \\
%		med\_bias.pro
%	}}}
	\put(0.0000,-3.0000){\framebox(8.0000,2.0000)[c]{\shortstack[c]{
		BIAS SUBTRACTION \\
		All frames \\
		new\_img\_reduction.pro
	}}}
	\put(4.0000,-3.0000){\vector(0,-1){1.0000}}
	\put(0.0000,-6.5000){\framebox(8.0000,2.5000)[c]{\shortstack[c]{
		FIX BAD PIXELS/COLUMNS \\
		All frames \\
		new\_img\_reduction.pro
	}}}
	\put(4.0000,-6.500){\vector(0,-1){2.5000}}
	\put(-1.0000,-7.500){\vector(1,0){5.000}}
	\put(-8.0000,-8.500){\framebox(7.000,2.500)[c]{\shortstack[c]{
		MEDIAN FLAT \& MOSAIC \\
		Up to 12 domeflat frames \\
		med\_domeflat.pro\\
		align\_ccd.pro
	}}}
	\put(9.0000,-7.500){\vector(-1,0){5.0000}}
	\put(9.0000,-8.5000){\framebox(7.0000,2.5000)[c]{\shortstack[c]{
		TRACE FIBRES \\
		Median mosaiced domeflat \\
		imacs\_online.pro \\
		trace\_inspect.pro
	}}}
	\put(0.0000,-11.0000){\framebox(8.0000,2.0000)[c]{\shortstack[c]{
		SPECTRAL EXTRACTION \\
		All frames \\
		imacs\_online.pro
	}}}
	\put(4.0000,-11.0000){\vector(0,-1){1.0000}}
	\put(0.0000,-14.0000){\framebox(8.0000,2.0000)[c]{\shortstack[c]{
		FLAT FIELDING \\
		All frames \\
		final\_reduction.pro
	}}}
	\put(4.0000,-14.0000){\vector(0,-1){2.5000}}
	\put(-1.0000,-15.000){\vector(1,0){5.0000}}
	\put(-8.0000,-16.000){\framebox(7.0000,2.0000)[c]{\shortstack[c]{
		DISPERSION SOLUTION \\
		Arcs \\
		final\_reduction.pro
	}}}
	\put(9.0000,-15.000){\vector(-1,0){5.0000}}
	\put(9.0000,-16.000){\framebox(7.0000,2.0000)[c]{\shortstack[c]{
		WAVELENGTH CALIBRATION \\
		Science frames \\
		final\_reduction.pro
	}}}
	\put(0.0000,-18.500){\framebox(8.0000,2.0000)[c]{\shortstack[c]{
		COSMIC RAY REJECTION \\
		Science frames \\
		final\_reduction.pro
	}}}
	\put(4.0000,-18.500){\vector(0,-1){1.0000}}
	\put(0.0000,-21.500){\framebox(8.0000,2.0000)[c]{\shortstack[c]{
		SKY SUBTRACTION \\
		Science frames \\
		final\_reduction.pro
	}}}
	\put(4.0000,-21.500){\vector(0,-1){1.0000}}
	\put(0.0000,-24.500){\framebox(8.0000,2.0000)[c]{\shortstack[c]{
		STACK EXPOSURES \\
		Science frames \\
		stack\_cubes.pro
	}}}
	\put(4.0000,-24.500){\vector(0,-1){1.0000}}
	\put(0.0000,-27.500){\framebox(8.0000,2.0000)[c]{\shortstack[c]{
		DERIVE FLUX CALIBRATION \\
		Standard stars \\
		imacs\_flux\_cal.pro
	}}}
	\put(4.0000,-27.500){\vector(0,-1){1.0000}}
	\put(0.0000,-30.500){\framebox(8.0000,2.0000)[c]{\shortstack[c]{
		APPLY FLUX CALIBRATION \\
		Science frames\\
		apply\_flux\_cal.pro
	}}}
\end{picture}
\vspace{-.75in}
\caption{Schematic flow-chart summarising the data reduction process.}
\label{fig:flow}
%% end of flow output
\end{centering}
\end{figure*}

Fig.~\ref{fig:flow} shows a schematic of the reduction steps needed to produce a final datacube. The routines used to do each step are also shown. All processing was done using specific scripts written in the Interactive Data Language (IDL) by the authors.

\subsubsection{Raw IMACS-IFU data}

\begin{figure*}
		\begin{centering}
		\includegraphics[scale=0.17]{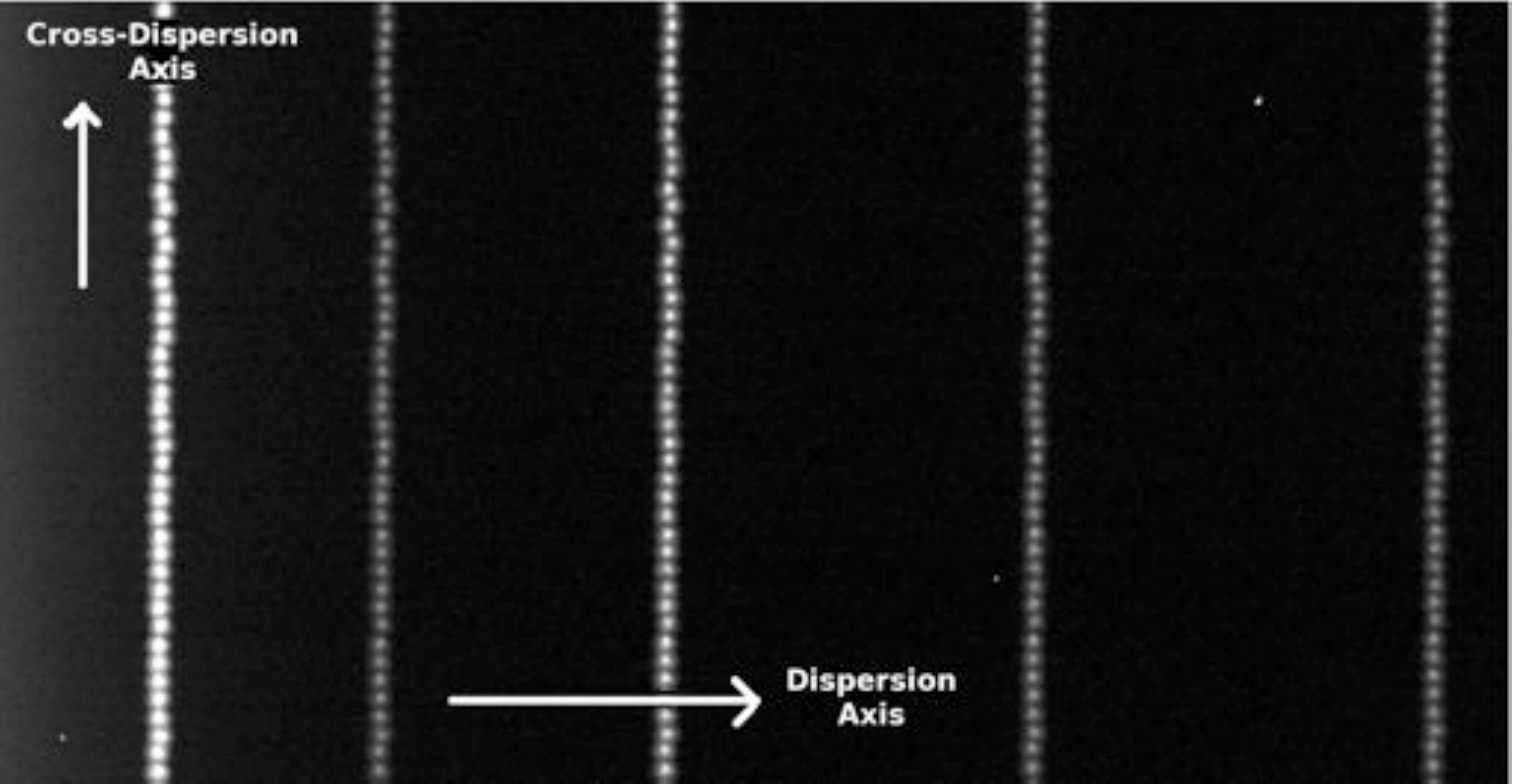}
		\caption{Greyscale image of a section of IMACS-IFU raw data. The dispersion axis is 
				the \textit{x}-axis in this case, while the cross-dispersion (`spatial') axis 
				is along the \textit{y}-axis.}
		\label{fig:raw}
		\end{centering}
	\end{figure*}

As in the majority of fibre-fed spectrographs, IMACS-IFU raw data consists of a number of spectra distributed along one axis of a two-dimensional frame. Fig.~\ref{fig:raw} shows an example of a raw IMACS-IFU data frame. This is a small section of a raw, 60 second arc exposure with He--Ne--Ar calibration lamps turned on. Each spectrum is distributed across the image, in the dispersion axis. At each wavelength, the spectra are also spread perpendicular to the dispersion axis, in the `cross-dispersion' axis. Spectra are separated by approximately 5 pixels along the cross-dispersion axis, limiting the cross-contamination (or cross-talk) between neighbouring fibres. The spectra are not perfectly aligned along the dispersion or cross-dispersion axes due to the instrument itself, the instrument setup, and telescope flexure effects. The transmission of individual fibres also varies due to tensions, misalignments, and intrinsic physical differences \citep{sanch06}, which cause the light from each fibre to enter the spectrograph with slight relative offsets. 

\begin{figure*}
		\begin{centering}
		\includegraphics[scale=0.60]{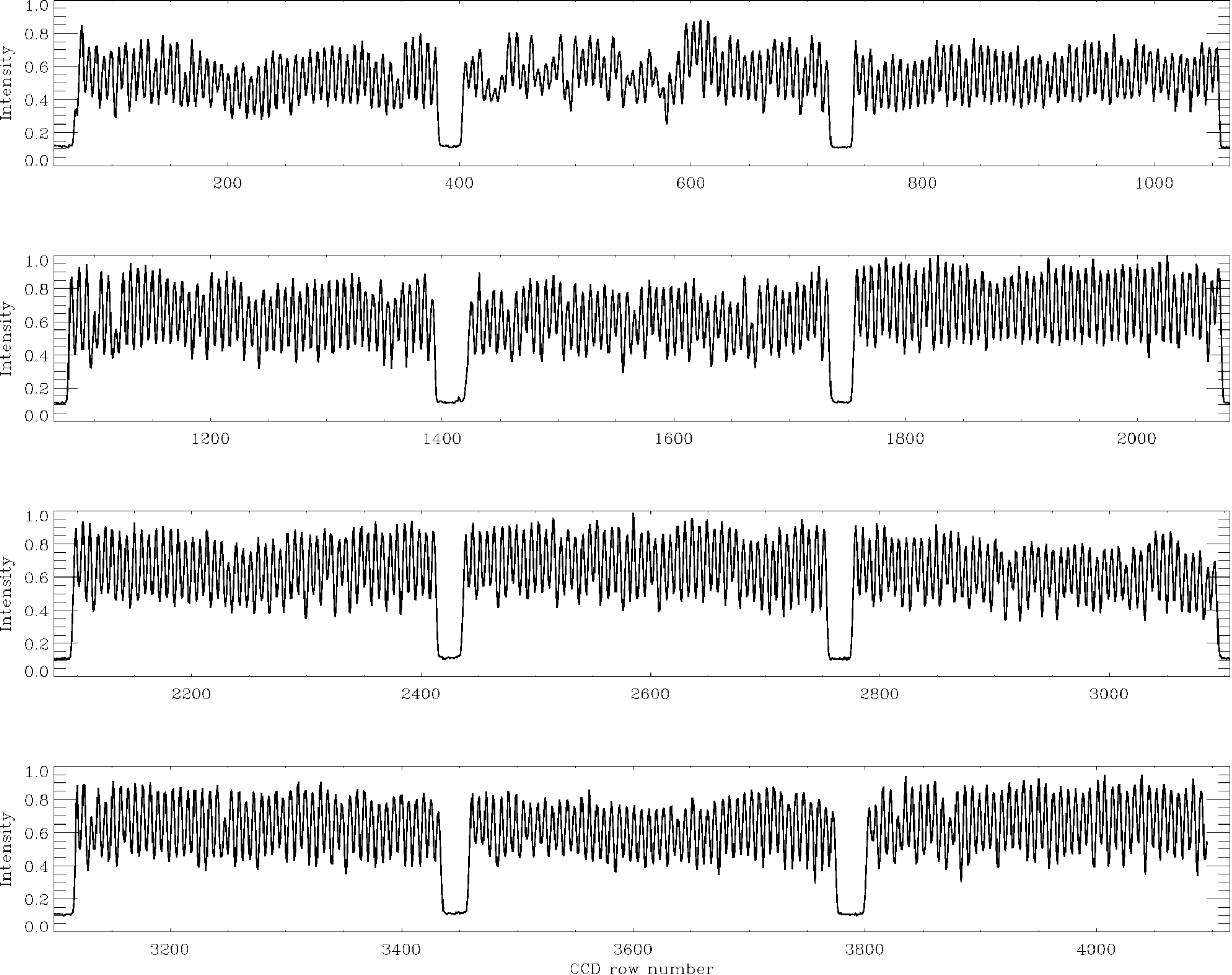}
		\caption{Cross-dispersion cut through a domeflat image illustrating the typical fiber-to-fiber throughput variation.}
		\label{fig:cross-disp}
		\end{centering}
	\end{figure*}

A cross-dispersion cut through a raw dome flat exposure can illustrate the variations in fibre response. Fig.~\ref{fig:cross-disp} shows such a cut through a single CCD chip at approximately 6500~\AA. The plot shows all 12 blocks of 50 fibres, of which 6 blocks are object fibres, and 6 blocks are sky fibres. Each block is separated by approximately 20 pixels that are not directly illuminated. Assuming each peak to be approximately Gaussian, this plot highlights the effect of cross-talk between adjacent spectra, which has to be accounted for along with scattered light, when extracting the spectra.

The IMACS-IFU consists of two fields. One field is centred on the target object, while the other field is located 1 arcmin away from the object field, to collect uncontaminated sky emission only. 

All fibres are arranged in blocks of 50. The blocks along the pseudo-slit then come alternately from the \textit{object} fibres and \textit{sky} fibres to avoid large changes in spectrograph behaviour when the background is to be subtracted. The distribution of the two fields on the detector also follows this pattern, allowing for straightforward sky subtraction via the `mean-sky' method or interpolated sky values.

\subsubsection{Basic Imaging Reduction}

As a starting point, all raw frames were subjected to standard CCD reduction processes. All images were debiassed using the bias strips at the top and side of the frame. In the cross-dispersion direction the bias level was approximately constant in each case. In the dispersion direction, however, some structure was observed, but successfully removed. The overscan regions were then removed from the bias-subtracted images.

Next, a number of flat-field direct-image frames were studied to identify bad pixels and columns on the CCD. These bad pixels were then replaced with the average value of the six neighbouring pixels in the dispersion direction. A further correction was also required for CCD chip number 8, which suffers from a large number of saturated pixels in one area of the CCD. These were also replaced with the average value of the neighbouring pixels in the wavelength direction.

\subsubsection{CCD Mosaicing}
\label{subsec:ccd-mos}

\begin{figure*}
		\begin{centering}
		\includegraphics[scale=0.4]{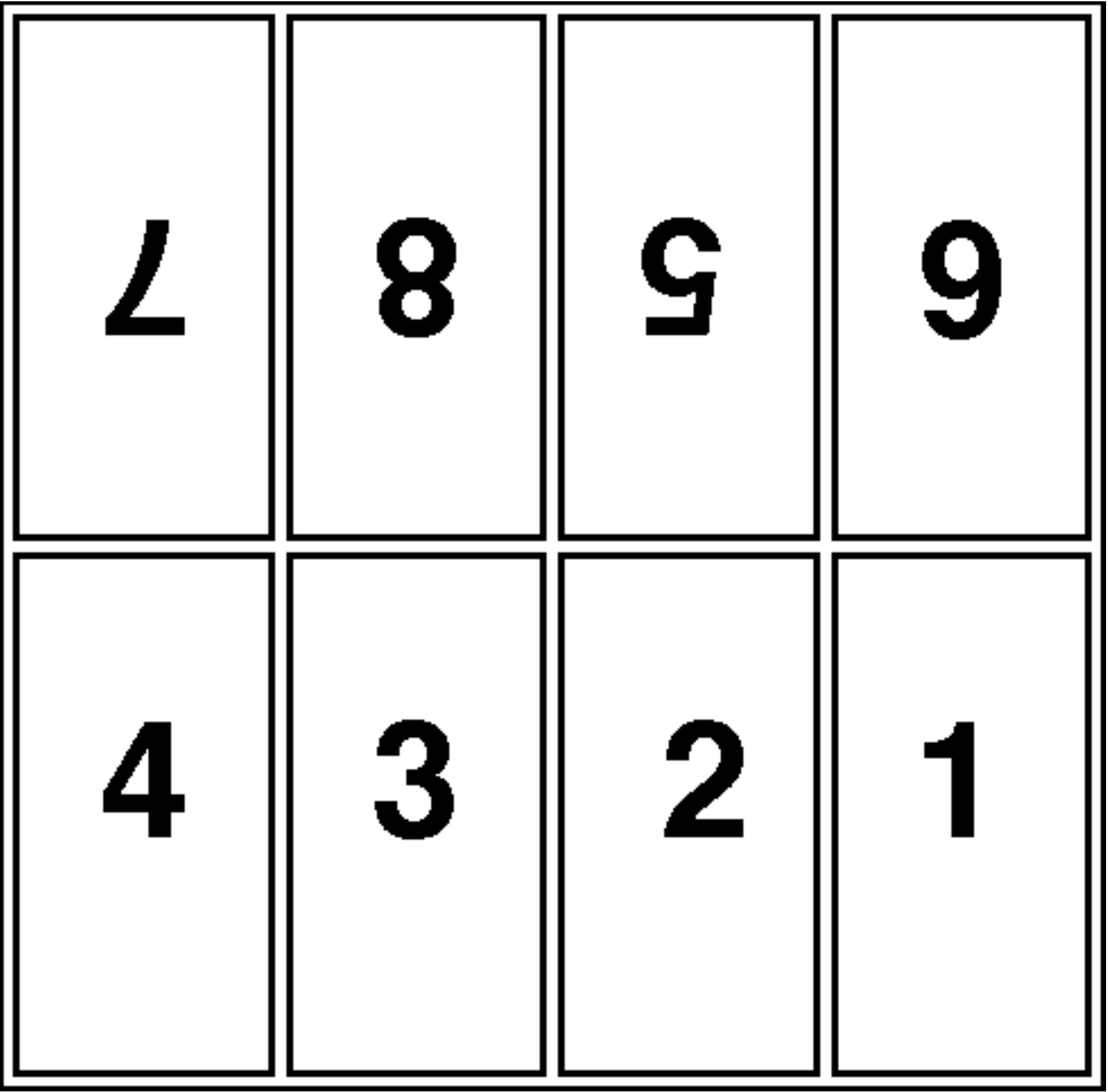}
		\caption{IMACS CCD chip configuration. The dispersion axis is in the `x'-direction, and 
			wavelength increases from left-to-right.}
		\label{fig:ccd-chips}
		\end{centering}
	\end{figure*}

Once the individual CCD frames were corrected for bad pixels they were mosaiced together to form a single image displaying all fibres over the full wavelength range. The configuration of the CCD chip setup is displayed in Fig.~\ref{fig:ccd-chips}. Of importance here is that the `top' row of CCD chips need to be flipped in both the `$x$'- and `$y$'-directions (or alternatively rotated 180\textdegree), to result in wavelength increasing left-to-right. Doing this on a flat-field frame first, however, showed that this process alone does not provide an accurate mosaic, since the illuminated fibres were not aligned at the chip boundaries. To overcome this problem, a cross-correlation of the last column of one CCD chip with the first column of the next CCD chip was performed to derive offsets (in the cross-dispersion direction) between neighbouring CCD chips. The CCD frames were then shifted accordingly to align the fibres over the full wavelength range. These offsets could be applied to correctly align the CCD chips in all subsequent observations for that night.

\subsubsection{Spatial Calibration}

In order to perform an accurate spectral extraction, the location of each spectrum on the CCD needs to be determined. This requires all fibres to be illuminated with a continuum source, so that they can be traced across the whole wavelength range. 

As implied by \S\ref{subsec:ccd-mos} and Fig.~\ref{fig:cross-disp}, the optical fibres are resolved from each other in a domeflat. The location of the spectra can therefore be found for each column on the CCD by comparing the intensity at each row along the column with that of neighbouring pixels in the same column. Each peak in the cross-dispersion direction marks the centre of each fibre. Doing this for each column then traces the fibres across the CCD. 

Creating the `trace masks' is done within the \texttt{imacs\_online} package. The algorithm used to do the trace is described in detail in \S4.1 of \cite{sanch06}, and basically searches pixels along a column, locating the pixels that satisfy a maximum criterion relative to adjacent pixels. The user can specify certain input parameters, such as the distance (in pixels) between adjacent maxima and the number of adjacent pixels to include in the search for maxima. The algorithm also makes it possible to use more than one column to look for the peaks. This is beneficial to avoid cosmic rays and to increase S/N. The number of columns to be used can also be specified by the user.

The tracing of the peak intensity along the dispersion axis is done by searching for more maxima around the original location within a certain window, specified by the user. This is an iterative process, starting in the original column, and continuing across the CCD. The result is a trace mask containing the central locations of each fibre, at each pixel in the dispersion direction. The user specifies the distance between adjacent maxima, and they must also account for the large gaps between the blocks of 50 fibres. A visual inspection of the trace mask superimposed on the domeflat is therefore often required to check that the input parameters are satisfactory, and that the gaps are being dealt with accordingly, as this can be difficult to implement automatically.

Due to the weight of the IFU itself, IMACS-IFU suffers from flexure throughout a night of observing. As a result, a unique trace mask is required for an accurate extraction of each observation. Domeflats were only taken at the start of each night, so fresh trace masks at different stages of the night could not be created. However, by overplotting the trace mask created for a domeflat at the start of the night on a calibration arc frame, it was found that the general pattern of the fibres across the CCD was approximately constant, and that there was only a shift in the cross-dispersion direction, so one set of trace masks could be created for each night, and just shifted (in the `\textit{y}'-direction) accordingly.

Shifting a trace mask for use on a set of arc frames was straightforward, as there are many bright emission lines to see whether the trace mask is passing through the peak emission. In the science frames, however, this is more difficult, as the S/N is much lower, and the locations of the fibres cannot be seen directly. To overcome this problem, the arc frames taken before and after the science frames were used to limit the amount of signal lost in the extraction process.

Our galaxy observations were typically four exposures of 1800~s each. In the majority of cases, the trace mask is stable over the observing time. In the remaining cases the post-observation arc required shifting only a few pixels.

\subsubsection{Extraction of Spectra}

After tracing the location of the spectra on the detector, it is then possible to extract the spectra---i.e., extract the flux corresponding to the 1200 spectra at each pixel along the dispersion axis. This is again done in \texttt{imacs\_online}, and involves co-adding the flux within a certain aperture around the location of the spectral peaks defined by the trace mask, and storing the resulting spectra in a two-dimensional image. The \textit{x}-axis of the resulting image remains as the original dispersion axis, while the \textit{y}-axis translates to the ordering of the spectra along the pseudo-slit -- i.e.,, each row contains a spectrum corresponding to a particular point in the FOV. The extraction works to minimize the effects of cross-talk, while maximizing the recovered flux. This utilises a technique developed for \texttt{P3d}, known as \textit{Gaussian-suppression}, as described in \S4.3 of \cite{sanch06}. 

\begin{figure*}
		\begin{centering}
		\includegraphics[scale=0.7]{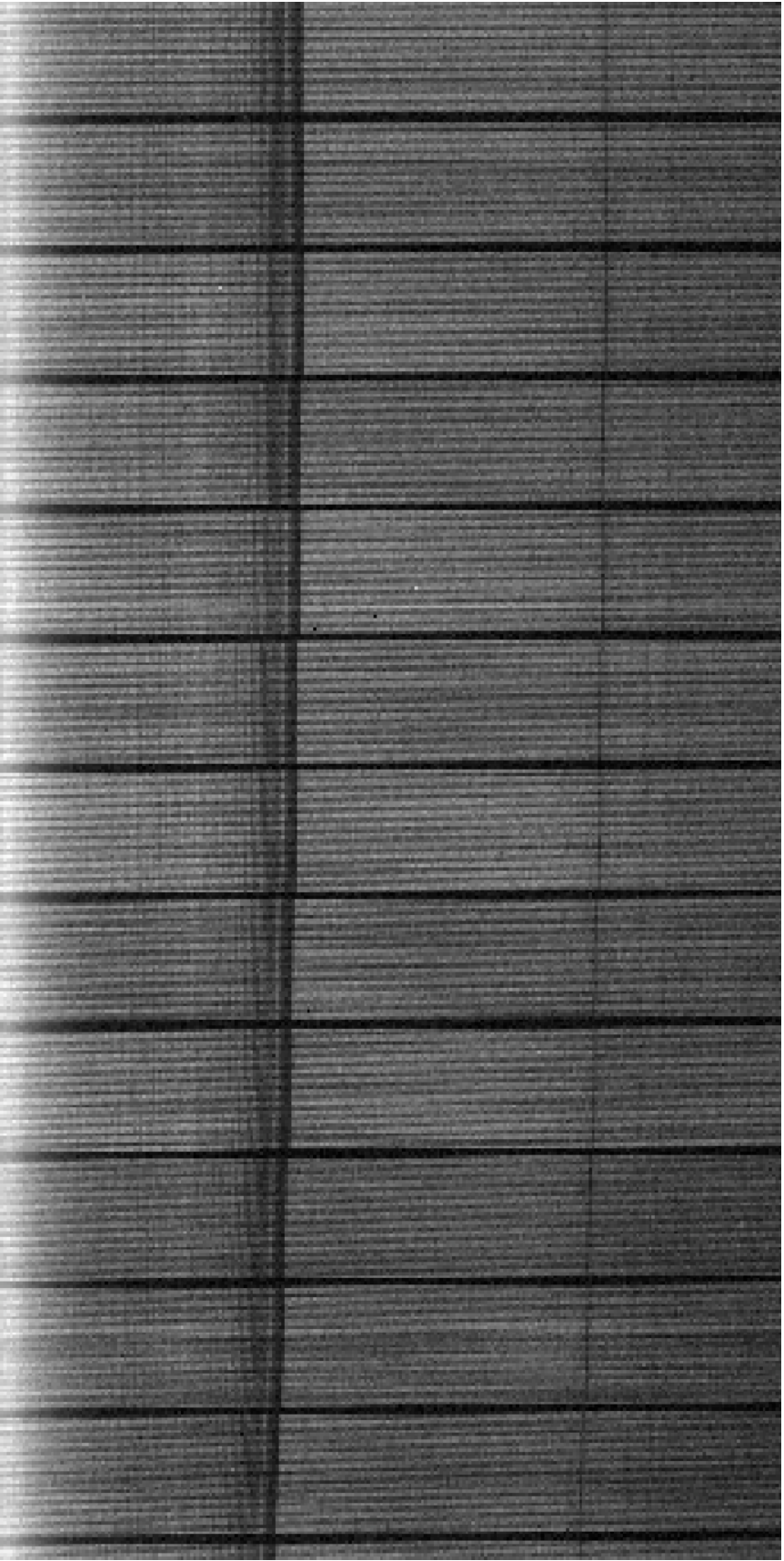}
		\caption{Greyscale image of a sky flat frame prior to spectra extraction.}
		\label{fig:sky}
		\end{centering}
	\end{figure*}

\begin{figure*}
		\begin{centering}
		\includegraphics[scale=0.7]{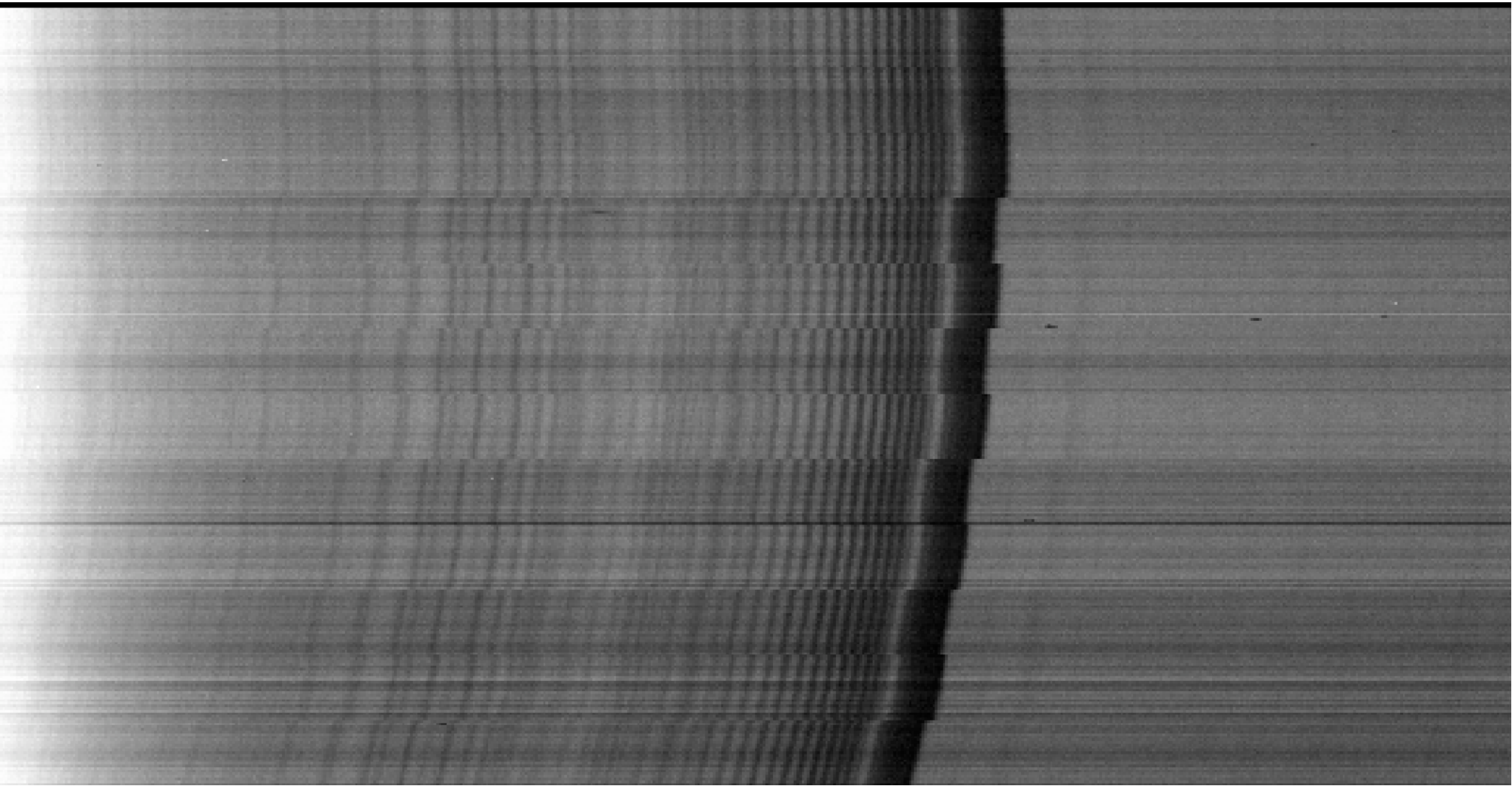}
		\caption{Greyscale image of an extracted sky flat. Each row now represents a unique fibre/spectrum.}
		\label{fig:sky-ext}
		\end{centering}
	\end{figure*}

Fig.~\ref{fig:sky} shows a twilight flat before extraction, while Fig.~\ref{fig:sky-ext} shows the same exposure after spectral extraction. In these frames, emission features are brighter sources, and absorption features appear darker. The separation of the fibre blocks can be seen in the skyflat before extraction, but not in the extracted image.

\subsubsection{Flat-Fielding}

As with CCD imaging, flat-fielding is required to correct for pixel-to-pixel variations. In the dispersion direction, this corresponds to the detector response as a function of wavelength (the quantum efficiency, QE), while in the cross-dispersion direction this translates to fibre-throughput variations. This variation of fibre-to-fibre response, as seen in Fig.~\ref{fig:cross-disp}, can also be seen in the extracted skyflat image of Fig.~\ref{fig:sky-ext}. Due to the effects of flexure it is not possible to flat-field the raw science frames, as the locations of the fibres vary with each telescope pointing. Instead, flat-fielding is done on extracted data, using an extracted domeflat exposure. The extracted domeflat is first normalised to unity and then divided out of all science frames. This procedure not only corrects for QE and fibre-throughput variations but also any relative CCD scalings not corrected during bias subtraction.

\subsubsection{Wavelength Calibration}
\label{subsec:wav-cal}

\begin{figure*}
		\begin{centering}
		\includegraphics[scale=0.2]{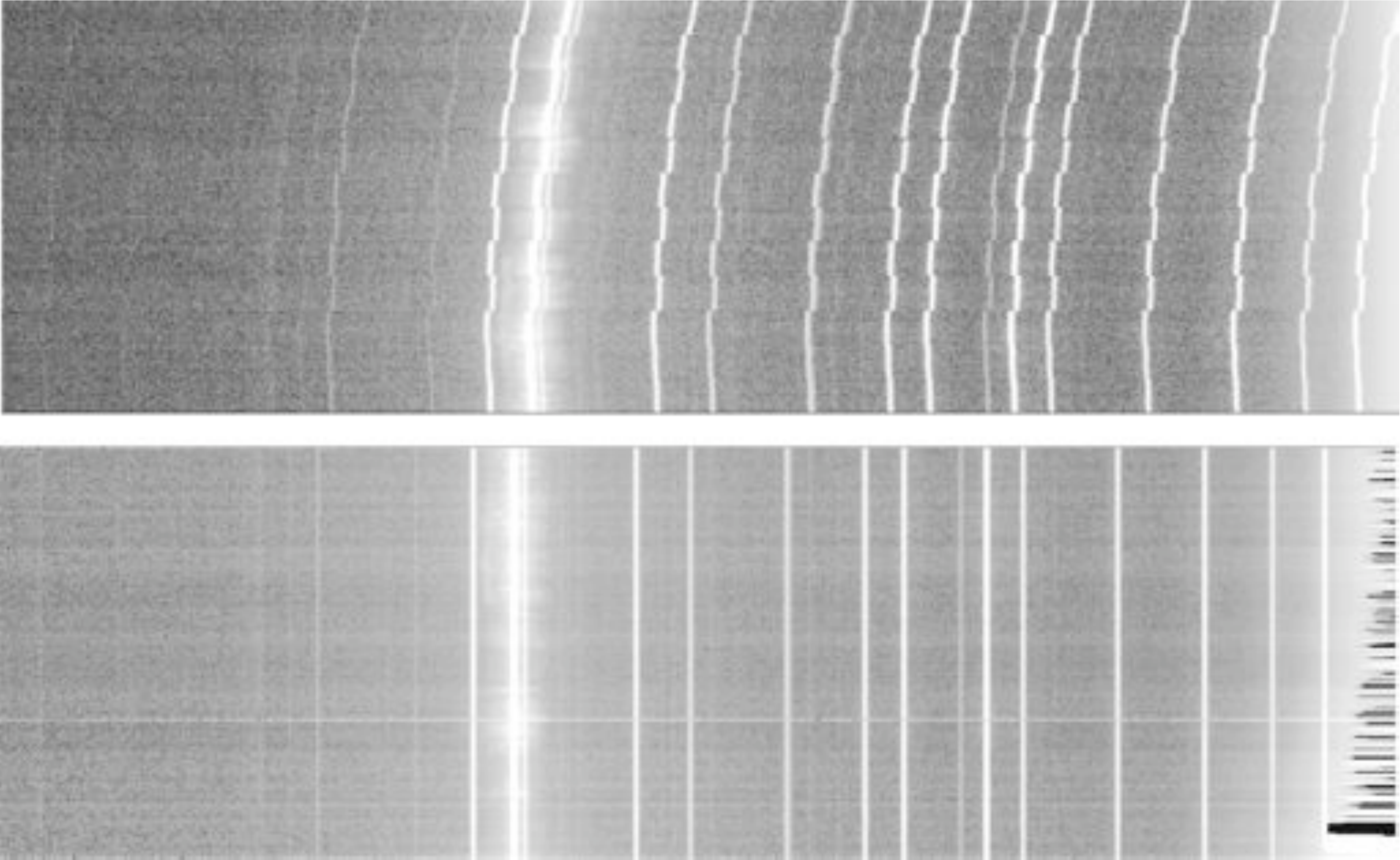}
		\caption{Example of an extracted calibration arc lamp frame before (\textit{upper panel}), 
					and after (\textit{lower panel}) wavelength calibration.}
		\label{fig:calib-arc}
		\end{centering}
	\end{figure*}

Grating spectrographs cause distortions in the entrance slit, leading to the spectral image being curved on the CCD \citep{meab84}. Fibre-fed integral-field units suffer from additional distortions due to the placing of the fibres at the slit. The upper panel of Fig.~\ref{fig:calib-arc} shows both of these effects. The overall curvature can be seen in the cross-dispersion direction, and there are also varying shifts in the dispersion direction. As a result, these distortions must be corrected fibre-to-fibre before deriving a common wavelength solution.

He--Ne--Ar arc calibration frames were therefore first used to correct for these distortions, and then used to determine the wavelength solution. The relative fibre-to-fibre offsets in the dispersion direction were found by tracing the location of the peak of one arc emission-line along the cross-dispersion axis. The fibre-to-fibre offsets are then applied in order to shift the lines to a common reference. 

The wavelength solution is determined by the identification of a number of known arc emission lines. The corrected arc spectra are then transformed to a linear wavelength coordinate system, by assuming a polynomial transformation. This transformation is then stored in an ASCII file to be applied to the science frames. The accuracy of the dispersion solution depends on the selected order of the polynomial 
transformation (typically chosen to be three or four), the number of identified arc lines, and the coverage of these lines across the wavelength range.

The lower panel of Fig.~\ref{fig:calib-arc} shows the wavelength calibrated arc frame: wavelength increases from left-to-right. These images have also been flat-fielded, so the bright line approximately a third of the way up this image is a broken fibre. In the dispersion direction, as many as 50 pixels per CCD chip were lost in wavelength calibration as a result of the image being curved on the CCD.

\subsubsection{Cosmic-ray Rejection}

Cosmic-ray rejection was of critical importance for our data set. Science exposures were of the order of 1800~s resulting in a large number of cosmic rays. In IFS, care needs to be taken in removing cosmic rays, as emission lines can easily be mistaken for cosmic rays. The most effective method of cosmic-ray rejection was found to be after the spectra were extracted, by comparing adjacent spectra \citep{swin03}. 

\begin{figure*}
		\begin{centering}
			\includegraphics[scale=0.45]{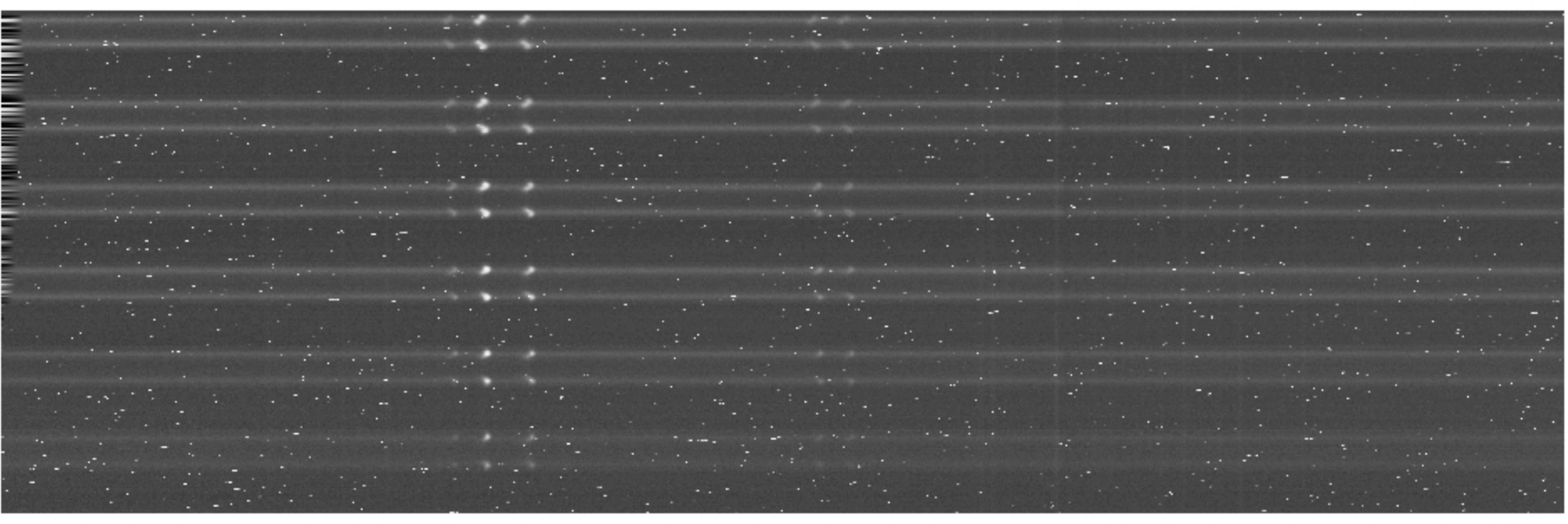} \\
			\vspace{4mm}
			\includegraphics[scale=0.45]{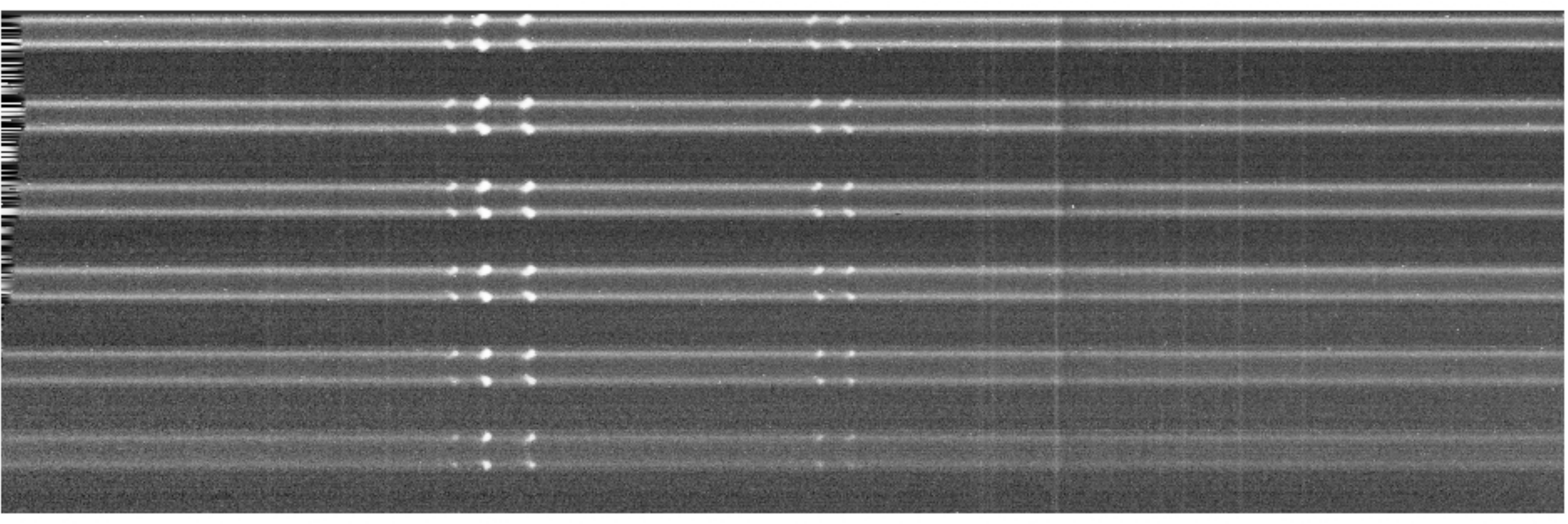} \\
			\vspace{4mm}
			\includegraphics[scale=0.45]{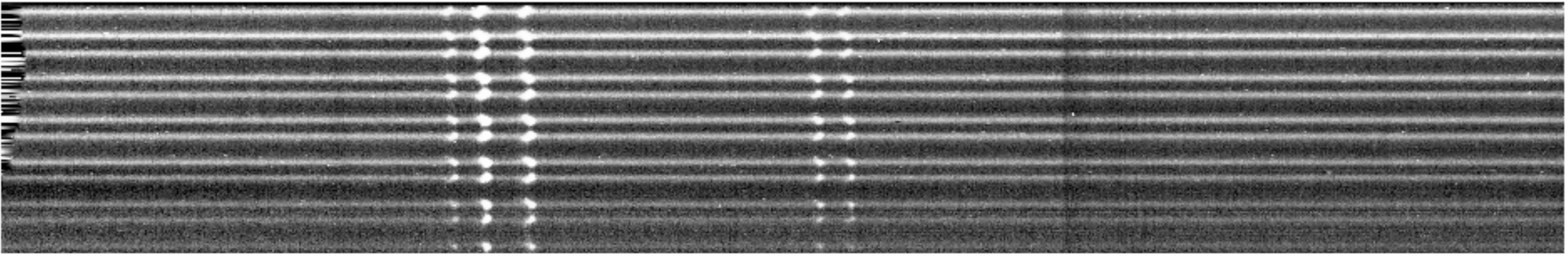} \\
			\caption{\textit{Top panel}: Example of an extracted, wavelength calibrated science frame. 
				\textit{Middle panel}: Science frame following cosmic-ray rejection. 
				\textit{Bottom panel}: The sky-subtracted science frame.}
			\label{fig:CRR}
		\end{centering}
	\end{figure*}

Fig.~\ref{fig:CRR} shows an example of a science frame before and after cosmic-ray rejection. This example is one CCD frame of the Seyfert 2 galaxy NGC\,5740, in the wavelength range $6375--7100$~\AA. This frame also demonstrates the ordering of the fibres on the detector---i.e., in a sequence of 50 object spectra followed by 50 sky spectra followed by 50 object spectra, and so forth. The peaks in the object spectra correspond to emission lines in the galaxy. The lines that can be seen are \nii\ (doublet), \ha\ and \sii\ (doublet). Sky lines can also be seen as the emission that appears in both the object fibres and sky fibres, illustrating the importance of sky subtraction.

\subsubsection{Sky Subtraction}

On the IMACS CCD, object and sky fibres are distributed in blocks of 50 in the cross-dispersion direction with the intention of allowing simple sky subtraction via the mean-sky method (i.e., for each wavelength slice (column), subtract the average of the 50 pixels in one block of sky fibres from the adjacent block of object fibres.

When using this method, however, the sky was found to be systematically over- then under-subtracted in each neighbouring block, resulting in a `striping' effect in the reconstructed two-dimensional images. This is in part due to the fact that the sky-background in the sky-fibres is not guaranteed to be exactly representative of the background in the object fibres. In addition, scattered light in the optics, estimated to be of the order of 1--2\% by \cite{dress11}, contributes to a more varying background.

A number of different ways to estimate the sky background at the location of the object fibres were tried, but the most robust method was to interpolate across all sky fibres, as the sky was found to be brighter in the fibres towards the centre of the CCD compared to the outermost fibres. A model of the sky in the cross-dispersion direction was therefore derived for each wavelength slice, to produce a modelled sky field over all wavelengths. The sky was then smoothed over a few pixels in the wavelength direction, and subtracted from the science frame. An example of a sky-subtracted frame is also shown in Fig.~\ref{fig:CRR}.

\subsubsection{Image Stacking}

As our science observations were multiple exposures of 1800~s, they need to be stacked to form one, combined image. A number of our observations also made use of dithering, and as a result multiple exposures could not be simply added together, as the target would fall on different fibres in each different exposure. Typically we dithered by \das04, which is twice the spatial resolution of the IFU. The spatial offsets can be seen when viewing the reconstructed two-dimensional images. We therefore transformed the two-dimensional frames into datacubes using the mapping information given in \cite{schmoll04}. 

The result is a $25 \times 24$ pixel image for each wavelength slice. Given the shape of the lenslets (hexagonal), and the number of pixels in a two-dimensional image, there was a significant error in terms of finding the galaxy centres. We therefore oversampled in both the `\textit{x}'- and `\textit{y}'-directions to transform to a $50 \times 48$ square grid of \das01 $\times$ \das01 pixels, for each wavelength slice. 

\begin{figure}
		\begin{centering}
		\includegraphics[scale=0.24]{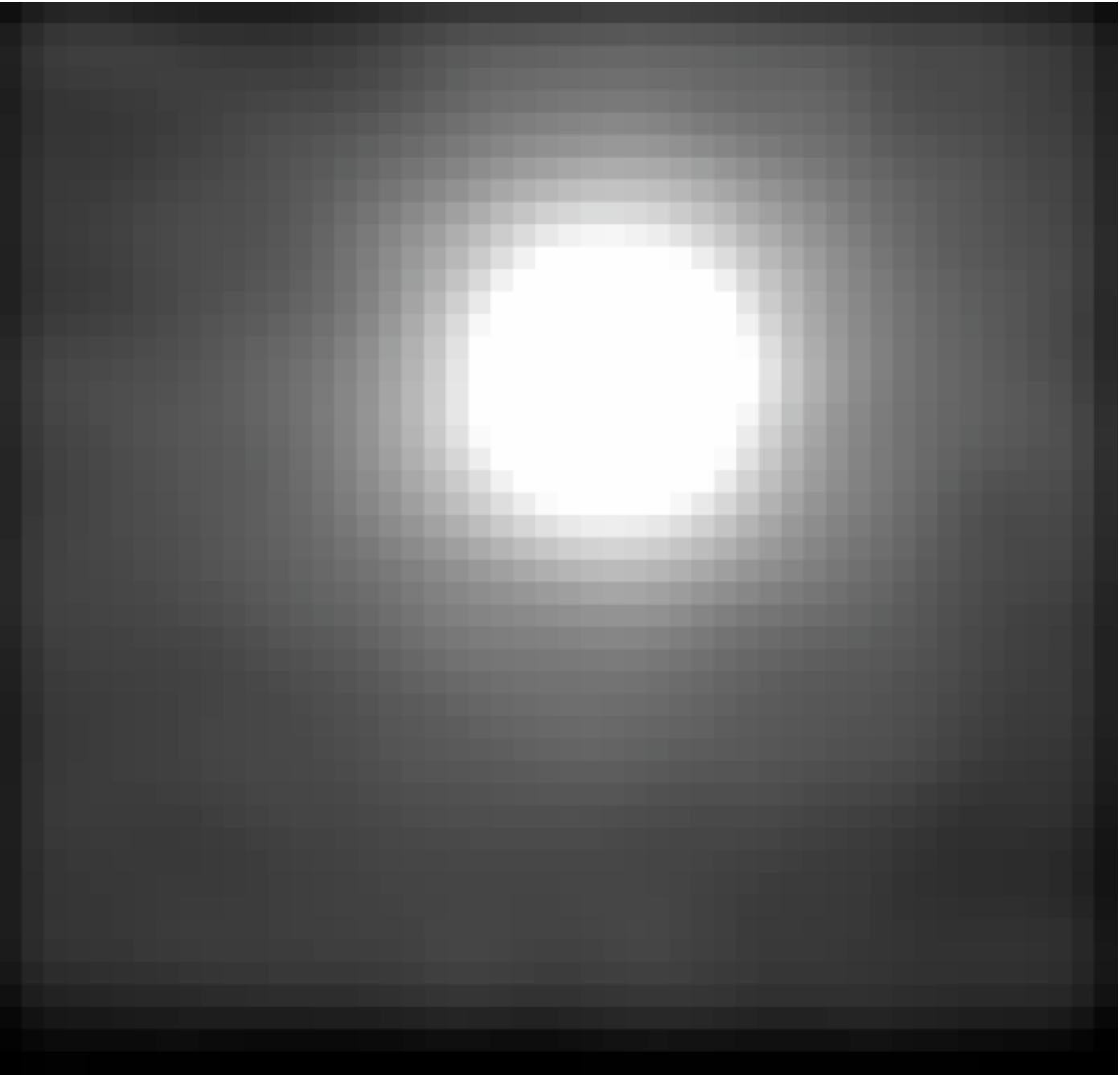}
		\hspace{4mm}
		\includegraphics[scale=0.24]{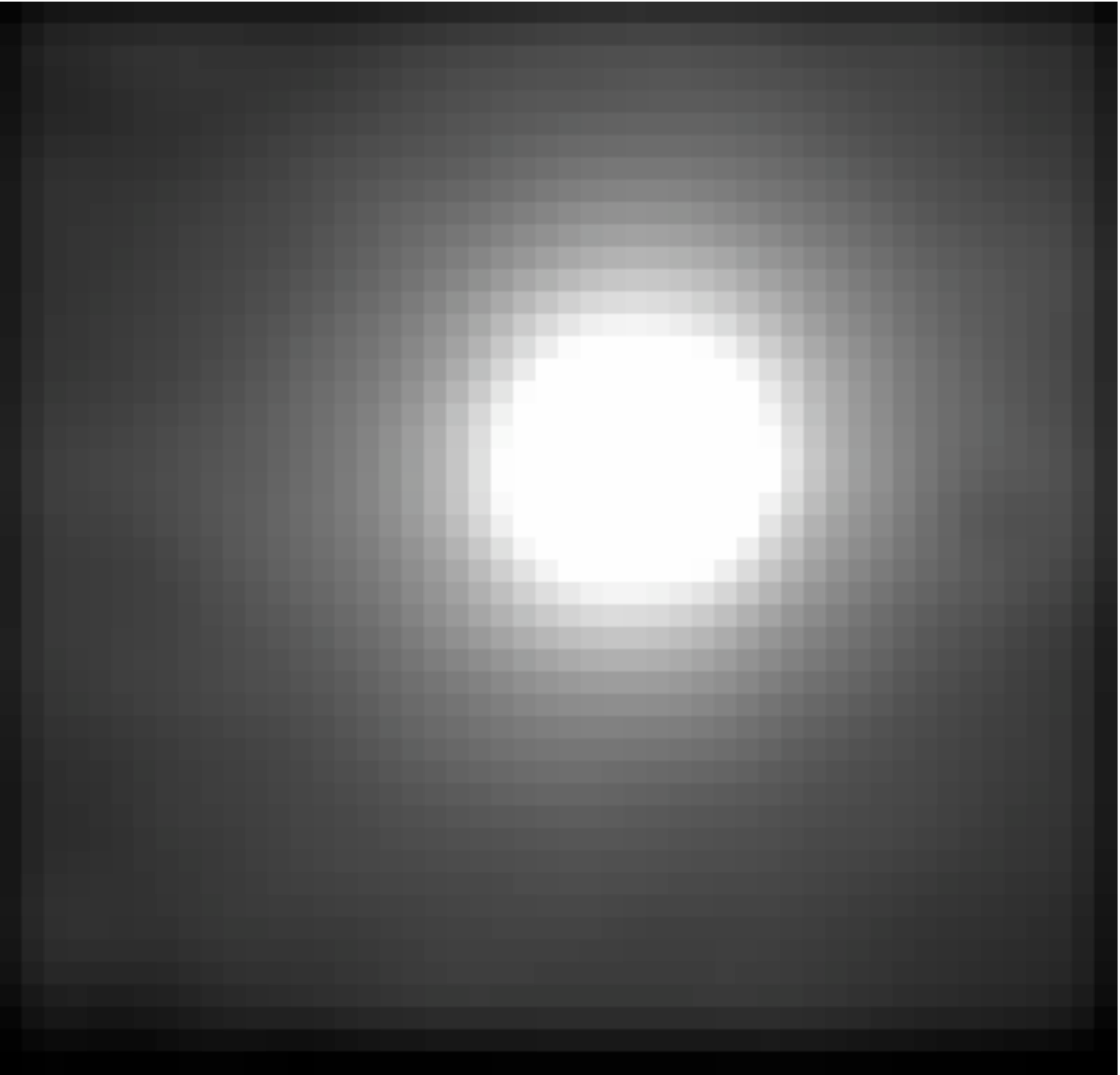}
		\caption{Re-constructed 2d images of 2 exposures of UGC~05226.}
		\label{fig:2d-im}
		\end{centering}
	\end{figure}

The effects of dithering can easily be seen by viewing different exposures at a common wavelength, or sum of wavelengths (Fig.~\ref{fig:2d-im}). Summing pixel-to-pixel over the full wavelength range can provide a high-S/N image in which a `photometric' centre can be found through profile fitting. The multiple images were then shifted so that the photometric peaks were aligned in all frames. The exposures were normalised to one second, and a median datacube created. At this point, our pipeline also allows for a weighting term to be applied to each data set, to account for seeing conditions, for example. In this paper, however, seeing conditions remained approximately constant through a set of galaxy exposures, so all weights were set equal to one.

\subsubsection{Flux Calibration}

\begin{figure}
		\begin{centering}
		\includegraphics[scale=0.32, angle=90]{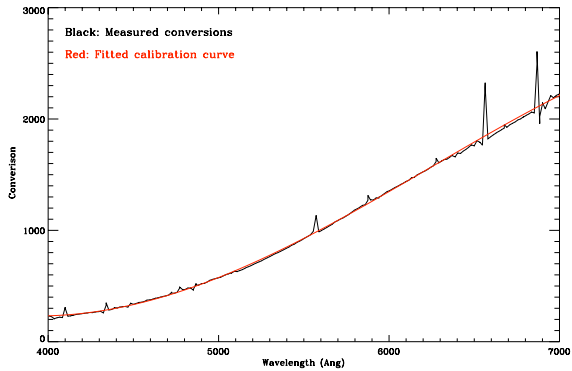}
		\caption{Example of calibration curve derived for the standard star HR3454.}
		\label{fig:cal-curve}
		\end{centering}
	\end{figure}

The final element of the reduction pipeline is spectroscopic flux calibration. A number of standard stars were observed during each observing run and reduced as described above. An integrated spectrum of a standard star was then created by summing all spectra in an aperture covering the whole star---typically greater than 3\as. An absolute flux calibration was then found using the spectrophotometric flux calibration tables of \cite{hamuy92}. The integrated IMACS spectrum is divided by the standard spectrum, to provide a table of conversions as a function of wavelength. Absorption features were masked out, and a third-order polynomial fit to the conversion data, to provide an absolute calibration curve, such as that shown in Fig.~\ref{fig:cal-curve}, which can be applied directly to the science observations to convert to an $F_{\lambda}$ flux scale.

\subsubsection{Spatial Binning}

It is very common for IFU data to be locally averaged to maximise the S/N, at the expense of spatial resolution. Averaging is generally done by either \textit{smoothing} or \textit{binning}. The Voronoi two-dimensional-binning method of \cite{capp03} performs adaptive spatial binning of two-dimensional data to reach a constant S/N per bin. To maximise the possibility of extracting the stellar kinematics from our data, all datacubes were rebinned using the Voronoi method. This method required a noise datacube, for which used the non-sky-subtracted datacube. A reconstructed two-dimensional continuum image and a reconstructed two-dimensional noise image were then created using the full wavelength range, and excluding emission-lines, and two-dimensional binning performed on the resulting image, to achieve a constant S/N of 60~\AA$^{-1}$~pixel$^{-1}$. Each wavelength slice in the datacube was then binned according to these results to maximise the probability of deriving reliable stellar kinematics.

\subsection{Derivation of Stellar and Gaseous Distributions and Kinematics}

To derive the stellar kinematics, we used the maximum penalized likelihood formalism developed by \cite{capp04}. The \textit{penalized pixel-fitting} method (pPXF), treats the line-of-sight velocity distribution (LOSVD) as a Gauss-Hermite series, and fits the kinematic parameters ($V_{*}$, $\sigma_{*}$, $h_{3}$, ..., $h_{n}$) simultaneously, while adding an adjustable penalty term to the $\chi^{2}$. An advantage of this method is that emission lines and bad pixels can be excluded from the fit---a property that is of particular importance when fitting to Seyfert spectra. 

The pPXF algorithm finds the best fit to all galaxy spectra by convolving a template stellar spectrum with the corresponding LOSVD. The success of pPXF is therefore sensitive to the input stellar templates. An optimal template is produced for each spectrum in the datacube, thus the input stellar spectra must cover a range of metallicities and ages, to account for any potential variation in the properties across the field of view. Libraries of observed or synthetic stellar spectra are therefore often used as the input stellar templates. Previous studies have shown, however, that reliable stellar kinematics can be successfully derived over the spectral range $4800-5400$~\AA\ using the Mg\,$b$ doublet and \feii\ absorption features. Although IMACS-IFU has a much broader wavelength coverage, the continuum fit is very sensitive to slight inaccuracies in CCD relative normalisation. As a result, we restricted pPXF to the range $4800-5400$~\AA. An initial pPXF fit was performed on a median galaxy spectrum to select a subset of the input stellar spectra, which was then used to derive the kinematics for the whole datacube.

A number of libraries were available, such as the simple stellar population (SSP) models of \citealt{vazd99} (hereafter V99), the Indo-U.S. Coud{\'e} Feed Spectral Library \citep{valdes04}, the Medium-resolution Isaac Newton Telescope library of empirical spectra (MILES; \citealt{sanchblazq06}), and an IMACS-IFU library of observed velocity standards. For comparison with previous work (e.g., SAURON; \citealt{sarzi06,dumas07}), the stellar kinematics presented in this paper are those derived from the V99 synthetic stellar library. The V99 library covers the spectral range $4800--5400$~\AA\ with a resolution of 1.8~\AA\, which is well-matched to the IMACS-IFU spectra (FWHM$_{IMACS} \sim 1.6$~\AA\ at $5000$~\AA). The observed Indo-U.S. Library, of higher spectral resolution (FWHM 1.0~\AA), was also tested and found to be in good agreement with the kinematics derived using the V99 stellar models.

To estimate the errors in recovering the LOSVD from the optimal template, Monte Carlo simulations were performed to create synthetic spectra by adding Poisson noise to the template spectra, and fitting for ($V_{*}$,$\sigma_{*}$). The resulting rms scatter was found to be of the order of a few percent. The scatter in the recovered parameters, however, puts only a lower limit on the uncertainty in deriving the kinematics from a given template, as it does not include the more dominant uncertainty due to template mismatch. Errors are therefore considered to be in the range 5--10\%.

The Gas AND Absorption Line Fitting (GANDALF) algorithm of \cite{sarzi06}, extended pPXF to also derive the gas kinematics from the emission lines. Assuming the emission lines to be Gaussian, the line strength, mean position (velocity) and width (velocity dispersion), can be calculated. GANDALF was performed simultaneous to pPXF, so was also restricted to the wavelength range of approximately $4800--5400$~\AA, thus constraining only the \hb, \oiii\, and \none\ emission-line kinematics. Beyond the operating range of GANDALF, however, there is still a wealth of data. In particular, in the range $6000--7000$~\AA, there lies \nii, \ha\ and \sii\ emission lines. Kinematics from \ha, \nii\ and \sii\ emission were therefore subsequently derived independent of GANDALF, through single-Gaussian fitting. 

In the range where the stellar kinematics were extracted, the optimal stellar template derived from pPXF was subtracted from the observed spectrum, removing contamination from any underlying stellar absorption---this is of particular importance for recoving the true \hb\ kinematics as this line is expected to be a blend of emission and absorption that cannot be assumed to have the same kinematic profiles. The spectrum in the range 6000--7000\AA\ lacks stellar absorption features---analogous to Mg~\textit{b} or \feii---to allow an independent estimate of any \ha\ stellar absorption, so the line intensity relative to the local continuum was used, as is standard in narrowband imaging. The total \ha\ flux density may therefore be an underestimate (e.g., \citealt{charl02}).

The best amplitude, mean velocity and velocity dispersion, were therefore extracted for all the emission lines in each spectrum. The velocity dispersions were corrected for the instrumental resolution, which were estimated to be 35~\kms\ at $\sim$6500~\AA\ and 42~\kms\ at $\sim$5000~\AA.

\section{Observed IMACS-IFU Kinematics}

\begin{figure}
		\begin{centering}
		\includegraphics[scale=0.88]{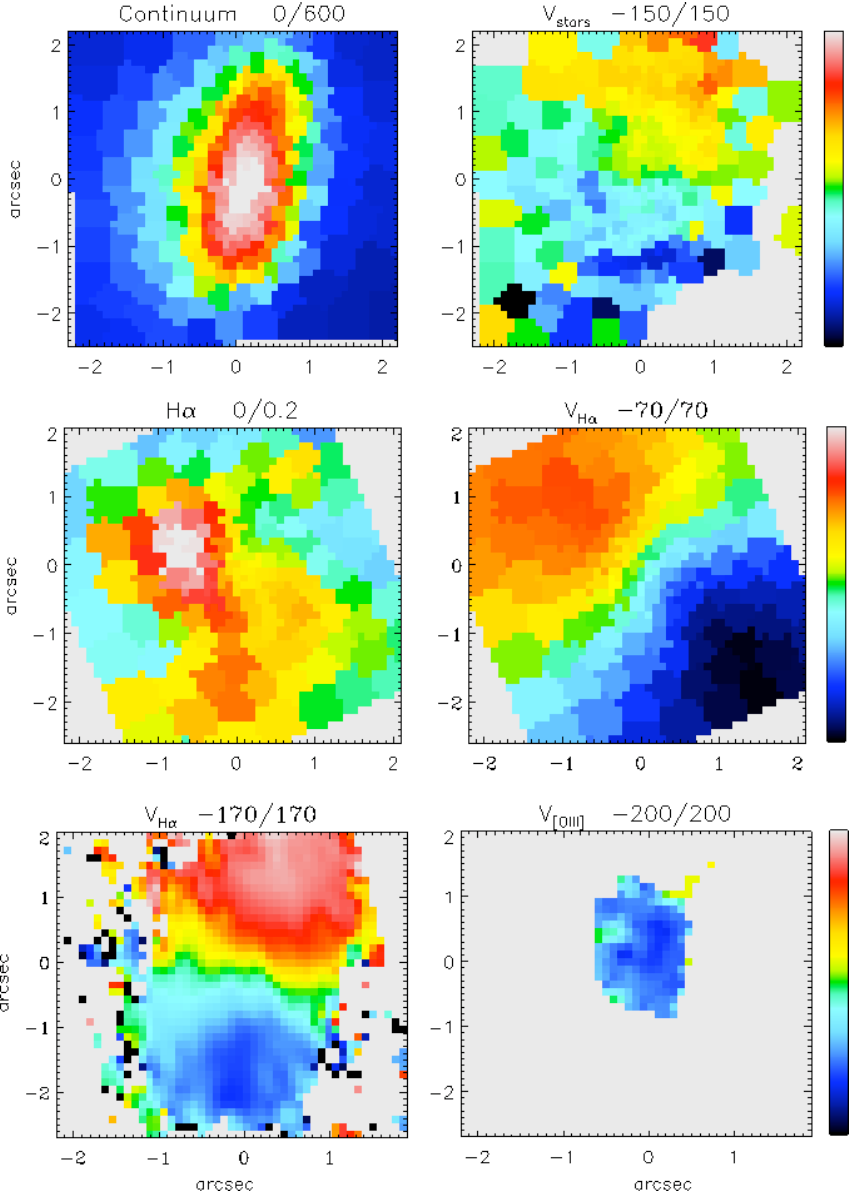}
		\caption{Examples of the features exhibited in the IMACS-IFU kinematic maps. 
				\textit{Top row}: Photometric nuclear bar revealed by the continuum distribution (left) 
					and the corresponding stellar rotation field (right) for the Seyfert galaxy NGC~5740.
				\textit{Middle row}: \ha\ distribution (left) and velocity field (right) of the control galaxy Mrk~1404, revealing a 
					complex star-formation pattern undergoing uniform rotation.
				\textit{Bottom row}: Gas velocity fields for the Seyfert~2 galaxy SDSS~J033955.68--063237.5. 
					The \ha\ velocity field (left) suggests a uniform rotating gas disk, while the \oiii\ velocity field (right) 
					suggests the presence of an outflow component.
				}
		\label{fig:map-egs}
		\end{centering}
	\end{figure}

This section is dedicated to the presentation of the IMACS-IFU moment maps. A variety of structures are revealed in the IMACS kinematic maps, examples of which are shown in Fig.~\ref{fig:map-egs}. Features include: elongated stellar continuum light profiles suggestive of \textbf{photometric nuclear bars} (top left in Fig.~\ref{fig:map-egs}); \textbf{complex star formation regions} (centre left); \textbf{stellar and gas rotation} (top right, middle right and bottom left); highly ionised gas with significantly blue-shifted velocities with respect to systemic, indicative of \textbf{outflow} (bottom right); and other distorted velocity structures. 

\begin{figure*}
	\begin{centering}
	\includegraphics[scale=0.9]{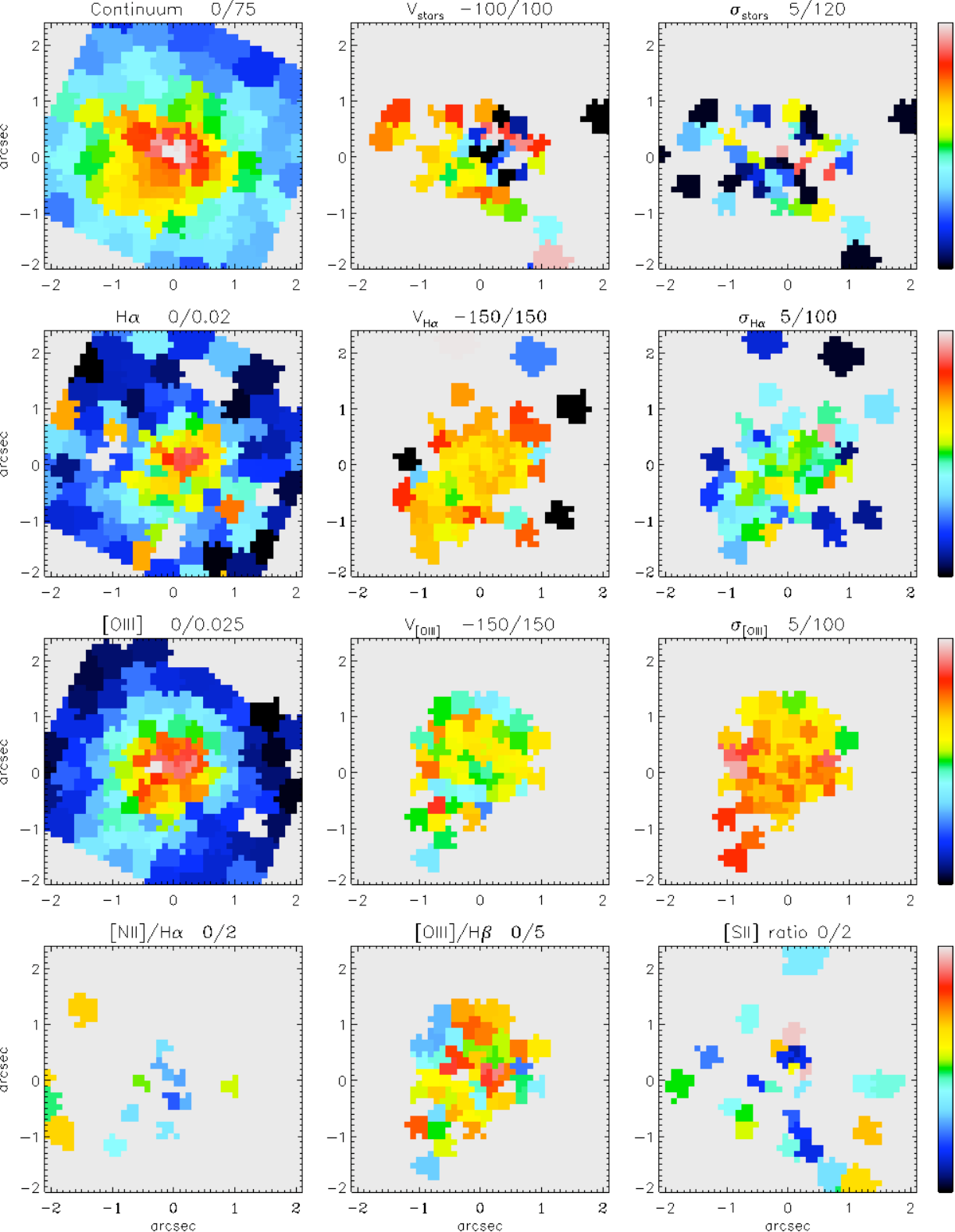}
	\caption{IMACS-IFU maps for active galaxy SDSS~J023311.04-074800.8. Cosmology corrected scale: 581 pc/arcsec. 
		\textit{Top row, from left-to-right}: Reconstructed continuum image; mean stellar velocity, 
			$V_{stars}$; stellar velocity dispersion, $\sigma_{stars}$.
		\textit{Second row}: \ha\ distribution; \ha\ velocity; \ha\ velocity dispersion.
		\textit{Third row}: \oiii\ distribution; \oiii\ velocity; \oiii\ velocity dispersion.
		\textit{Bottom row}: \nii/\ha\ ratio map; \oiii/\hb\ ratio; \siil/\siib\ ratio. Fluxes are in units of $F_{\lambda}$ ($10^{-17}$ erg~cm$^{-2}$~s$^{-1}$~\AA$^{-1}$), and velocities given in \kms.} 
	\label{fig:Maps-J023311}
	\end{centering}
\end{figure*}

\begin{figure*}
	\begin{centering}
	\includegraphics[scale=1.110]{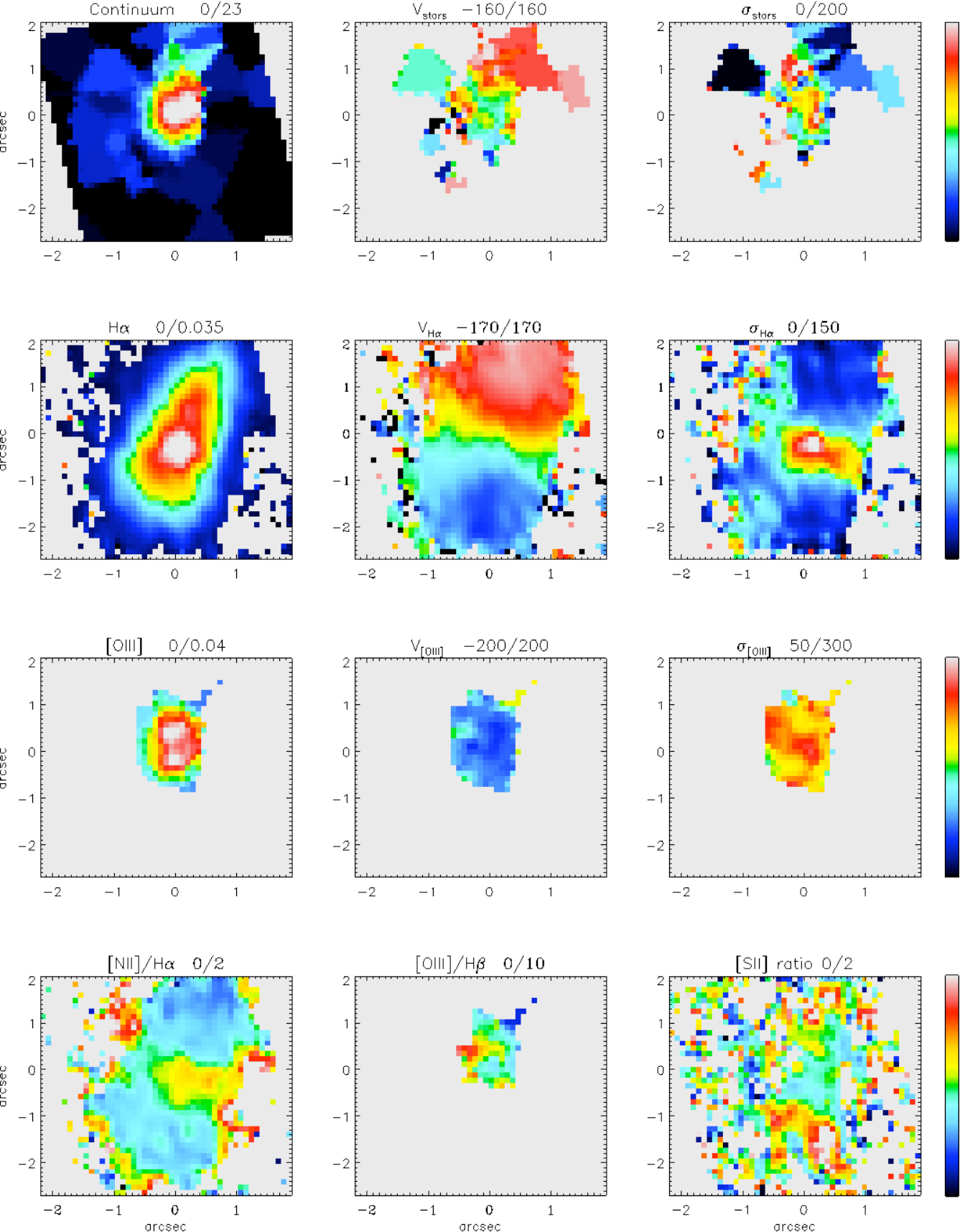}
	\caption{IMACS-IFU maps for active galaxy SDSS~J033955.68--063237.5. Cosmology corrected scale: 592 pc/arcsec. 
			See Fig.~\ref{fig:Maps-J023311} for description of maps plotted.}
	\label{fig:Maps-J033955}
	\end{centering}
\end{figure*}

\begin{figure*}
	\begin{centering}
	\includegraphics[scale=1.110]{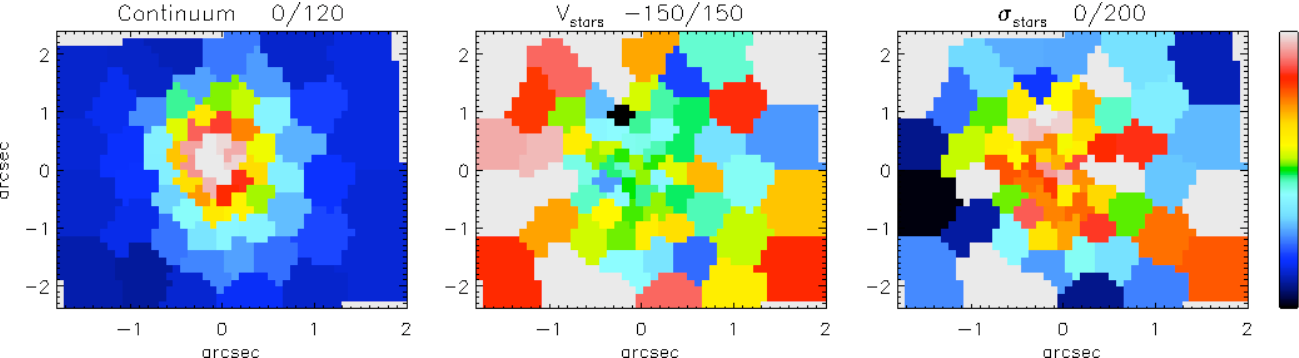}
	\caption{IMACS-IFU maps for control galaxy SDSS~J082323.42+042349.9. Cosmology corrected scale: 604 pc/arcsec. 
		\textit{From left-to-right}: Reconstructed continuum image; mean stellar velocity, $V_{stars}$; stellar velocity dispersion, $\sigma_{stars}$, in \kms. Fluxes are in units of $F_{\lambda}$ ($10^{-17}$ erg~cm$^{-2}$~s$^{-1}$~\AA$^{-1}$), and velocities given in \kms.} 

	\label{fig:Maps-J082323}
	\end{centering}
\end{figure*}

\begin{figure*}
	\begin{centering}
	\includegraphics[scale=1.110]{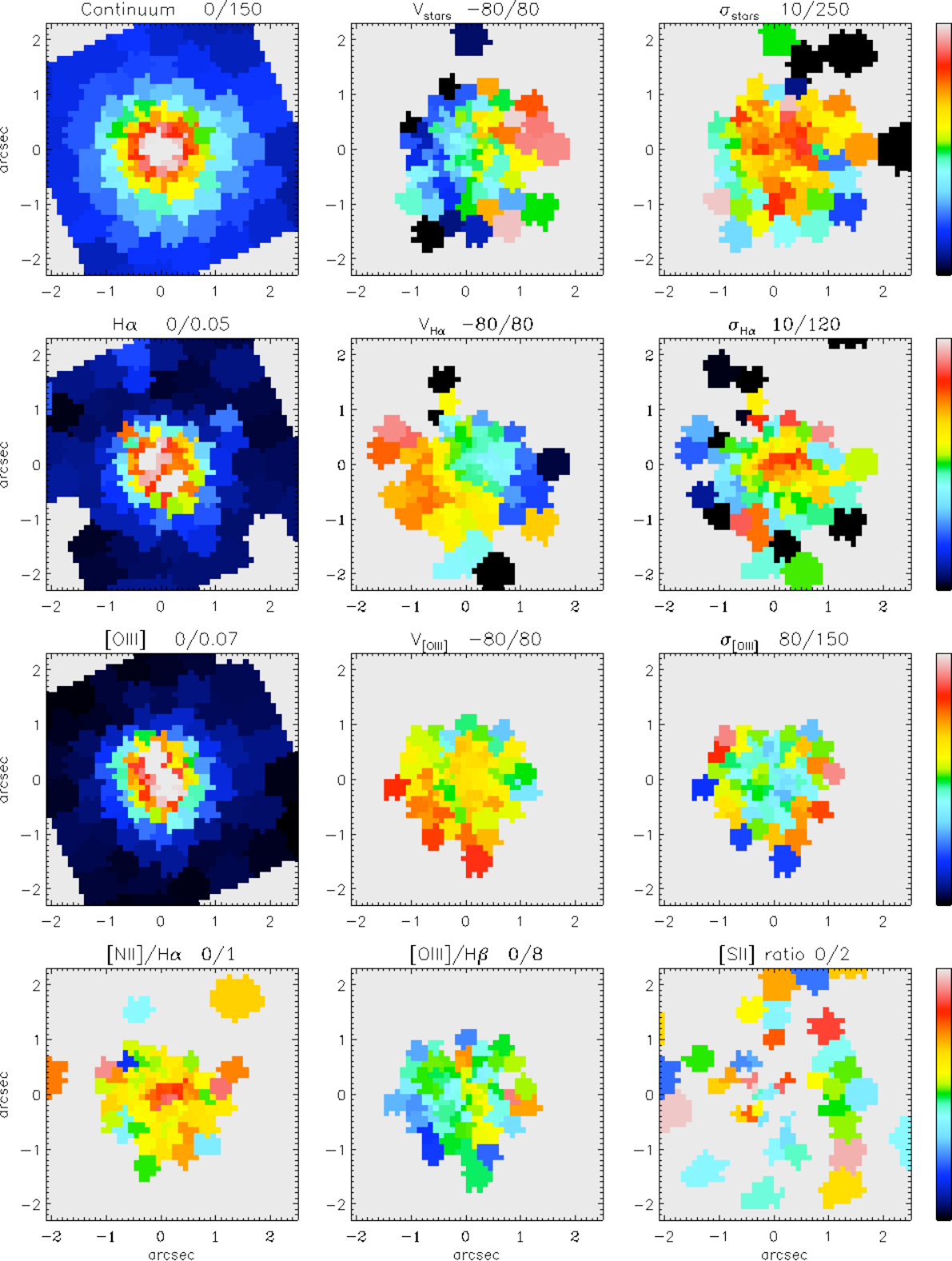}
	\caption{IMACS-IFU maps for active galaxy SDSS~J024440.23--090742.4. Cosmology corrected scale: 446 pc/arcsec. 
			See Fig.~\ref{fig:Maps-J023311} for description of maps plotted.}
	\label{fig:Maps-J024440}
	\end{centering}
\end{figure*}

\begin{figure*}
	\begin{centering}
	\includegraphics[scale=1.110]{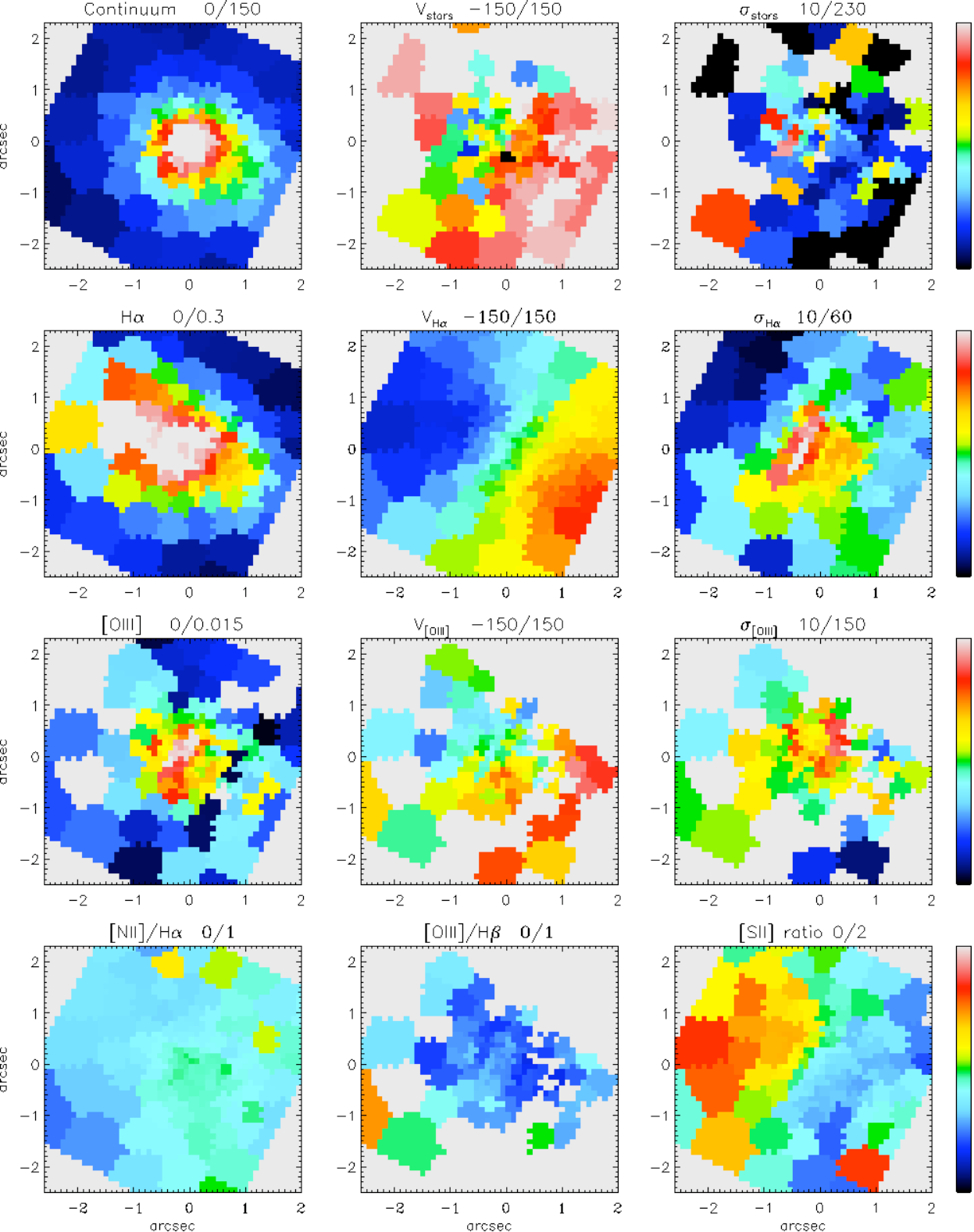}
	\caption{IMACS-IFU maps for control galaxy SDSS~J015536.83--002329.4. Cosmology corrected scale: 418 pc/arcsec. 
			See Fig.~\ref{fig:Maps-J023311} for description of maps plotted.}
	\label{fig:Maps-J015536}
	\end{centering}
\end{figure*}

\begin{figure*}
	\begin{centering}
	\includegraphics[scale=1.110]{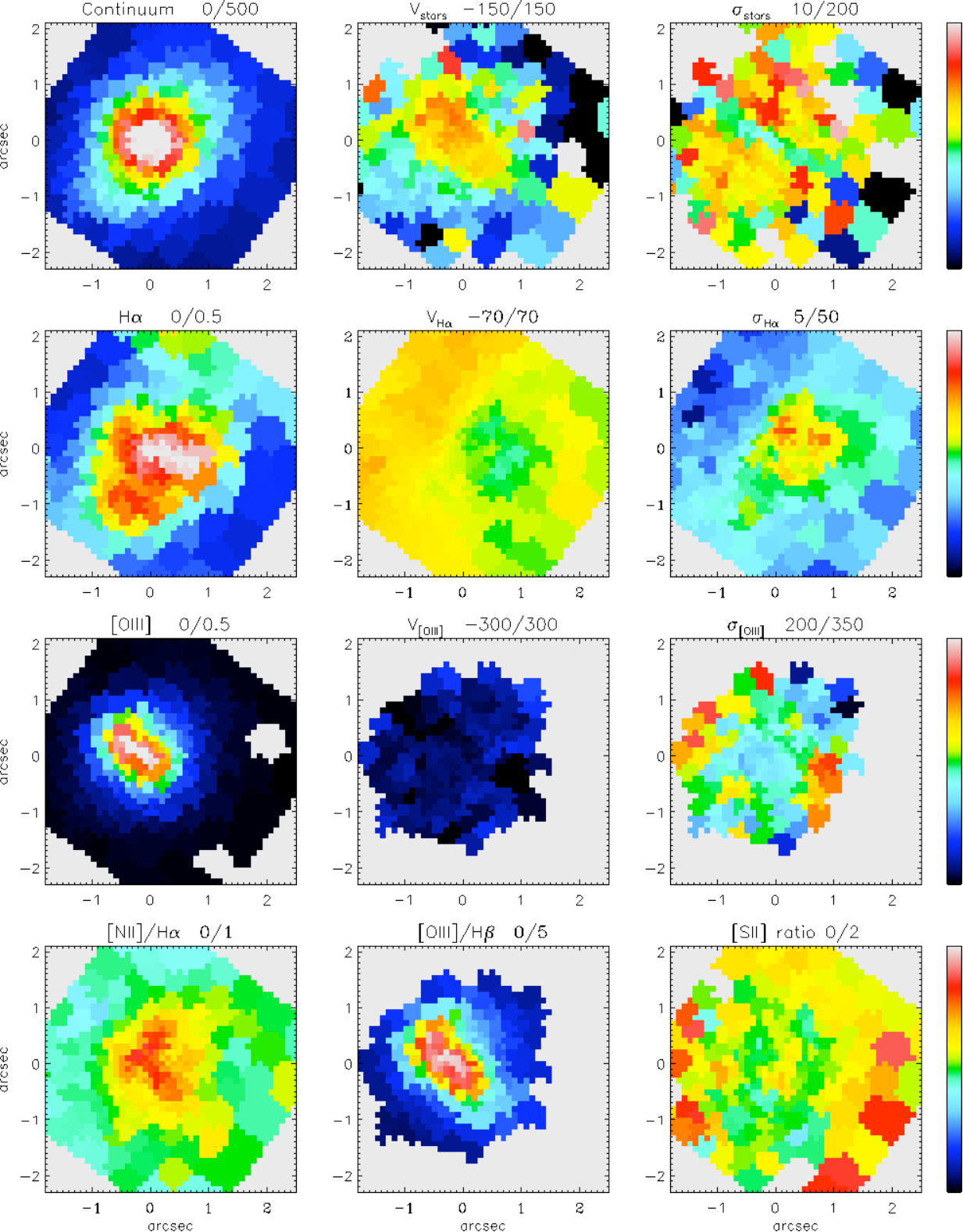}
	\caption{IMACS-IFU maps for active galaxy Mrk~609. Cosmology corrected scale: 650 pc/arcsec. 
			See Fig.~\ref{fig:Maps-J023311} for description of maps plotted.}
	\label{fig:Maps-Mrk609}
	\end{centering}
\end{figure*}

\begin{figure*}
	\begin{centering}
	\includegraphics[scale=1.110]{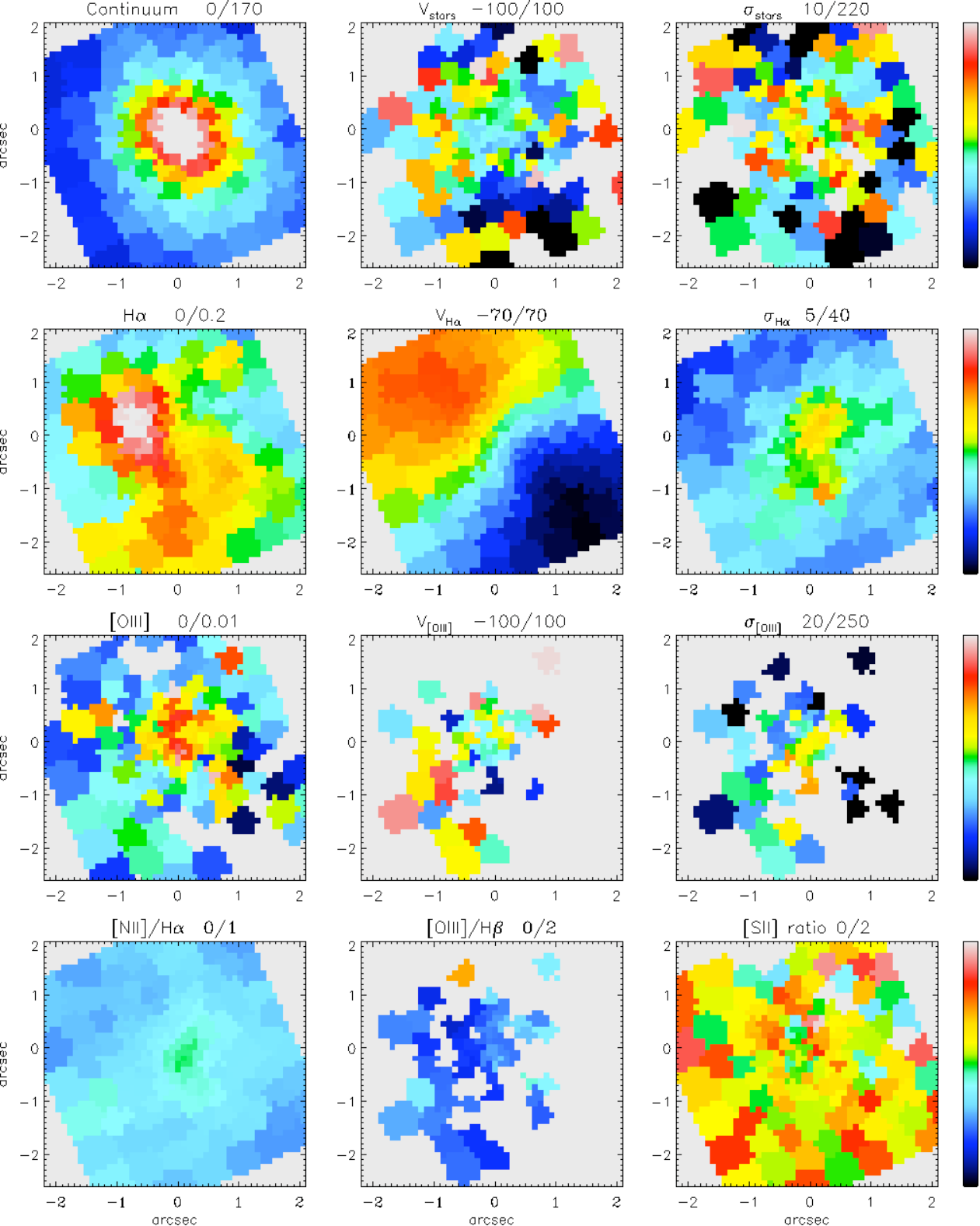}
	\caption{IMACS-IFU maps for control galaxy Mrk~1404. Cosmology corrected scale: 677 pc/arcsec. 
			See Fig.~\ref{fig:Maps-J023311} for description of maps plotted.}
	\label{fig:Maps-Mrk1404}
	\end{centering}
\end{figure*}

\begin{figure*}
	\begin{centering}
	\includegraphics[scale=1.110]{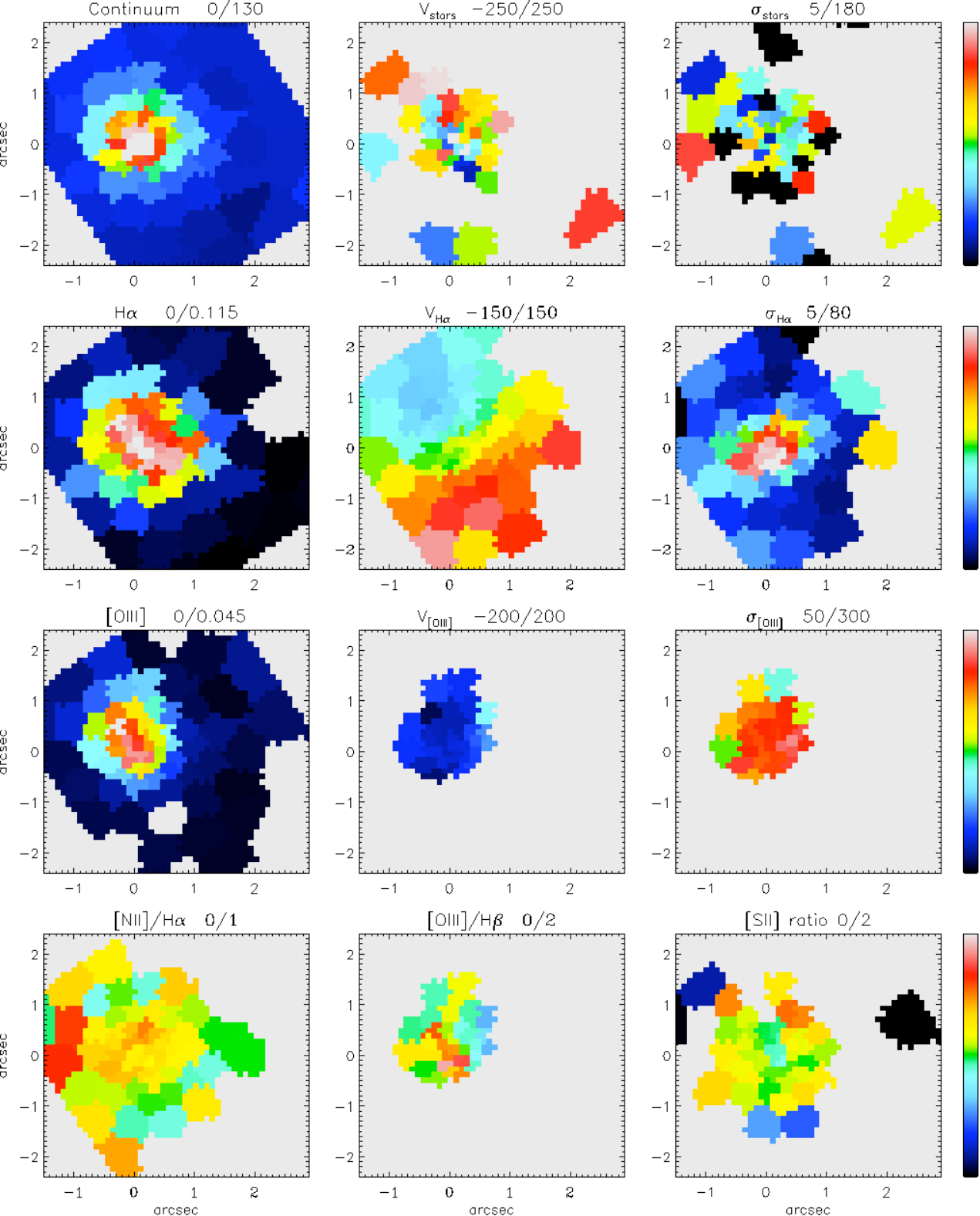}
	\caption{IMACS-IFU maps for active galaxy SDSS~J034547.53--000047.3. Cosmology corrected scale: 679 pc/arcsec. 
			See Fig.~\ref{fig:Maps-J023311} for description of maps plotted.}
	\label{fig:Maps-J034547}
	\end{centering}
\end{figure*}

\begin{figure*}
	\begin{centering}
	\includegraphics[scale=1.110]{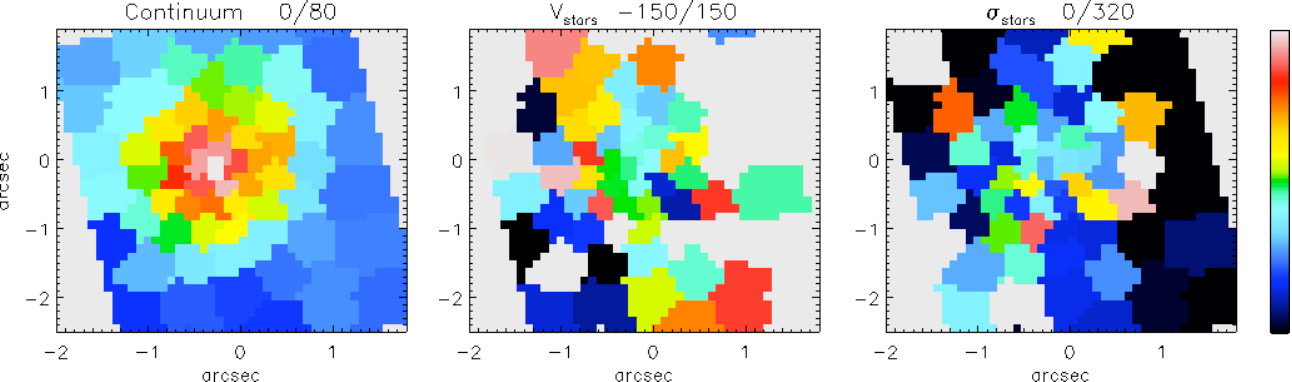}
	\caption{IMACS-IFU maps for control galaxy SDSS~J032519.40--003739.4. Cosmology corrected scale: 679 pc/arcsec. 
			See Fig.~\ref{fig:Maps-J082323} for description of maps plotted.}
	\label{fig:Maps-J032519}
	\end{centering}
\end{figure*}

\begin{figure*}
	\begin{centering}
	\includegraphics[scale=1.110]{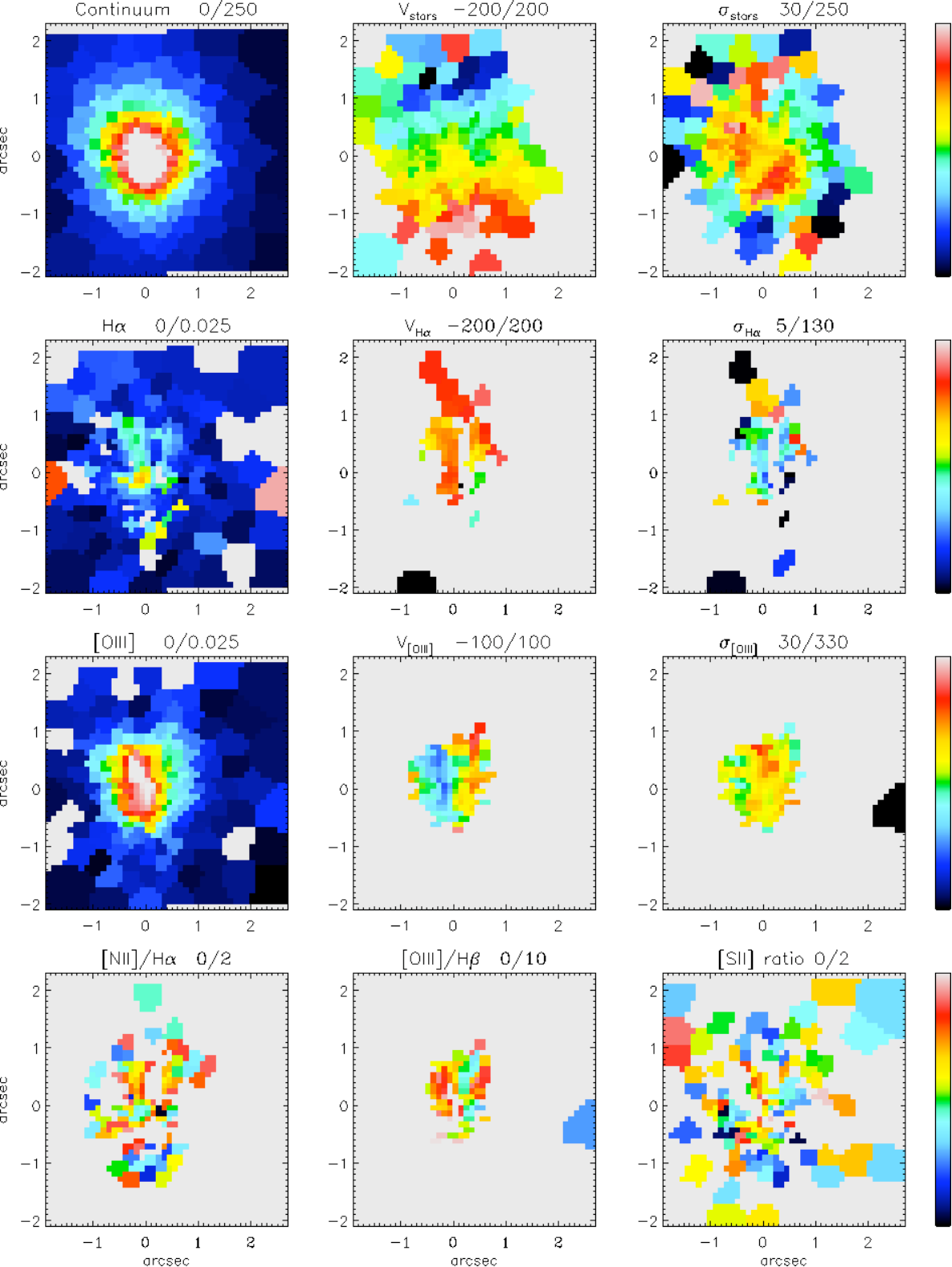}
	\caption{IMACS-IFU maps for active galaxy SDSS~J085310.26+021436.7. Cosmology corrected scale: 693 pc/arcsec. 
			See Fig.~\ref{fig:Maps-J023311} for description of maps plotted.}
	\label{fig:Maps-J085310}
	\end{centering}
\end{figure*}

\begin{figure*}
	\begin{centering}
	\includegraphics[scale=1.110]{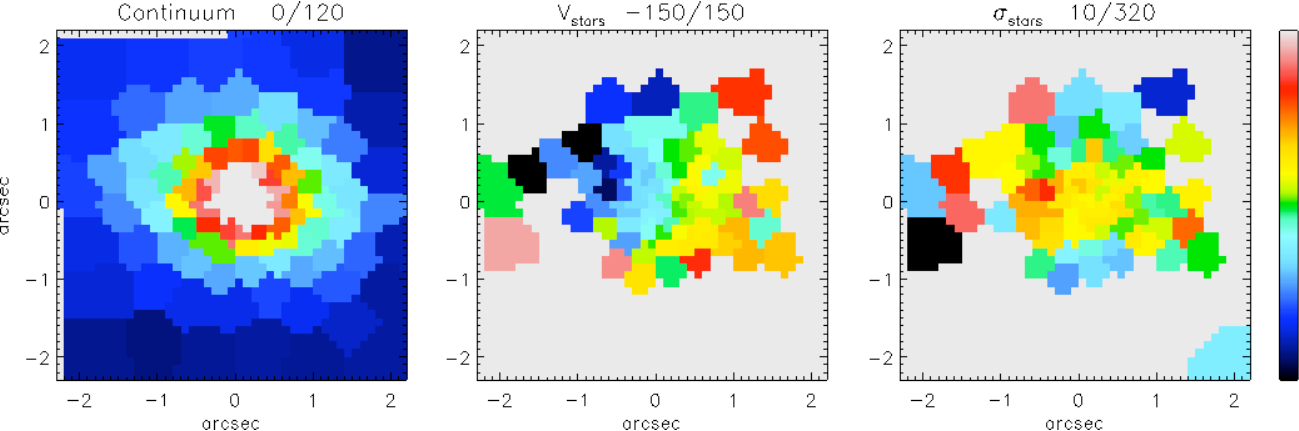}
	\caption{IMACS-IFU maps for control galaxy Mrk~1311. Cosmology corrected scale: 677 pc/arcsec. 
			See Fig.~\ref{fig:Maps-J082323} for description of maps plotted.}
	\label{fig:Maps-Mrk1311}
	\end{centering}
\end{figure*}

\begin{figure*}
	\begin{centering}
	\includegraphics[scale=1.110]{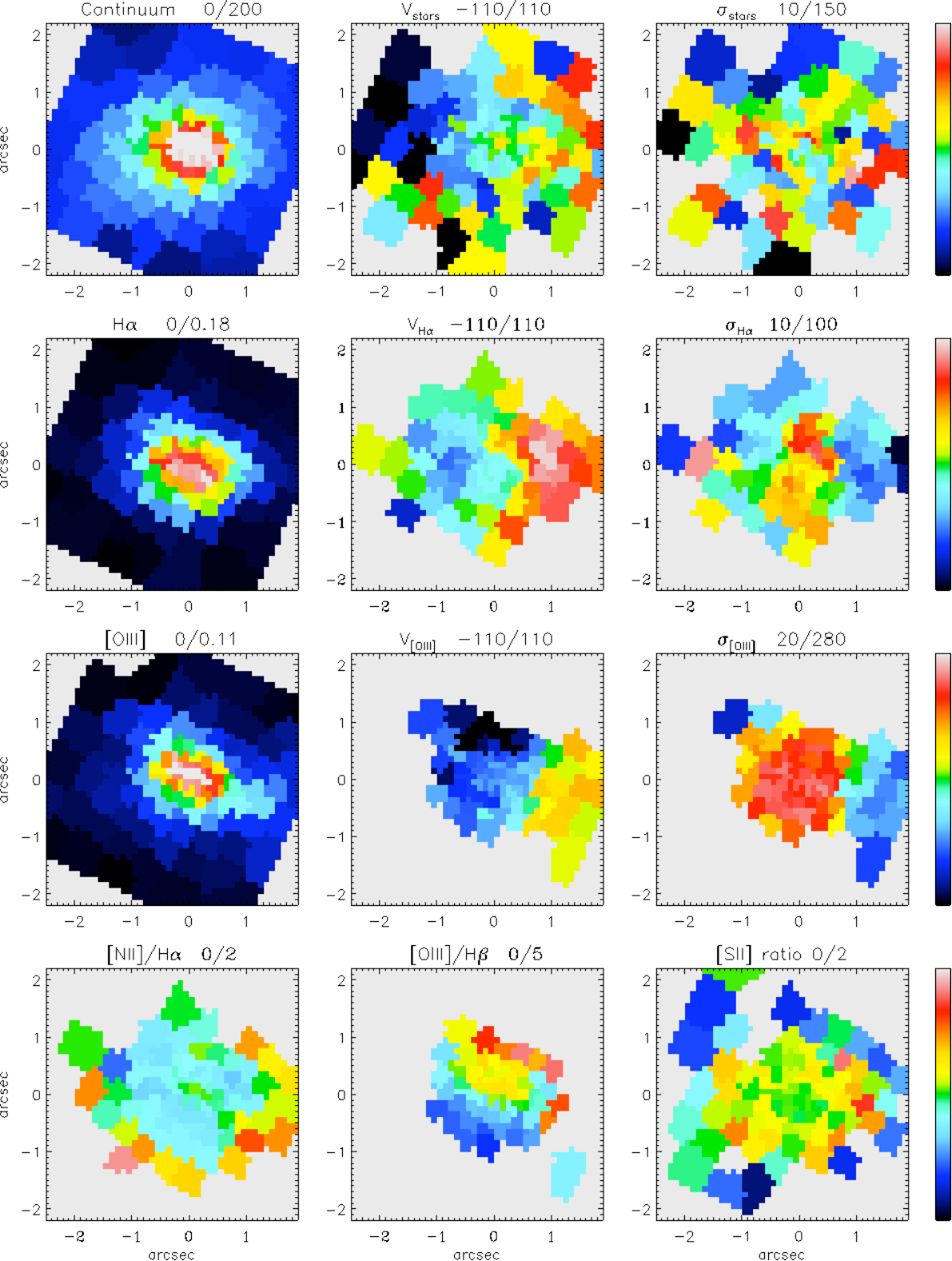}
	\caption{IMACS-IFU maps for active galaxy CGCG~005-043. Cosmology corrected scale: 569 pc/arcsec. 
			See Fig.~\ref{fig:Maps-J023311} for description of maps plotted.}
	\label{fig:Maps-CGCG005-043}
	\end{centering}
\end{figure*}

\begin{figure*}
	\begin{centering}
	\includegraphics[scale=1.110]{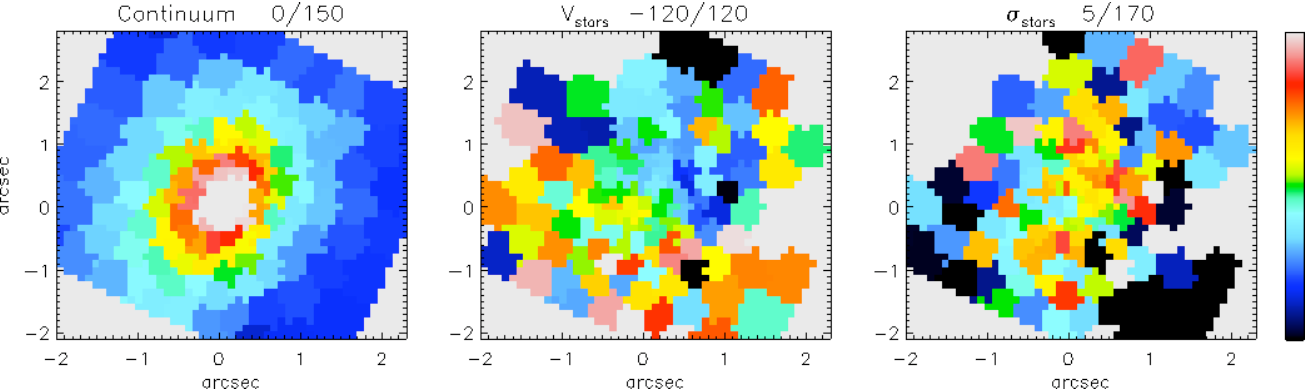}
	\caption{IMACS-IFU maps for control galaxy SDSS\,J104409.99+062220.9. Cosmology corrected scale: 594 pc/arcsec. 
			See Fig.~\ref{fig:Maps-J082323} for description of maps plotted.}
	\label{fig:Maps-J104409}
	\end{centering}
\end{figure*}

\begin{figure*}
	\begin{centering}
	\includegraphics[scale=1.110]{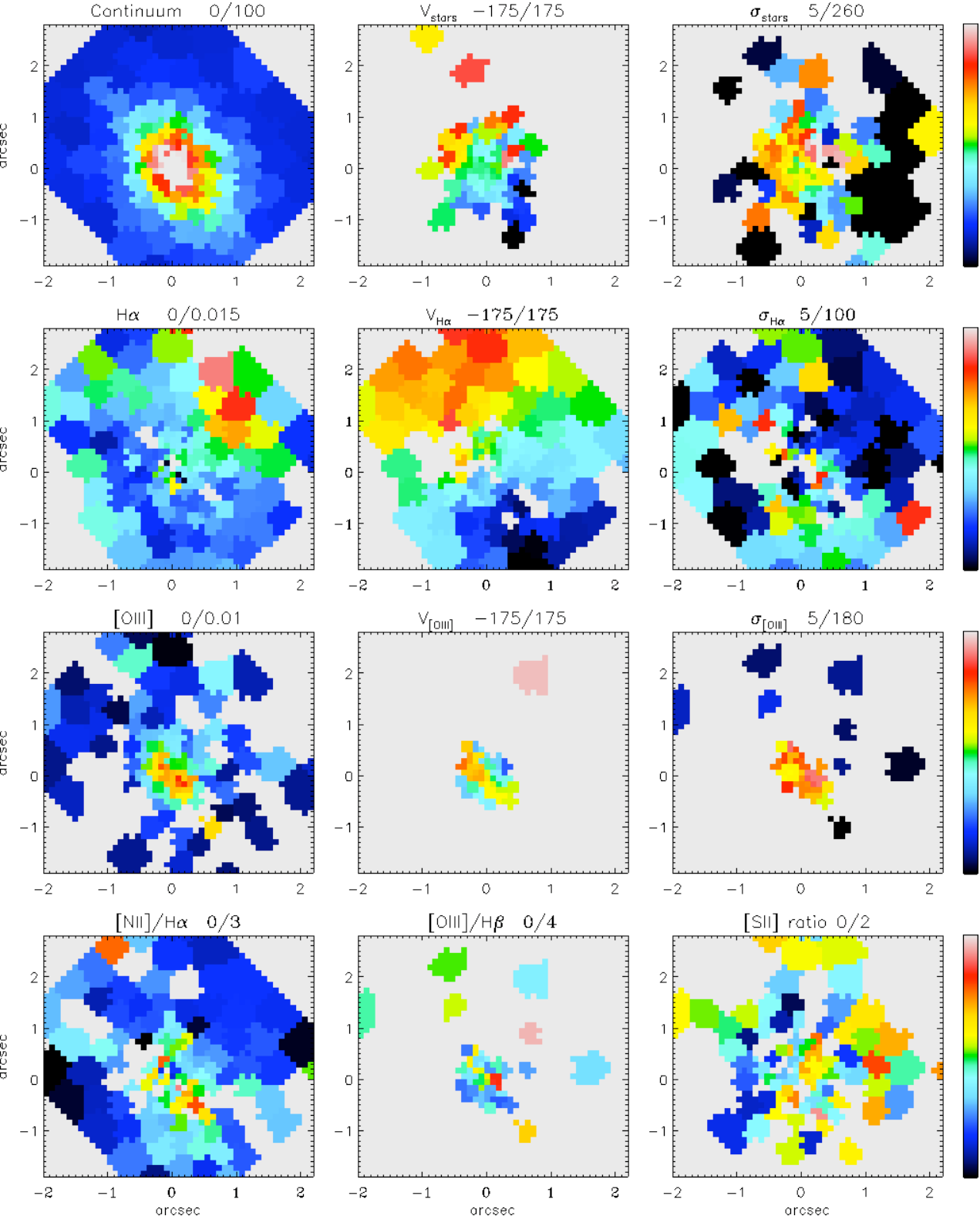}
	\caption{IMACS-IFU maps for active galaxy SDSS~J090040.66--002902.3. Cosmology corrected scale: 788 pc/arcsec. 
			See Fig.~\ref{fig:Maps-J023311} for description of maps plotted.}
	\label{fig:Maps-J090040}
	\end{centering}
\end{figure*}

\begin{figure*}
	\begin{centering}
	\includegraphics[scale=1.110]{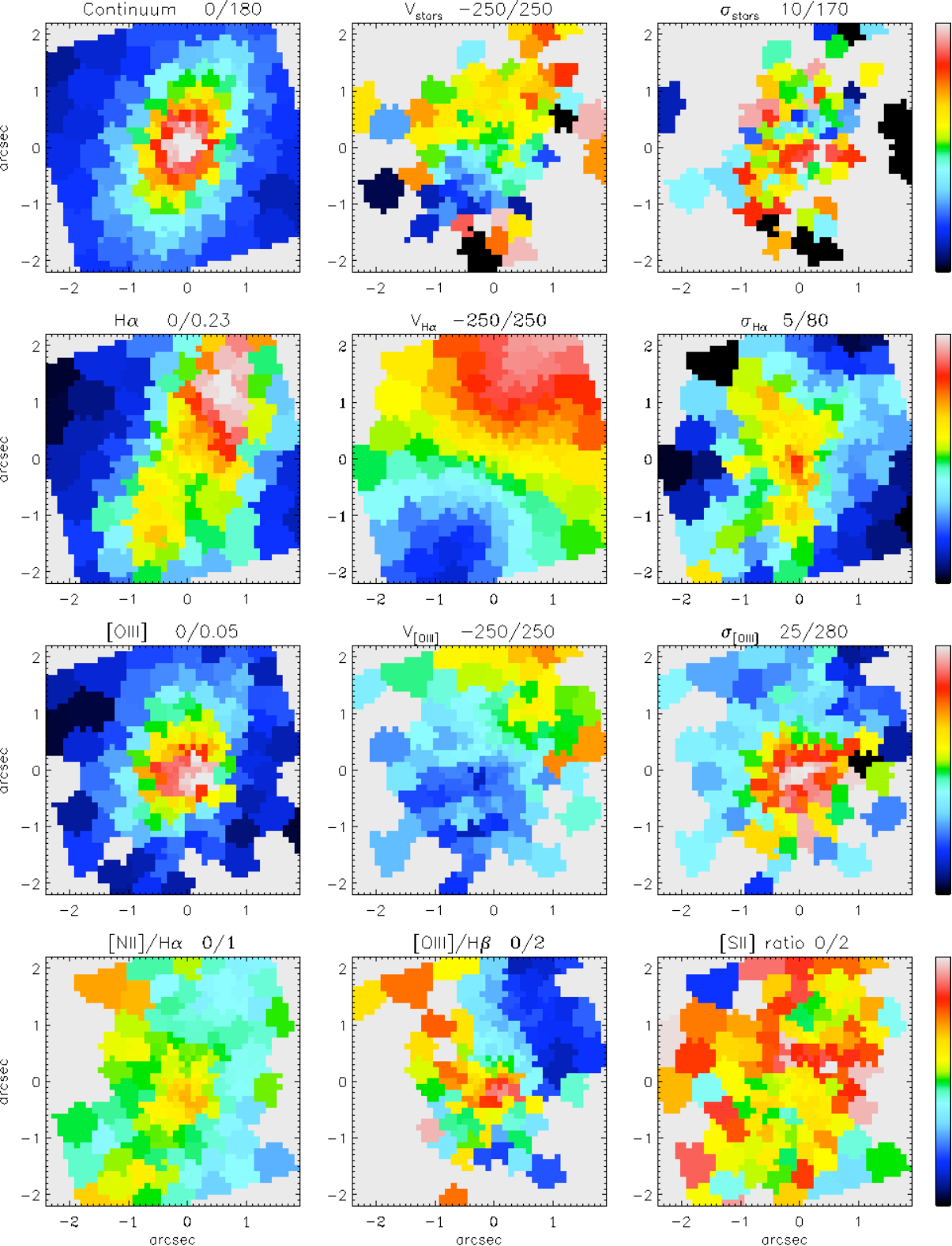}
	\caption{IMACS-IFU maps for active galaxy CGCG~050-048. Cosmology corrected scale: 745 pc/arcsec. 
			See Fig.~\ref{fig:Maps-J023311} for description of maps plotted.}
	\label{fig:Maps-CGCG050-048}
	\end{centering}
\end{figure*}

\begin{figure*}
	\begin{centering}
	\includegraphics[scale=1.110]{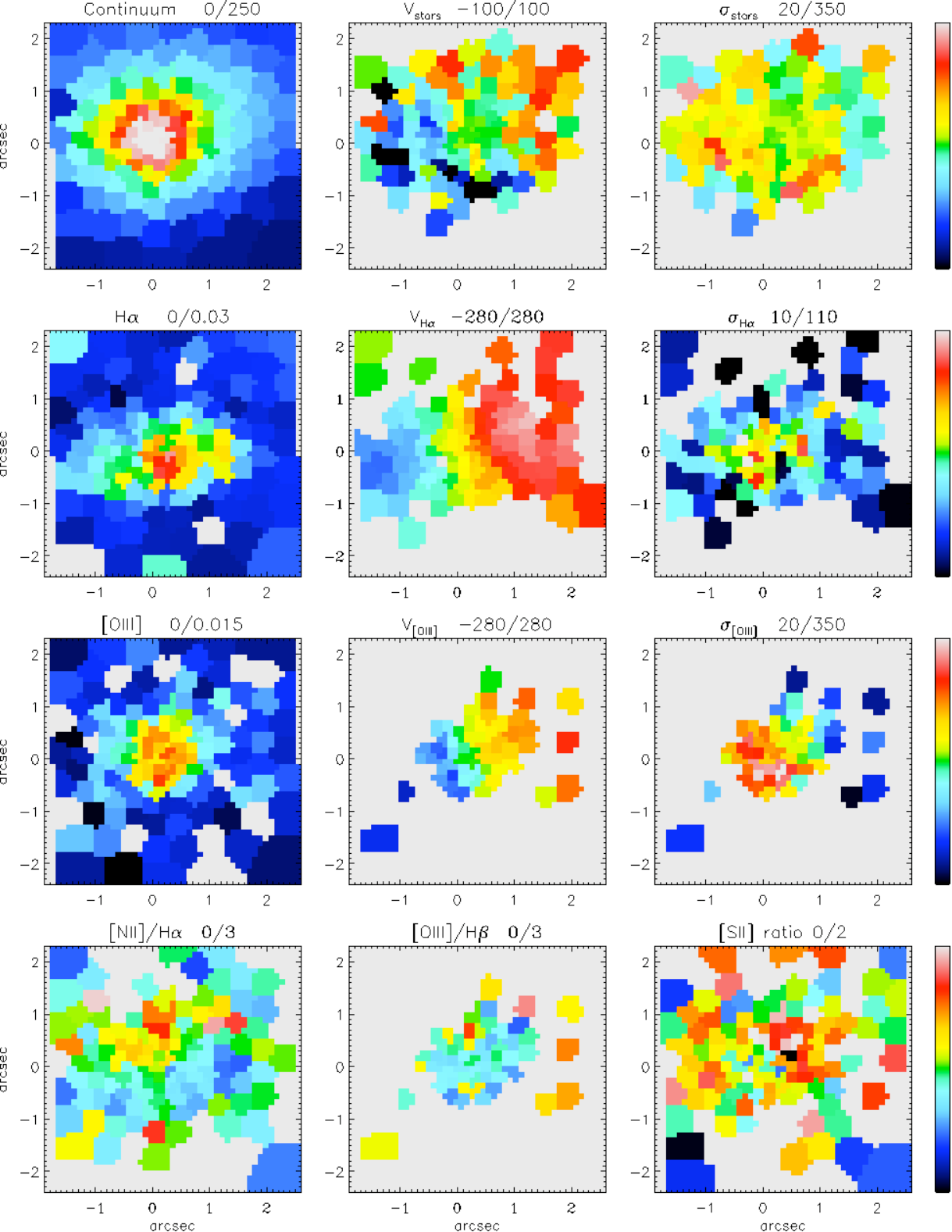}
	\caption{IMACS-IFU maps for active galaxy ARK~402. Cosmology corrected scale: 368 pc/arcsec. 
			See Fig.~\ref{fig:Maps-J023311} for description of maps plotted.}
	\label{fig:Maps-ARK402}
	\end{centering}
\end{figure*}

\begin{figure*}
	\begin{centering}
	\includegraphics[scale=1.110]{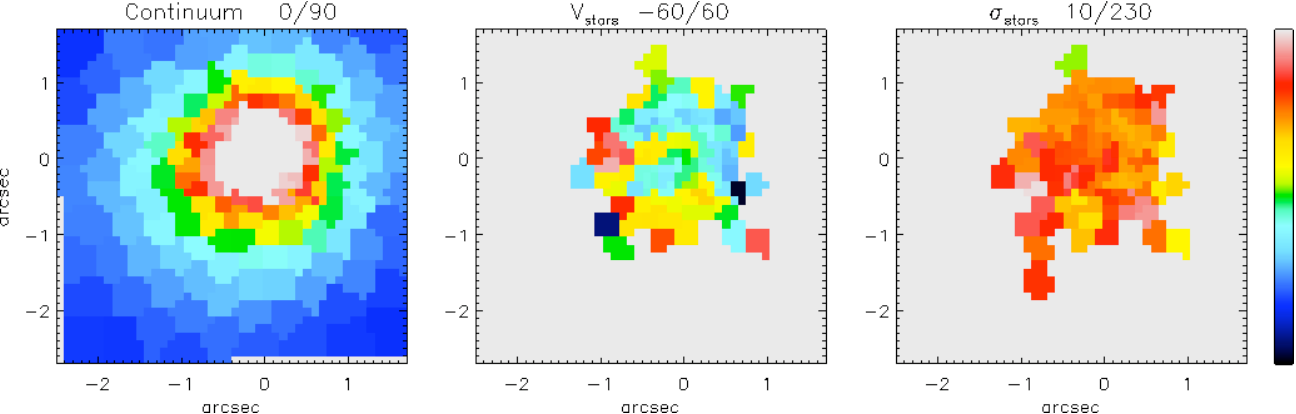}
	\caption{IMACS-IFU maps for control galaxy UGC~05226. Cosmology corrected scale: 346 pc/arcsec. 
			See Fig.~\ref{fig:Maps-J082323} for description of maps plotted.}
	\label{fig:Maps-UGC05226}
	\end{centering}
\end{figure*}

\begin{figure*}
	\begin{centering}
	\includegraphics[scale=1.110]{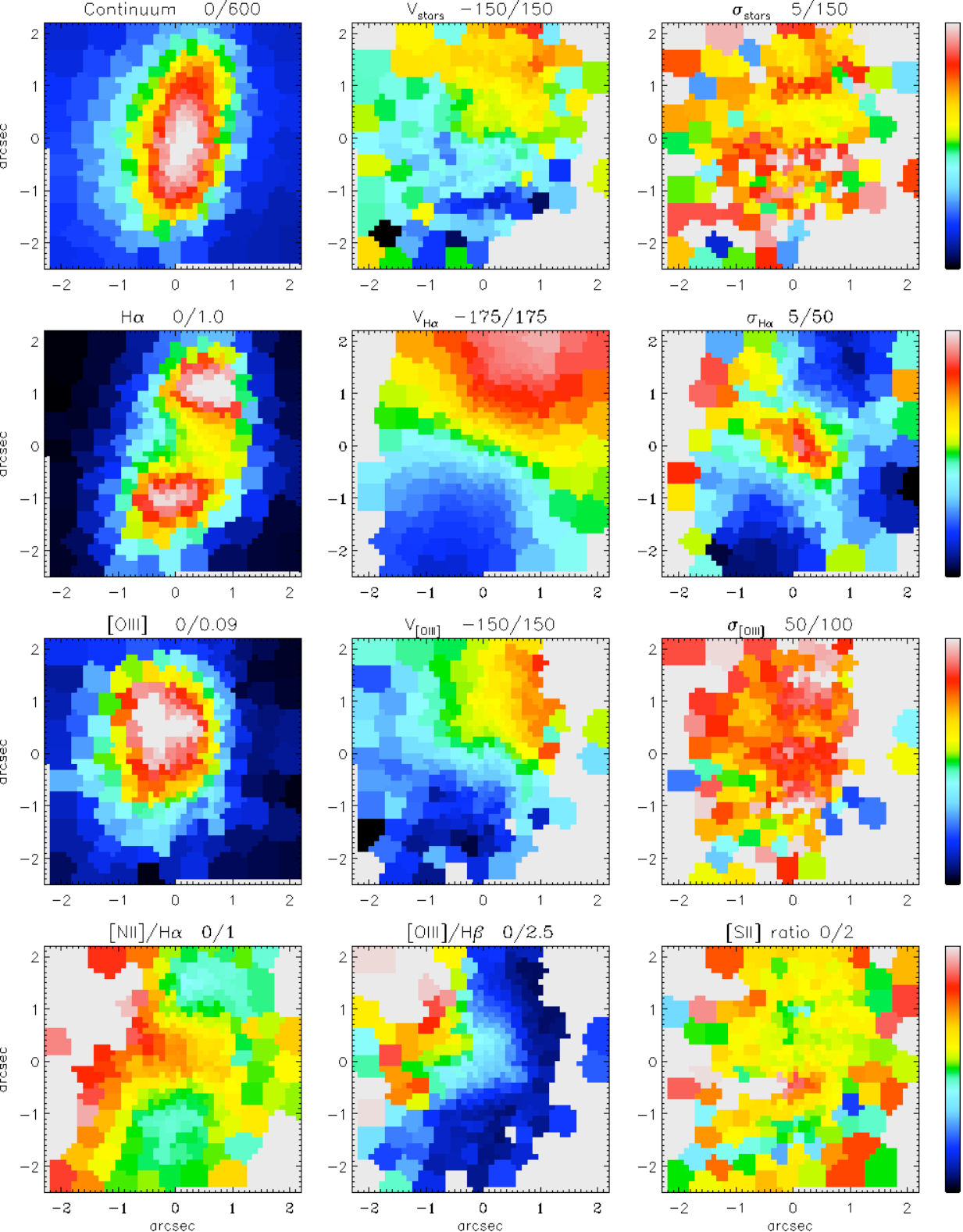}
	\caption{IMACS-IFU maps for active galaxy NGC~5740. Cosmology corrected scale: 118 pc/arcsec. 
			See Fig.~\ref{fig:Maps-J023311} for description of maps plotted.}
	\label{fig:Maps-NGC5740}
	\end{centering}
\end{figure*}

\begin{figure*}
	\begin{centering}
	\includegraphics[scale=1.110]{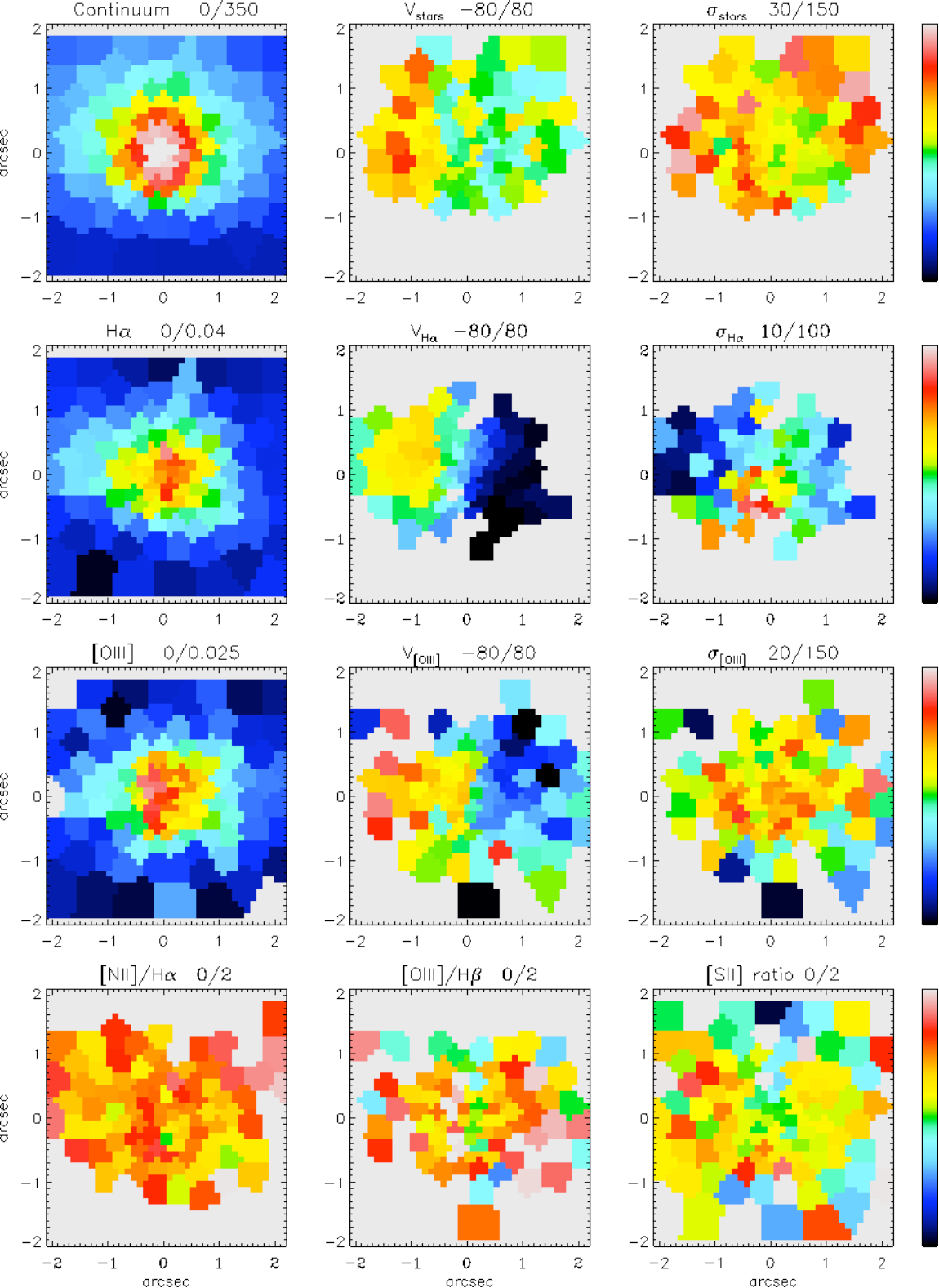}
	\caption{IMACS-IFU maps for active galaxy NGC~5750. Cosmology corrected scale: 126 pc/arcsec. 
			See Fig.~\ref{fig:Maps-J023311} for description of maps plotted.}
	\label{fig:Maps-NGC5750}
	\end{centering}
\end{figure*}

\begin{figure*}
	\begin{centering}
	\includegraphics[scale=1.110]{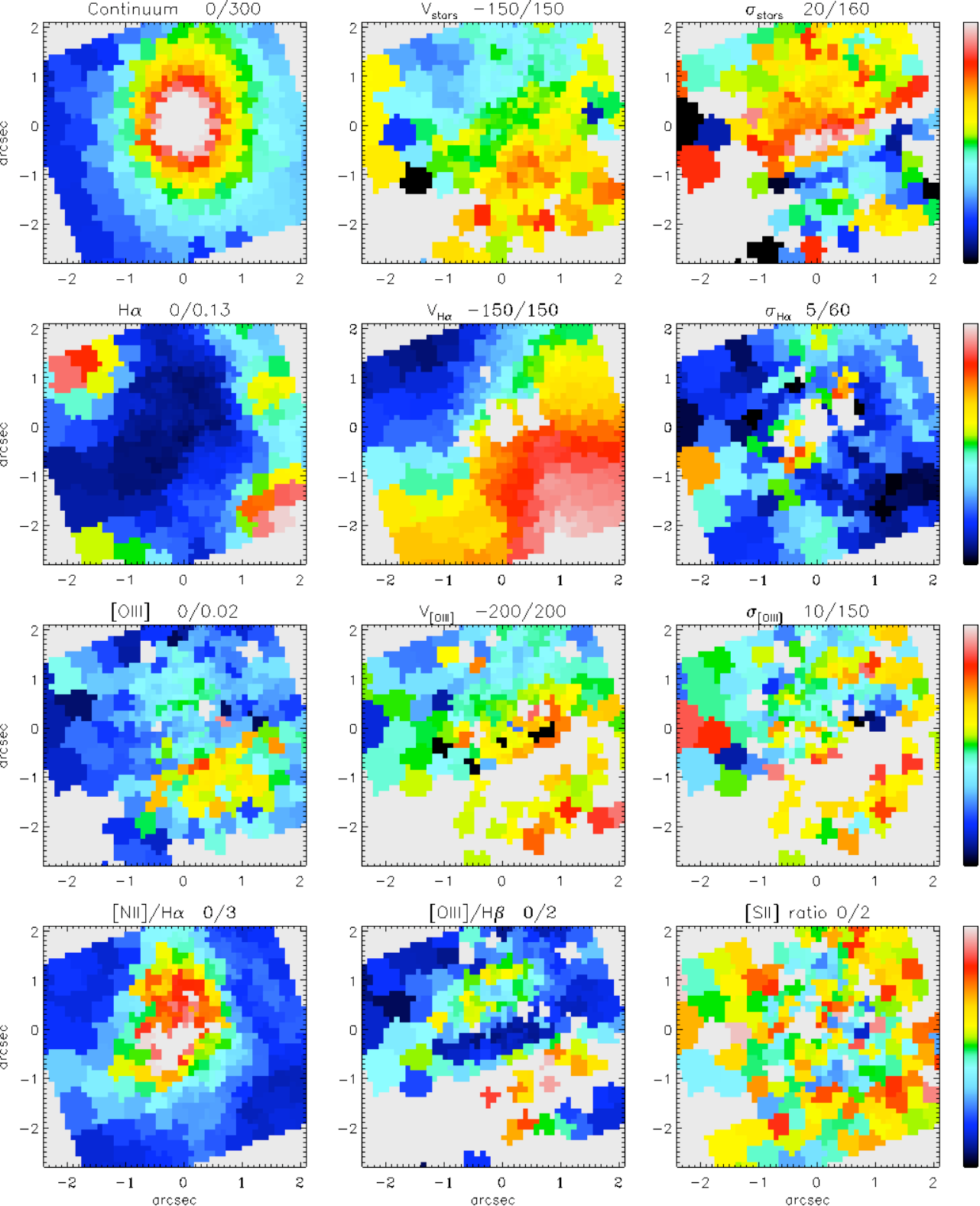}
	\caption{IMACS-IFU maps for composite galaxy NGC~5806. Cosmology corrected scale: 103 pc/arcsec. 
			See Fig.~\ref{fig:Maps-J023311} for description of maps plotted.}
	\label{fig:Maps-NGC5806}
	\end{centering}
\end{figure*}

\begin{figure*}
	\begin{centering}
	\includegraphics[scale=1.110]{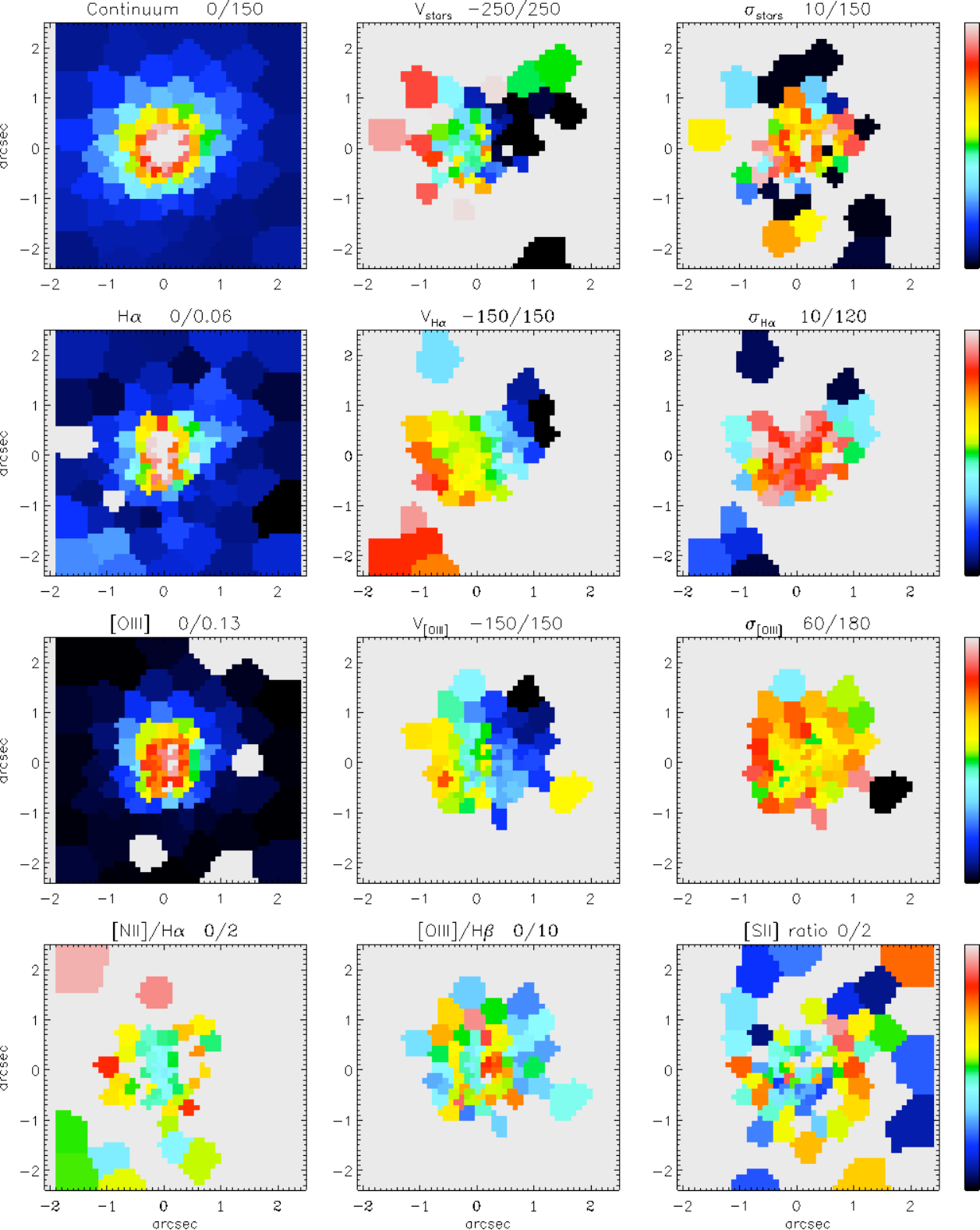}
	\caption{IMACS-IFU maps for active galaxy SDSS~J150126.67+020405.8. Cosmology corrected scale: 816 pc/arcsec. 
			See Fig.~\ref{fig:Maps-J023311} for description of maps plotted.}
	\label{fig:Maps-J150126}
	\end{centering}
\end{figure*}

\begin{figure*}
	\begin{centering}
	\includegraphics[scale=1.110]{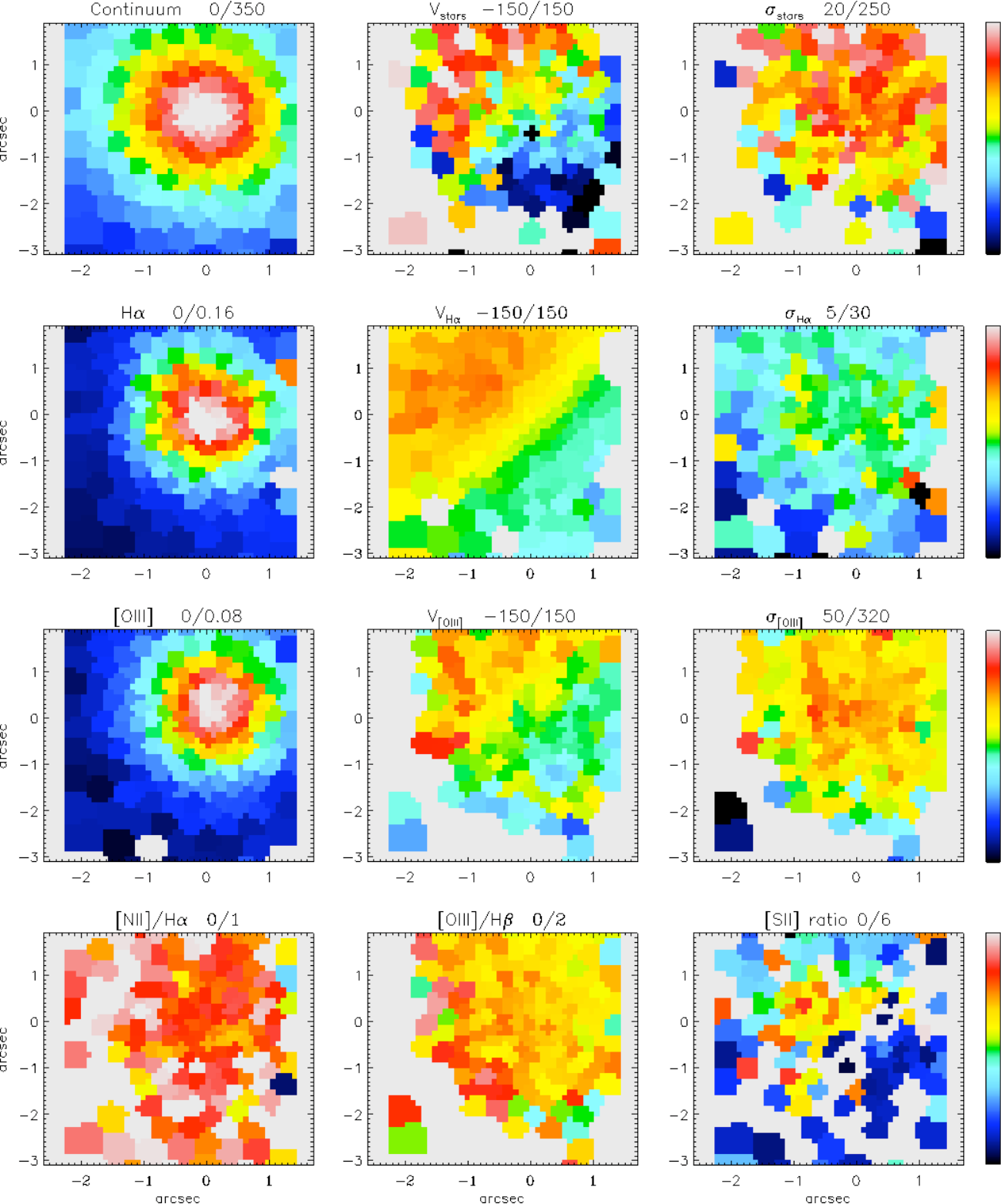}
	\caption{IMACS-IFU maps for active galaxy NGC~6500. Cosmology corrected scale: 192 pc/arcsec. 
			See Fig.~\ref{fig:Maps-J023311} for description of maps plotted.}
	\label{fig:Maps-NGC6500}
	\end{centering}
\end{figure*}

\begin{figure*}
	\begin{centering}
	\includegraphics[scale=0.375, angle=90]{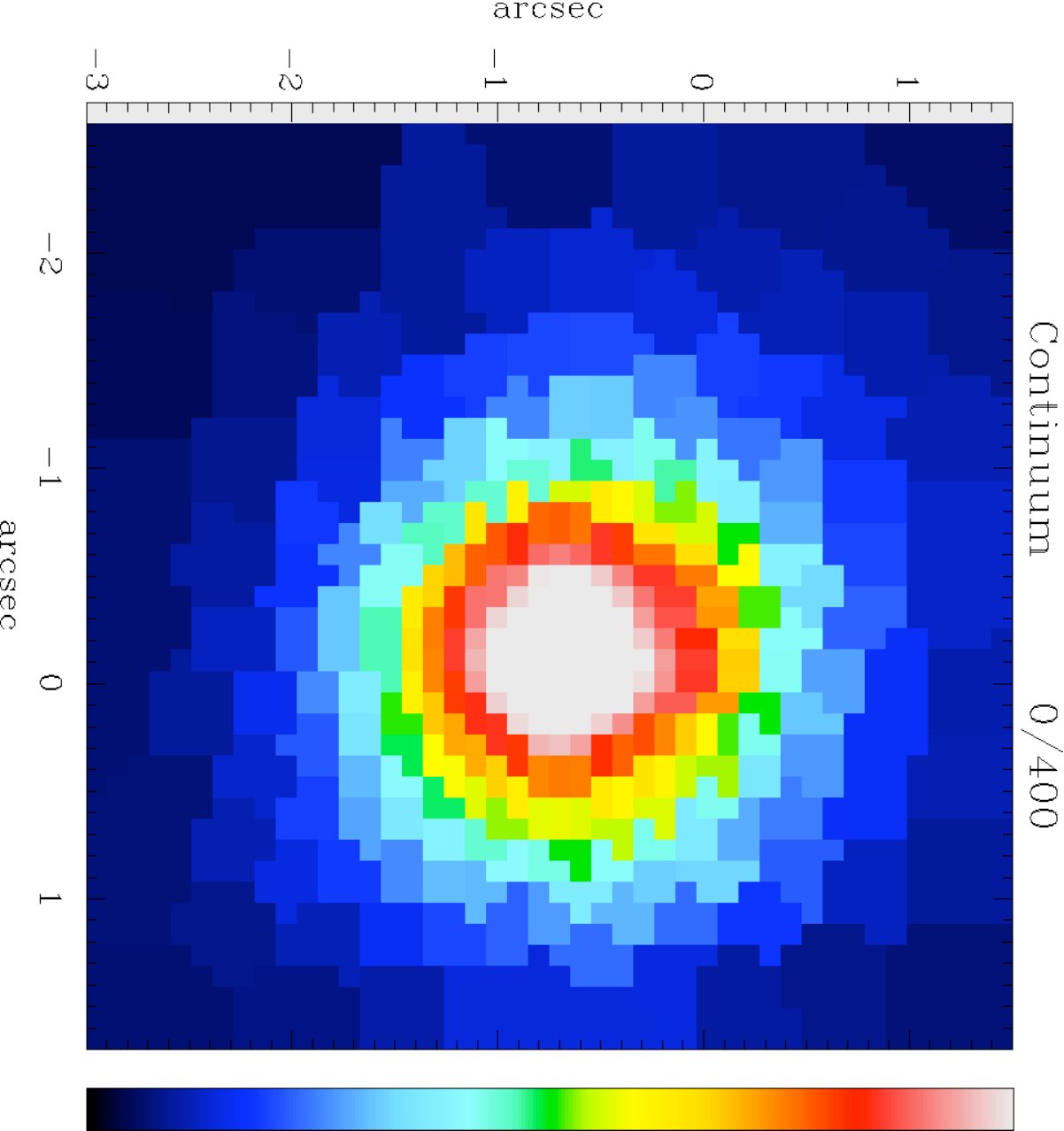}
	\vspace{0.4in}
	\includegraphics[scale=0.4, angle=90]{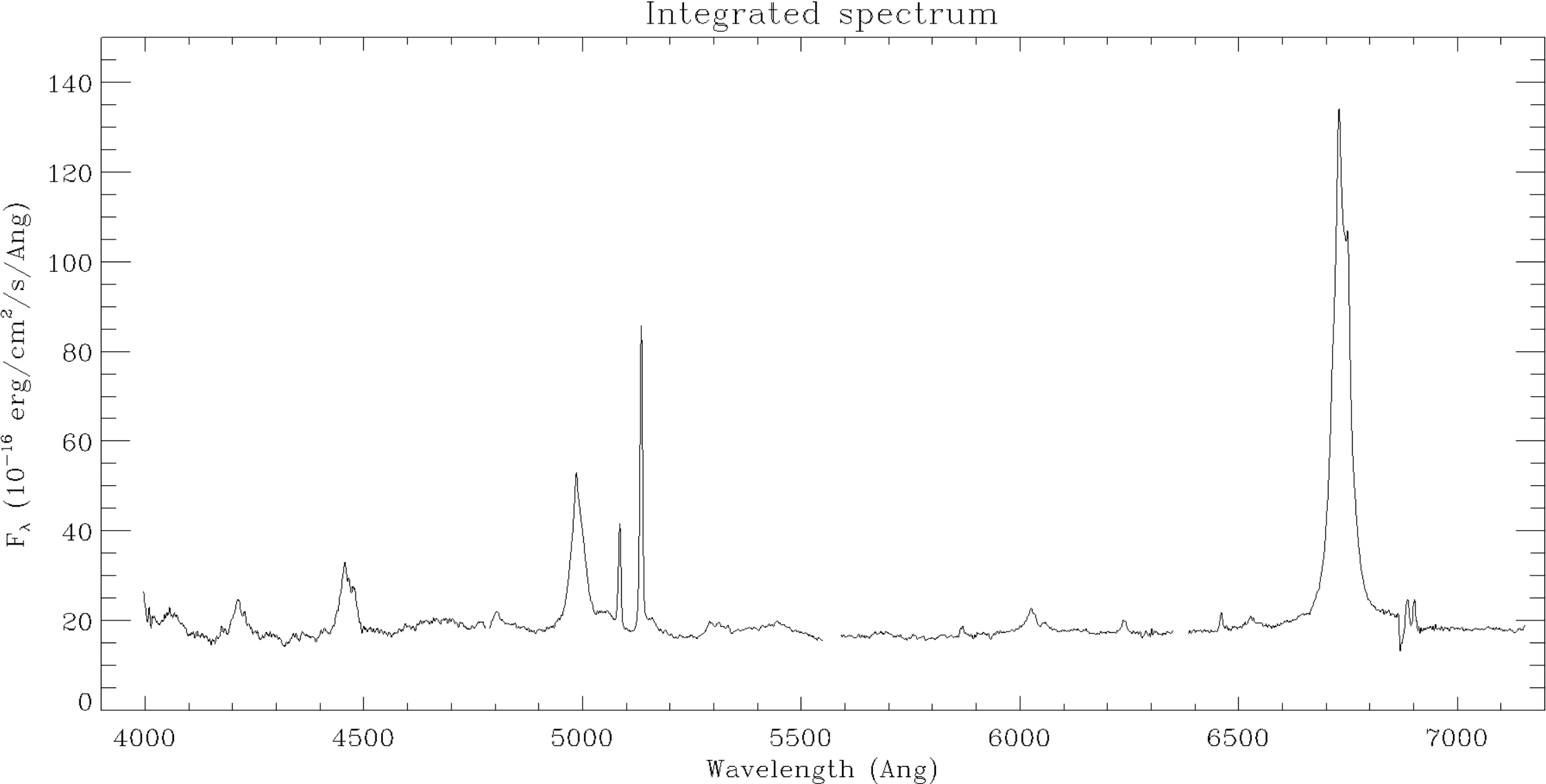} \\
	\caption{Continuum image and integrated spectrum of active galaxy ESO\,399-IG\,020. Cosmology corrected scale for the continuum image is 471 pc/arcsec.}
	\label{fig:ESO399}
	\end{centering}
\end{figure*}

\begin{figure*}
	\begin{centering}
	\includegraphics[scale=1.110]{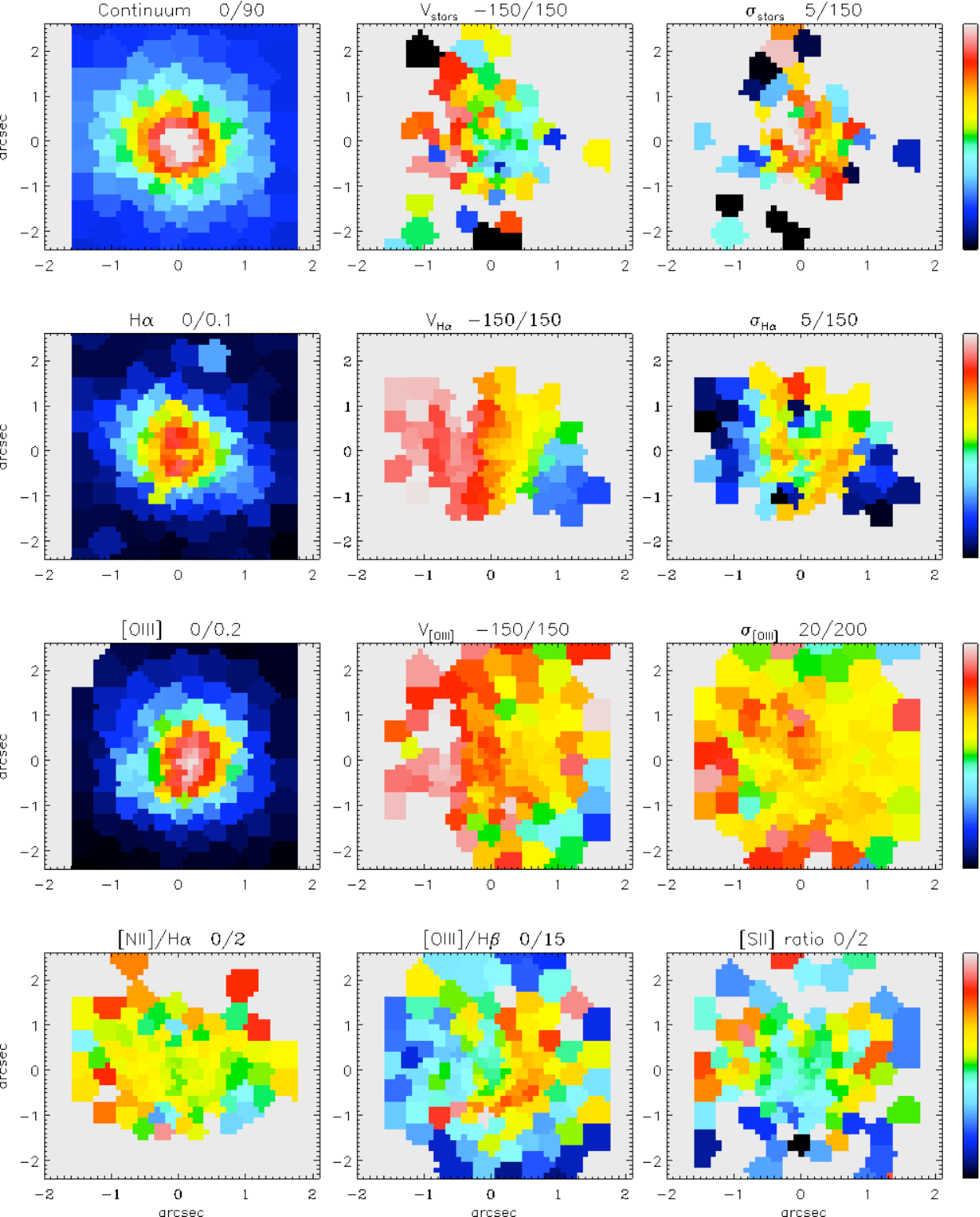}
	\caption{IMACS-IFU maps for active galaxy SDSS~J215259.07--000903.4. Cosmology corrected scale: 510 pc/arcsec. 
			See Fig.~\ref{fig:Maps-J023311} for description of maps plotted.}
	\label{fig:Maps-J215259}
	\end{centering}
\end{figure*}

\begin{figure*}
	\begin{centering}
	\includegraphics[scale=1.110]{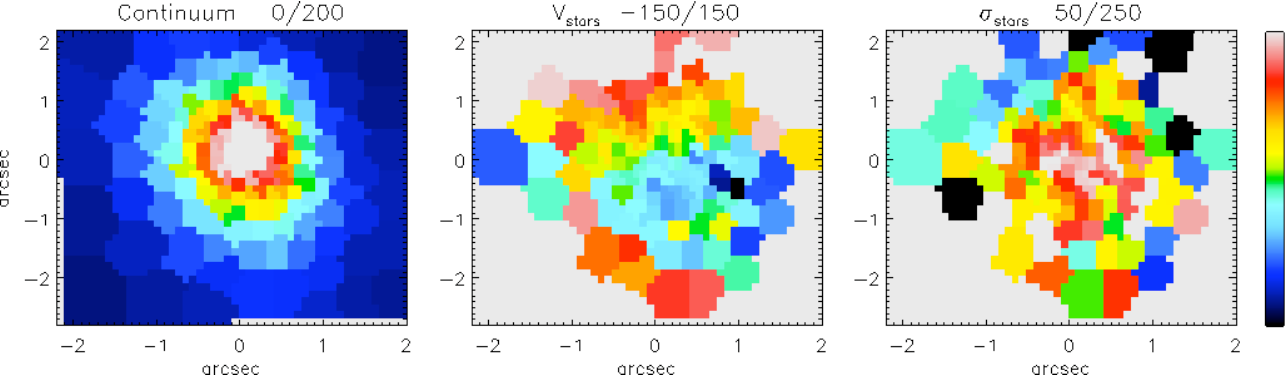} \\
	\caption{IMACS-IFU maps for control galaxy SDSS~J203939.41--062533.4. Cosmology corrected scale: 531 pc/arcsec.
			See Fig.~\ref{fig:Maps-J082323} for description of maps plotted.}
	\label{fig:Maps-J203939}
	\end{centering}
\end{figure*}

Figures \ref{fig:Maps-J023311}--\ref{fig:Maps-J203939} display the stellar and ionised gas distributions and kinematics of the 17 Seyferts and 10 control galaxies analysed. The maps are presented in the order they are described in \S\ref{sec:notes-galaxies}. For each Seyfert galaxy, and two of the control galaxies we show the stellar continuum distribution, the \oiiil\ and \ha\ emission-line distributions, the velocity and velocity-dispersion fields of the stars and ionised gas (\ha\ and \oiii), the \oiii/\hb\ and \nii/\ha\ line-ratio maps, and finally the \sii\ doublet-ratio map. For the remaining control galaxies, which contain no ionised gas, we only show the stellar continuum distribution, and the velocity and velocity-dispersion fields of the stars. All maps are oriented so that North is up, and East is left. Fluxes are in units of $F_{\lambda}$ ($10^{-17}$ erg~cm$^{-2}$~s$^{-1}$~\AA$^{-1}$), and velocities are given in kilometres-per-second. A signal-to-noise cut of a minimum of three was applied to all flux and velocity maps. 

The range of values plotted in each map is given above the map. $V_{stars}$, $V_{H\alpha}$ and $V_{[OIII]}$ are given relative to the systemic velocity derived from the stellar velocity fields, and is given in \kms. The map is plotted on the same scale, where the (0,0) coordinates mark the peak of the stellar continuum flux. The cosmology corrected  scale for each galaxy is given in the caption for each figure.\footnote{These values were taken from the NASA/IPAC Extragalactic Database, http://nedwww.ipac.caltech.edu/, and are based on the following cosmology: $H_{0}=73$\,\kms\,Mpc$^{-1}$; $\Omega_{matter}=0.27$; $\Omega_{vacuum}=0.73$.}
	
A brief description of the observed structures and kinematics, including references to any previous work on these objects are presented in 
\S\ref{sec:notes-galaxies}.

\section{Notes On Individual Galaxies}
\label{sec:notes-galaxies}
This section describes the properties of the individual galaxies in the sample with reference to their derived intensity, velocity, dispersion and line-ratio maps presented in Figures \ref{fig:Maps-J023311} to \ref{fig:Maps-J203939}. The galaxies are grouped into pairs of active and control, following Table 2. Seyferts which share a control galaxy are grouped together, with the control following afterwards so as to avoid repetition.

\subsection{Pairs One and Four}
\subsubsection{Active Galaxy: SDSS~J023311.04--074800.8}

SDSS~J023311.04--074800.8 is an S0 galaxy classified as a broad-line AGN in the Hao AGN catalogue. The IMACS continuum map (top left in Fig.~\ref{fig:Maps-J023311}) shows a central peak with a slight extension to the southeast (SE). The stellar velocity field shows no obvious rotation.

The \oiii\ and \ha\ emission line distributions roughly coincide, although the \ha\ emission is much weaker. There is no noticeable rotation in the ionised gas, but the gas dispersion velocity reaches around 100\,\kms. The \oiii/\hb\ line ratio increases toward the centre, consistent with an increase in ionising potential.

\subsubsection{Active galaxy: SDSS~J033955.68--063237.5}

SDSS~J033955.68--063237.5 is an S0 galaxy at a redshift of 0.031. The IMACS continuum map (Fig.~\ref{fig:Maps-J033955}) reveals a weak but concentrated stellar nucleus. The \ha\ emission is extended northwest (NW) of the photometric centre. The stellar velocity field shows signs of rotation, with the stellar velocity dispersion showing a possible increase towards the centre.

The \ha\ velocity field shows clear signs of rotation, with a kinematic position angle (P.A.) along the direction of elongation of the \ha\ emission. The \ha\ velocity dispersion peaks at approximately 150\,\kms\ in the galactic centre. The distribution of \oiii\ emission is also complex, with a double-peaked nucleus in the N--S direction. The \oiii\ velocity field reveals no rotation, but instead is blue shifted by $\sim150$\kms\ with respect to the systemic velocity, suggesting an outflow from the nuclear region. Interestingly, the fact that the \ha\ velocity field is symmetric about the minor axis, suggests that the outflowing \oiii\ component has had little effect on the rotating disk, and so must be directed perpendicular to the disk. The \oiii\ velocity dispersion is much higher than that of the stars and \ha\, but also increases towards the centre. The \oiii/\hb\ and \nii/\ha\ ratios also rises towards the centre, typical of Seyfert galaxies.

\subsubsection{Control galaxy: SDSS~J082323.42+042349.9}

SDSS~J082323.42+042349.9 is the control galaxy for both SDSS~J023311.04--074800.8 and SDSS~J033955.68--063237.5. The stellar nuclear region is uniformly concentrated (Fig.~\ref{fig:Maps-J082323}). The stellar velocity field shows no evidence of rotation. The velocity dispersion shows a general increase towards the centre. In contrast to the active galaxies, this control contains no ionised gas to a $3\sigma$ limiting flux density, so the corresponding emission-line maps are not shown.

\subsection{Pair Two}
\subsubsection{Active galaxy: SDSS~J024440.23--090742.4}

SDSS~J024440.23--090742.4 is an Sa galaxy classified as a type 2 Seyfert. The continuum distribution is regular, while the stellar velocity map (Fig.~\ref{fig:Maps-J024440}) shows a hint of low-level rotation ($V_{rot}< 100$\,\kms). The stellar velocity dispersion is high, and increases to around 200\,\kms in the centre. 

The \ha\ emission is faint and concentrated in the inner 500 pc. The \oiii\ emission is slightly brighter than the \ha\ emission and similarly concentrated, although a marginal elongation in the NE--SW direction can be seen. There is a hint of rotation in all maps, but the P.A. of the line of nodes for \ha\ and stellar velocity fields are misaligned by 180\textdegree. The \ha\ velocity dispersion shows and increase towards the centre, similar to that of the stars, but the \oiii\ velocity dispersion remains approximately constant at $\sim110$~\kms. The \oiii/\hb\ ratio is high, while the \nii/\ha\ ratio remains relatively low across the nucleus.

\subsubsection{Control galaxy: SDSS~J015536.83--002329.4}

The starburst galaxy SDSS~J015536.83--002329.4 is the control galaxy for SDSS~J024440.23--090742.4. The stellar continuum distribution is centrally concentrated, although a marginal elongation in the southwest (SW) direction can be seen (Fig.~\ref{fig:Maps-J015536}). Since the continuum is weak, there is no obvious rotation in the stellar velocity field, and the stellar velocity dispersion is irregular. 

The gas distribution is extended northeast (NE) in both \ha\ and \oiii. The \ha\ line strength is significantly larger than the \oiii\ strength, resulting in a low, but approximately uniform \oiii/\hb\ ratio across the field. Unlike the stellar velocity field, both of the gas velocity fields show evidence for rotation, resulting in velocities of up to 130~\kms\ within 1~kpc. The \ha\ dispersion velocity also increases towards the centre. The \nii/\ha\ shows a slight gradient across the kinematic major axis, although the difference is only $\sim0.2$. The \sii\ doublet ratio also shows an increase along the kinematic major axis, with the SW region seemingly of higher density than the NE region.

\subsection{Pair Three}
\subsubsection{Active galaxy: Mrk~609}

Mrk~609 is a starburst/Seyfert composite galaxy. There is a strong Seyfert 1-like nucleus that appears somewhat extended, and weak, broad H-recombination lines are seen in the optical spectrum, resulting in an intermediate Seyfert classification, 1.5--1.8 \citep{oster81,good90}. UV/optical ratios and X-ray spectra \citep{pappa02} show slight extinction toward the nucleus. On larger scales, observations show no signs of a bar \citep{crenshaw03,deo06}, although a more complex morphology emerges in the nuclear region, where two spiral arms connect to an elongated stellar structure. VLT-SINFONI observations reveal clumpy hydrogen recombination emission that peaks where the nuclear bar meets the spiral arms, and also along the minor axis \citep{zuther07}.

The stellar continuum map (Fig.~\ref{fig:Maps-Mrk609}) shows a slight elongation in the SE--NW direction. There is no overall rotation observed in the stellar velocity field ($V_{stars}$), while the stellar velocity dispersion ($\sigma_{stars}$) rises slightly towards the centre.

Strong \ha\ emission is consistent with VLT-SINFONI observations, with multiple peaks occurring in the SE--NW direction. In contrast, the \oiii\ emission is more concentrated, centered on the continuum peak and elongated in the NE-SW direction. The \ha\ velocity field shows signs of low-level rotation, with a slight velocity gradient north--south. In contrast to the low velocities observed in the \ha\ field, the \oiii\ velocity map is dominated by blueshifted emission up to 300~\kms\ across the central kiloparsec. These observed velocities are the result of prominent blue-wings seen in the \oiii\ emission lines, which could be attributed to an outflow component, in addition to a narrower, low-level rotation component, similar to that seen in the \ha\ velocity map.

The gas velocity dispersion is also high, but drops by around 50~\kms in the very centre. The \oiii/\hb\ line ratio rises towards the centre where the AGN dominates.

\subsubsection{Control galaxy: Mrk~1404}

The S0 galaxy Mrk~1404 is the control galaxy for Mrk~609. The optical spectrum exhibits strong \ha\ and \nii\ emission lines, but weaker \hb\ and \sii\ lines. 

The stellar velocity field (Fig.~\ref{fig:Maps-Mrk1404}) shows no significant rotation. The \ha\ distribution reveals an asymmetric star-formation pattern, with the peak offset NE of the continuum peak. The \ha\ velocity reveals a regularly rotating disk with peak velocity $\sim$100~\kms\ at a 1.3~kpc radius. The \ha\ velocity dispersion field is low ($\sigma<50$~\kms) but also shows an increase towards the centre. The weak \oiii\ emission confirms the lack of highly ionised gas in the central region of Mrk~1404. No net rotation is obvious in the \oiii\ velocity field, but the gas velocity dispersion shows a slight increase towards the centre. Due to the lack of strong emission, the \oiii/\hb\ and \nii/\ha\ line ratio maps both show generally low ratios, consistent with its classification as an inactive star-forming galaxy. The \sii\ doublet ratio map reveals a slight decrease toward the centre, indicating an increase in (electron) density.

\subsection{Pair Five}
\subsubsection{Active galaxy: SDSS~J034547.53--000047.3}

SDSS~J034547.53--000047.3 is an isolated S0/a galaxy \citep{allam05}, classified as a narrow-line Seyfert (see Fig.~\ref{fig:Maps-J034547}). The stellar continuum map shows a centrally concentrated, but weak continuum. As a result of the weak continuum, the stellar kinematics were not well constrained, so the velocity field does not reveal any signs of rotation, or obvious pattern in the velocity dispersion map.

The \ha\ distribution appears to be elongated (NE--SW) in the inner few arcseconds. There is a clear rotation component in the \ha\ velocity field, while the velocity dispersion also shows an increase towards the centre. The \oiii\ distribution, of comparable strength to \ha, is also slightly elongated in the NE--SW direction, but not entirely coincident with the \ha\ distribution. The \oiii\ velocity field, however, shows no rotation. Instead, there is again evidence for a strong outflow component, with blue-shifted velocities up to 200~\kms. The high velocity dispersions here ($\sim$275~\kms) also supports this scenario. The \oiii/\hb\ map peaks at the photometric centre and decreases with increasing radius. The \sii\ ratio distribution reveals little structure.

\subsubsection{Control galaxy: SDSS~J032519.40--003739.4}

SDSS~J032519.40--003739.4 is the control galaxy for SDSS~J034547.53--000047.3. It is an early type, almost face-on galaxy. The IMACS maps (Fig.~\ref{fig:Maps-J032519}) show a regular nuclear stellar structure. Due to the weak continuum of SDSS~J032519.40--003739.4, the stellar kinematics were again not well constrained. As a result, no hint of a gradient can be seen in the stellar velocity field, and the stellar velocity dispersion is also dominated by noise. The ionised gas maps confirmed the absence of ionised gas, and have thus been omitted from the figure.

\subsection{Pair Six}
\subsubsection{Active galaxy: SDSS~J085310.26+021436.7}

SDSS~J085310.26+021436.7 is an S0a, broadline Seyfert galaxy. The SDSS spectrum for SDSS~J085310.26+021436.7 reveals a strong continuum, which the reconstructed IMACS continuum map (Fig.~\ref{fig:Maps-J085310}) shows to be distributed symmetrically across the inner kpc. 

The stellar velocity field reveals a regular rotation field with a kinematic P.A. of roughly 180\textdegree, which is consistent with the photometric P.A. of the outer disk. The stellar velocity dispersion increases in the galaxy centre, reaching a maximum of $\sim$250~\kms. 

The \ha\ emission is weak so the underlying velocity field is not well traced in the \ha\ velocity and dispersion maps. In contrast, the \oiii\ emission is strong and centrally concentrated, with marginal evidence for an N--S elongation in the central 300~pc. The \oiii\ velocity map shows possible evidence for rotation, with an E--W gradient, which is offset by approximately 90\textdegree\ to that of the stellar velocity field. The \oiii\ dispersion velocity is high at 200--300~\kms\ in the central 300~pc.

\subsubsection{Control galaxy: Mrk~1311}

Mrk~1311 (classification S0a) is the control galaxy for SDSS\,J085310.26+021436.7. The IMACS maps (Fig.~\ref{fig:Maps-Mrk1311}) shows a centrally peaked continuum emission with a hint of elongation in P.A.$\sim$80\textdegree\ and evidence for rotation with a kinematic major axis along the same P.A. The stellar dispersion is high, at $\sim$250~\kms, and rises toward the nucleus.

Ionised gas emission was not detected to a $3\sigma$ limiting flux density, so the corresponding emission-line maps are not shown.

\subsection{Pair Seven}
\subsubsection{Active galaxy: CGCG~005-043}

CGCG~005-043 is an S0-a, broad-line Seyfert galaxy. The IMACS continuum map (Fig.~\ref{fig:Maps-CGCG005-043}, top-left panel) shows a roughly circular distribution in the stellar core, with a possible east-west elongation. The stellar velocity field shows weak evidence of rotation, with the eastern side of the galaxy significantly more blue-shifted. The photometric major-axis P.A. is known to be 151.5\textdegree\ \citep{paturel05}, so this would suggest a photometric--kinematic misalignment of around 60\textdegree. 

The ionised gas distribution is similar in both \ha\ and \oiii, with both extended roughly east--west. The \ha\ velocity field shows an S-shaped zero velocity line, but the kinematic P.A. remains similar to that implied by the stars. A gradient can also be seen across the \oiii\ velocity field, but the kinematic centre is offset to the west compared to the stars and \ha\ field, with strong blueshifted \oiii\ emission-lines observed in the very centre. This could be due to outflow from the nucleus. The gas velocity dispersion is consequently high and increases peaks in the inner few hundred parsecs. The \sii\ ratio distribution shows a slight drop in the centre.

\subsubsection{Control galaxy: SDSS~J104409.99+062220.9}

SDSS~J104409.99+062220.9 is the control galaxy for CGCG~005-043. The continuum distribution for SDSS~J104409.99+062220.9 (see Fig.~\ref{fig:Maps-J104409}) is weak but shows a slight extension approximately north-south. No trends or gradients are identified in the stellar velocity or dispersion maps. Again, no detection of emission from ionised gas is made and the noise-dominated maps are omitted.

\subsection{Pair Eight and Fourteen}
\subsubsection{Active galaxy: SDSS~J090040.66--002902.3}

SDSS~J090040.66--002902.3 is an Sab, narrow-line Seyfert galaxy. The IMACS maps (Fig.~\ref{fig:Maps-J090040}), weak stellar continuum emission that is slightly elongated in the NE direction. Since the continuum is weak the stellar kinematics are not well constrained, but there is possibly stellar rotation with a kinematic P.A. of $\sim30$\textdegree. 

The \ha\ distribution is relatively strong in the outer regions of the FOV, but shows a clear void in the central 2\as, suggesting the presence of a circumnuclear star-forming ring. The \oiii\ distribution is weak but highly concentrated in the central region where the \ha\ flux drops. The \ha\ velocity map reveals clear evidence of rotation over the entire IMACS-IFU FOV, with a P.A. similar to that implied by the stellar velocity field. The ionised gas dispersion velocity, again shows a small increase towards the centre. The \oiii\ kinematics reveal no clear patterns. The \nii/\ha\ and \oiii/\hb\ line ratios are comparable, with both highest in the centre, where the AGN dominates.

\subsubsection{Active galaxy: CGCG~050-048}

CGCG~050-048 is a radio-loud narrow-line Seyfert galaxy. The NVSS image for CGCG~050-048 shows circular contours for the galaxy, but with a resolution of 45\as. The IMACS maps, of much finer spatial resolution, reveal significant structure in the inner 2\as\ (see Fig.~\ref{fig:Maps-CGCG050-048}). The continuum map reveals an elongated distribution extended along P.A. $\sim$30\textdegree.

The stellar velocity field of CGCG~050-048 shows signs of rotation, with the kinematic major-axis roughly oriented in the N--S direction. The stellar velocity dispersion map again increases towards the centre.

The \ha\ emission line map reveals an elongated distribution running north-to-south, but with the northern half being almost twice as bright as the southern half. The \ha\ velocity field shows clear evidence of rotation, with a P.A. of $-13$\textdegree\ that is consistent with the stellar velocity field. The \oiii\ flux map shows a more compact distribution. Rotation is again visible in the \oiii\ velocity field, although the zero velocity line is significantly offset from the centre of the galaxy, suggesting again that there may be an outflow component. The gas dispersion velocity shows a sharp increase to 280~\kms\ at the location of the peak \oiii\ emission, further suggesting outflow from the nucleus. The \oiii/\hb\ ratio shows an increase towards the centre, where the AGN dominates, and a strong decrease in the NW corner, where star formation dominates. The \sii\ doublet ratio is generally high across the whole field, indicating a low density.

\subsubsection{Control galaxy: MCG~+00-02-006}

MCG~+00-02-006 is the control for SDSS~J090040.66--002902.3 and CGCG~050-048. MCG~+00-02-006 is an Sab galaxy with a P.A. of 58.2\textdegree\ \citep{paturel05}. The SDSS~spectrum shows that the continuum is quite weak ($<40\times10^{-17}$ erg\,cm$^{-2}$\,s$^{-1}$\,\AA$^{-1}$). This, coupled with the fact that only a single exposure of 20 minutes was achieved for this galaxy due to telescope time lost to bad weather, resulted in a very low signal-to-noise level for MCG~+00-02-006. As a result, IMACS-IFU maps are not presented here.

\subsection{Pair Nine}
\subsubsection{Active galaxy: ARK~402}

The compact red object ARK~402 is a narrow-line Seyfert galaxy of morphological type S0a. The IMACS continuum distribution (Fig.~\ref{fig:Maps-ARK402}) is regular in the central 2\as, with a possible elongation in the direction of the photometric axis of the outer disk (98\textdegree). The stellar velocity field hints at rotation with a kinematic P.A. consistent with the P.A. of the outer disk. The stellar velocity dispersion is generally high ($\sigma_{stars} > 200$\kms) in the inner 4\as.

The ionised gas content is low, but reveals some structure. The \ha\ distribution is slightly elongated E--W, with a velocity field showing clear, fast rotation along that axis. The \oiii\ flux distribution is more concentrated in the centre than \ha, but not uniform. The \oiii\ velocity field also shows signs of rotation along the photometric major axis. The \oiii\ velocity dispersion shows a strong increase toward the centre.

\subsubsection{Control galaxy: UGC~05226}

UGC~05226 (classification S0) is the control for ARK~402. It has strong continuum emission (see Fig.~\ref{fig:Maps-UGC05226}), but shows no clear sign of rotation and the velocity dispersion remains high at 200~\kms\ across the central $\sim$350~pc. There is no clear sign of rotation in the stellar velocity field, but the stellar velocity dispersion remains above 200\,\kms in the inner few arcseconds. No line emission is detected in the circumnuclear region so the maps are not presented.

\subsection{Pairs Ten, Eleven and Twelve}
\subsubsection{Active Galaxy: NGC~5740}

NGC~5740 is a nearby, SAB(rs)b-type, narrow-line Seyfert galaxy. The IMACS continuum map (Fig.~\ref{fig:Maps-NGC5740}) and the \ha\ emission-line map, which shows two strong peaks, confirm the presence of a circumnuclear bar running approximately south-north. The stellar velocity map shows overall rotation, with a kinematic P.A. similar to the photometric P.A. of the outer disk ($\sim160$\textdegree), derived from SDSS~data. The stellar velocity dispersion shows a general increase towards the very centre. The \ha\ velocity map also shows an axisymmetric rotation field with a kinematic P.A. similar to the photometric P.A. 

The \oiii\ maps exhibit a more complexity. The \oiii\ emission-line distribution does not show the same bar-like feature that is observed in \ha\ and starlight. Firstly, the \oiii\ emission is less extended in the south of the nucleus, and in the northern half, the distribution tends to the east. The \oiii\ velocity field shows that the \oiii\ emission is rotating at velocities comparable to the stellar field. The kinematic P.A. of the \oiii\ velocity field, however, differs from the stellar kinematic P.A. by approximately 30\textdegree, suggesting the gas is more disturbed in the nucleus. The kinematic minor axis is also warped in the NE corner, which could be due to either an outflow in the NE corner or gas streaming into the nucleus. The \oiii\ dispersion velocity is low, but does increase to $\sim50$\,\kms\ in the centre. The \oiii/\hb\ line ratio shows a gradient from east to west, with \oiii\ dominating in the centre and east, suggestive of an outflow in this direction.

\subsubsection{Active galaxy: NGC~5750}

NGC~5750 is a nearby spiral galaxy classified as SB(r)0/a. The SDSS spectrum shows that it is a low-luminosity narrow-line Seyfert. The stellar continuum (Fig.~\ref{fig:Maps-NGC5750}) shows a possible extension E--W, confirming the presence of a nuclear bar.

The IMACS stellar velocity field shows tentative signs of low-level rotation. The kinematic major axis of the stellar velocity field is oriented at a P.A. of around 75\textdegree, which roughly coincides with the photometric major axis of 65\textdegree\ \citep{erwin05}. 

The \ha\ gas emission coincides with the peak in the stellar distribution, although it is quite weak. General rotation is clear in the \ha\ velocity field, although the velocity field is blueshifted, with the kinematic minor axis offset eastwards from the photometric center. The \oiii\ emission line distribution shows a more complex, clumpy structure within the central 2\as. The \oiii\ velocity field shows similar rotation to that in \ha, albeit without the apparent shift in kinematic center. The \oiii\ and \ha\ velocity dispersion maps shows an increase in the nucleus, reaching over 100\,\kms. Line ratio maps show no particular trend across the field.

\subsubsection{Active galaxy: NGC~5806}

NGC~5806 is a late-type, SAB(s)b galaxy, originally classified as a narrow-line Seyfert galaxy based on its SDSS~classification. \textit{HST}/NICMOS2 observations, however, re-classified it as a galaxy with concentrated nuclear star formation mixed with dust, based on its nuclear NIR- and optical/NIR-colour morphology \citep{carollo02}.

Very Large Array observations from the VHIKINGS survey \citep{cgm07}, revealed that the \hi\ distribution traced the optical disk well, while the \hi\ kinematics were consistent with global rotation aligned with the outer disk major axis. SAURON kinematics \citep{dumas07}, showed that the gas and stellar velocity fields (33\as\,$\times$40\as) also revealed regular rotation patterns aligned with the photometric major axis, with a velocity dispersion that rises in the nucleus. 

Elongated features in the nucleus were seen in the HST/NICMOS2 observations and SAURON observations. This is also evident from the IMACS stellar continuum map (Fig.~\ref{fig:Maps-NGC5806}), which shows elongation at a P.A. of roughly zero. The stellar velocity field in the inner arcsec shows global rotation at a P.A. of approximately  $-160$\textdegree---a 50\textdegree\ misalignment from the outer disk major axis. The stellar velocity dispersion shows a small increase in the nucleus.

The \oiii\ emission line distribution is somewhat weak, and irregular throughout the IMACS FOV. The \oiii\ velocity field, however, still shows tentative signs of rotation, but with a kinematic major axis even further offset from previous results (i.e., the outer disk major axis). 
The \ha\ distribution, however, shows almost no emission in the inner few arcseconds, but strong emission in the outskirts of the field. This emission is likely to be the inner edge of the nuclear star-forming ring at a radius $\sim3$\as\ found in \citealt{dumas07}. In addition, the \ha\ velocity field shows a departure from axisymmetry, with an S-shaped zero-velocity line. The \ha\ velocity dispersion remains low ($< 60$\,\kms) across the IMACS-IFU FOV. Finally, the \nii/\ha\ line ratio map shows a small increase toward the centre. The \oiii/\hb\ line ratio map generally remains quite low, again suggesting that the SDSS~classification as a narrow-line Seyfert maybe incorrect, and that NGC~5608 is at most, a composite galaxy \citep{west07}.

\subsubsection{Control galaxy: NGC~7606}

\begin{figure*}
	\begin{centering}
	\includegraphics[scale=1.110]{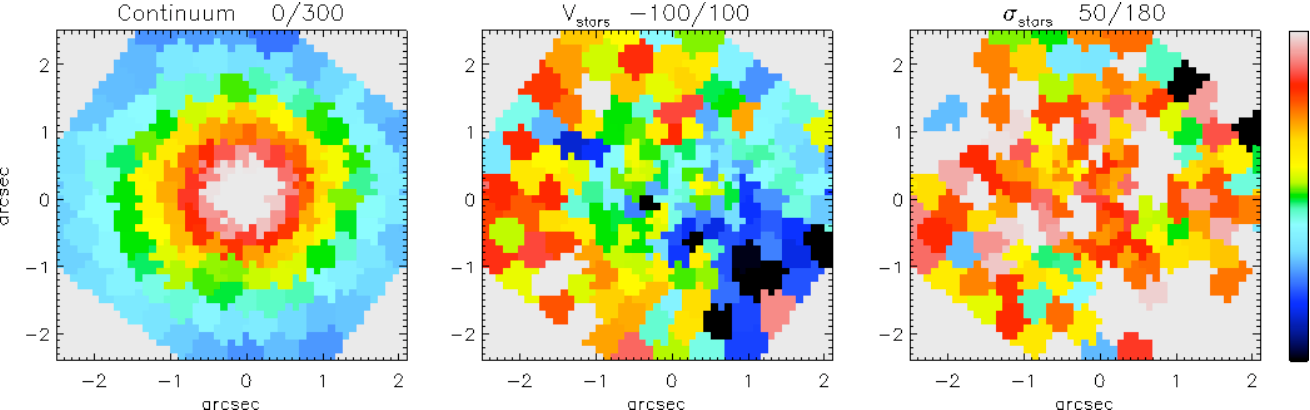}
	\caption{IMACS-IFU maps for control galaxy NGC~7606. Cosmology corrected scale: 124 pc/arcsec. 
			See Fig.~\ref{fig:Maps-J082323} for description of maps plotted.}
	\label{fig:Maps-NGC7606}
	\end{centering}
\end{figure*}

The SA(s)b star-forming galaxy NGC~7606 is the control for NGC~5806, NGC~5750 and NGC~5740. It is thought to have a nuclear bar, based on infrared ellipticity measurements \citep{karin07}. However, the IMACS continuum map (Fig.~\ref{fig:Maps-NGC7606}) reveals an almost circular distribution in the nucleus, suggesting that a stellar bar may not be present. The stellar velocity field reveals a low-level ($V_{rot}<80$\,\kms) rotation field, with a kinematic P.A. of approximately 75\textdegree. A $\sigma$-drop of $\sim60$\,\kms\ is observed in the inner 2\as\ of the stellar velocity dispersion map. 

\ha\ rotation curves derived from long-slit spectroscopy along the galactic major axis \citep{math92} show that the outer galaxy disk ($R > 100$\as) is rotating. The IMACS \ha\ and \oiii\ emission-line maps, however, show an absence of ionised gas in the central few hundred parsecs, and have therefore been omitted from Fig.~\ref{fig:Maps-NGC7606}.

\subsection{Pair Thirteen}
\subsubsection{Active galaxy: SDSS~J150126.67+020405.8}

SDSS~J150126.67+020405.8 is a broad-line Seyfert galaxy. SDSS data suggest that the galaxy is a morphological type S0, with a P.A. of 125\textdegree. The IMACS continuum distribution (Fig.~\ref{fig:Maps-J150126}) shows a symmetric central peak, with no signs of a nuclear bar. There is a slight gradient across the stellar velocity field (E--W direction) suggesting it may be rotating at a kinematic P.A. of approximately 100\textdegree, but this is inconclusive. The stellar dispersion velocity increases to around 150~\kms\ in the centre. 

The \oiii\ emission-line distribution shows a slight elongation N--S, and possibly some structure. The \ha\ emission is also extended N--S, but weaker than the \oiii. The \ha\ velocity field shows a more obvious gradient, confirming that the galaxy disk is rotating at a P.A. similar to the global photometric P.A.. The \oiii\ velocity field also shows signs of rotation at a similar P.A. to the other components. The \oiii\ velocity dispersion shows an increase towards the centre, reaching up to $\sim175$\,\kms. The \oiii/\hb\ line ratio is high and increases towards the centre of the galaxy, as expected for AGN.

\subsubsection{Control galaxy: SDSS~J122224.50+004235.6}

\begin{figure*}
	\begin{centering}
	\includegraphics[scale=1.110]{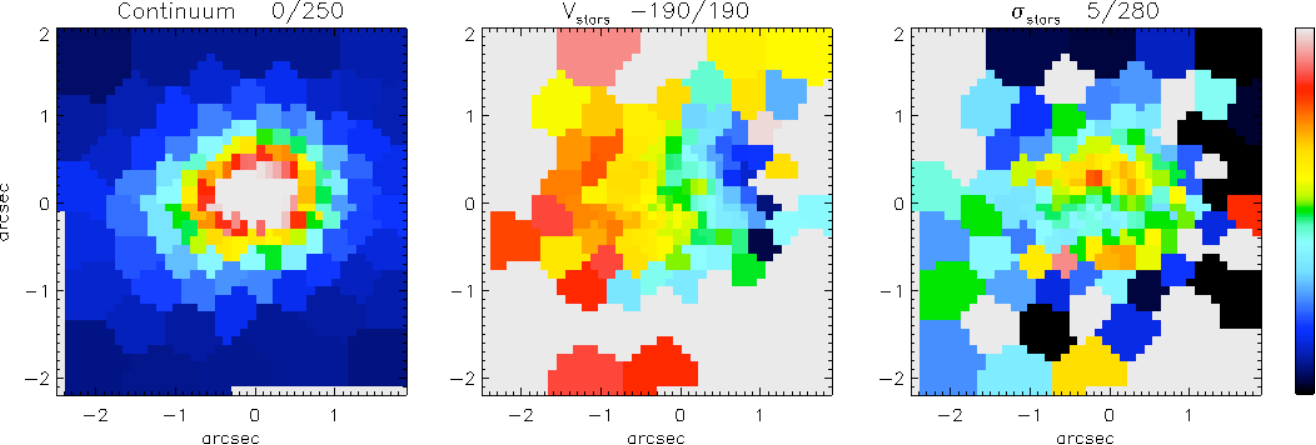}
	\caption{IMACS-IFU maps for control galaxy SDSS~J122224.50+004235.6. Cosmology corrected scale: 818 pc/arcsec.
			See Fig.~\ref{fig:Maps-J082323} for description of maps plotted.}
	\label{fig:Maps-J122224}
	\end{centering}
\end{figure*}

The control galaxy for SDSS~J150126.67+020405.8 is the S0 galaxy SDSS\,J122224.50+004235.6. The IMACS continuum map (Fig.~\ref{fig:Maps-J122224}) shows a uniform distribution of stars. There is evidence of rotation in the stellar velocity field, with a kinematic P.A. approximately aligned E--W. The velocity dispersion shows an increase towards the centre. No ionised has is detected in the 2.5-kpc field of view so the corresponding maps are omitted from Fig.~\ref{fig:Maps-J122224}.

\subsection{Pair Fifteen}
\subsubsection{Active galaxy: NGC\,6500}

The SAab galaxy NGC~6500 contains a LINER nucleus. 20-cm radio studies have revealed extended emission to the south-east, perpendicular to the galactic major axis \citep{humm83}, and on smaller scales (central arcsecond) the nuclear emission is extended along the major axis (P.A. $\sim 55$\textdegree; \citealt{unger89}). This complex radio structure has been interpreted as an outflow component along the galactic minor axis. 

No evidence for significant elongation is seen in the IMACS-IFU stellar continuum, \ha\ and \oiii\ intensity maps (Fig.~\ref{fig:Maps-NGC6500}), although a marginal elongation in P.A. 80\textdegree\ may be present. The velocity fields of the stars and gas all reveal rotation along the galactic major axis. The \ha\ and \oiii\ gas velocities at the photometric centre are redshifted with respect to that of the stars, suggesting a receding outflow component in addition to any galactic rotation. The increased ionised gas velocity dispersion in this region is consistent with this interpretation. The velocity dispersion maps of the gas and stars are also in agreement with each other, with an increase towards the centre.

The \nii/\ha\ and \oiii/\hb\ ratio maps reveal little structure, while the \sii\ ratio map shows a strong increase towards the centre, indicating a lower gas-density in the nucleus.

\subsubsection{Control galaxy:  NGC\,3731}

NGC\,3731 has been selected as the control galaxy for NGC\,6500, but it is yet to be observed.

\subsection{Pair Sixteen}
\subsubsection{Active galaxy: ESO\,399-IG\,020}

As can be seen from the Digitized Sky Survey (DSS) image in Fig.~\ref{fig:sdss}, ESO\,399-IG\,020 is the largest galaxy of an interacting system consisting of up to 3 galaxies. The presence of broad Hydrogen-Balmer emission lines in the spectra of ESO\,399-IG\,020 confirm its classification as a Seyfert 1 nucleus, but has so far prevented the extraction of the kinematics. As can be seen in the integrated IMACS spectrum shown in Fig.~\ref{fig:ESO399}, the width of the \ha\ emission line is unresolved from the \nii\ doublet. In addition, the \hb\ emission line is not consistent with a single Gaussian model, suggesting the presence of a narrow component in addition to the broad component.

As a result, in the operating range of GANDALF, masking the broad emission lines leaves little continuum spectra with which to fit the stellar kinematics. Higher angular resolution spectral imaging is required to determine the two-dimensional gas kinematics robustly. The detailed study of this object therefore lies beyond the scope of this paper.

\subsubsection{Control galaxy: IC\,1068}

The matched control galaxy for ESO 399-IG020, IC\,1068, is yet to be observed.

\subsection{Pair Seventeen}
\subsubsection{Active galaxy: SDSS\,J215259.07--000903.4}

SDSS\,J215259.07--000903.4 is a Seyfert 2 galaxy. The stellar continuum is relatively weak, but strong emission lines can be seen in the SDSS spectrum. The IMACS maps (Fig.~\ref{fig:Maps-J215259}) show an regular structure in the continuum image. The continuum is, however, quite weak, so there is only marginal evidence for rotation in the stellar velocity field and little systematic structure in the velocity dispersion map.
 
In contrast, the \ha\ velocity field provides evidence for galactic rotation, with a kinematic major axis in P.A. 70\textdegree. The \ha\ dispersion velocity increases towards the centre, but remains below 100\,\kms. The \oiii\ emission is very strong, and concentrated in the nucleus. The resulting \oiii\ velocity field shows less-obvious signs of rotation, but with a kinematic P.A. comparable to that of the \ha\ velocity field. The distribution of \ha\ and \oiii\ emission is well matched and the \oiii/\hb\ ratio increases towards the nucleus as the AGN dominates.

\subsubsection{Control galaxy: SDSS~J203939.41--062533.4}

SDSS~J203939.41--062533.4 is the control galaxy for SDSS~J215259.07--000903.4. The stellar continuum distribution (Fig.~\ref{fig:Maps-J203939}) is regular, with a bright central peak. The stellar velocity field suggests rotation, with the kinematic major axis roughly consistent with the photometric P.A. from SDSS observations. No significant ionised gas, \ha\ or \oiii, is detected in the IMACS-IFU maps, so they are omitted from the figure.

\section{Summary of Observed Features}

\subsection{Stellar Continuum Distributions}

Nineteen of the 27 galaxies show an axisymmetric, roughly circular distribution in the central 2\as. Six galaxies show more flattened isophotes, while CGCG~050-048 and NGC~5740 show strong elongations, reflecting the presence of a nuclear bar.

\subsection{Ionised Gas Distributions}

The ionised gas was mapped in both \oiii\ and \ha, via direct Gaussian fitting to the emission lines. \ha\ emission is detected to some level in all 15 Seyfert galaxies, but in only two of the nine controls analysed. \oiii\ emission, associated with active accretion, is also found in all of the Seyferts, and in one of the controls. A number of different properties were seen in either \ha\ or \oiii, or both. These can be summarised as follows:

	\begin{itemize}
		\item	Eight Seyfert galaxies (e.g., SDSS~J024440.23--090742.4; Fig.~\ref{fig:Maps-J024440}) show regular, centrally concentrated distributions in both \ha\ and \oiii. 
		\item	Mrk~609 and SDSS~J090040.66-002902.3 show round distributions in \oiii, but more complex \ha\ distributions. In Mrk~609 (Fig.~\ref{fig:Maps-Mrk609}) spiral features emerge, with the distribution extending to an excess of emission in the north of the FOV. 		SDSS~J090040.66--002902.3 (Fig.~\ref{fig:Maps-J090040}) shows a ring-like star-forming structure, with a radius of $\sim2$~\as. 
		\item	Spiral-like features can be seen in the nuclear \ha\ emission of the Seyfert galaxy NGC~5806 and the control galaxy Mrk~1404 (Figs.~\ref{fig:Maps-NGC5806} and \ref{fig:Maps-Mrk1404} respectively). In these cases the \oiii\ emission is also complex, with the distribution confined to the nuclei, but broken up. ARK~402 appears to show ring-like structures in both emission lines, with the \ha\ ring more broken-up. 
		\item	Four Seyferts and one control galaxy show extended, or bimodal distributions. SDSS~J033955.68--063237.5 (Fig.~\ref{fig:Maps-J033955}) reveals an elongated \ha\ distribution and a misaligned, bimodal \oiii\ distribution suggesting a significant dust presence. The \ha\ emission in SDSS~J034547.53--000047.3 (Fig.~\ref{fig:Maps-J034547}) reveals extended star formation (NE--SW) while the \oiii\ emission extends to the NE only. CGCG~050-048 (Fig.~\ref{fig:Maps-CGCG050-048}) shows compact \oiii\ emission in the nucleus, while the \ha\ emission is extended N--S, with the northern half significantly brighter. NGC~5740 (Fig.~\ref{fig:Maps-NGC5740}) shows extended star-formation in a nuclear bar, with two bright peaks north and south of the continuum centre. In between lies the location of the \oiii\ emission, which also extends north. The control galaxy SDSS~J015536.83--002329.4 (Fig.~\ref{fig:Maps-J015536}) shows extended 	emission in the SW--NE direction, with \ha, in particular, showing a strong excess in the NE. 
	\end{itemize}

\subsection{Ionised Gas Line Ratios}

As implied by our selection criteria, emission-line ratios trace the ionisation mechanism. Low \oiii/\hb\ and \nii/\ha\ ratios are characteristic of star formation, while high \oiii/\hb\ and \nii/\ha\ ratios are due to the presence of a harder ionising continuum, characteristic of AGN activity. In addition, the \sii\ doublet ratio is a good density indicator. 

As expected, the \oiii/\hb\ ratios are generally higher for the Seyfert galaxies: the maximum seen in the Seyferts is around 14, compared to a maximum of approximately two seen in the controls. Over the IMACS-IFU FOV, the majority of Seyfert galaxies show a general increase in \oiii/\hb\ toward the centre. An increase in \oiii/\hb\ is often accompanied by an increase in \nii/\ha\ ratio. 

The majority of the control galaxies contain little ionised gas; only two controls---Mrk\,1404 and SDSS\,J015536.83--002329.4---contained sufficient ionised gas to make meaningful maps from which to derive line ratios.

The \sii\ doublet ratio maps show very little structure for nine of the Seyferts and eight of the controls. A number of Seyferts (e.g., Mrk~609; Fig.~\ref{fig:Maps-Mrk609}) show a slight drop toward the centre, indicating an increase in (electron) density, while NGC~6500 (Fig.~\ref{fig:Maps-NGC6500}) shows very low \sii\ ratio values in the central few arcseconds. The control galaxy SDSS~J015536.83-002329.4 (Fig.~\ref{fig:Maps-J015536}), shows a gradient across the \sii\ field, with a decrease in electron density observed at the location of increased star formation in the north-east section.

\subsection{Stellar and Ionised Gas Kinematics}

The velocity maps discussed here present the line-of-sight velocity components, while the velocity 	dispersion maps represent the more random motions of the system.
	
\subsubsection{Stellar Kinematics}

Only 7 of the 26 galaxies analaysed show a gradient in their stellar velocity field consistent with rotation, with 4 more showing marginal gradients that require deeper observations for confirmation. These are: SDSS~J024440.23--090742.4, CGCG~005-043, SDSS~J090040.66--002902.3 and NGC~5750.

In the majority of galaxies (Seyferts and controls), the stellar velocity dispersion shows a general increase toward the centre. The remainder of the galaxies show more random dispersion distributions, as a result of the kinematics not being well constrained.

\subsubsection{Ionised Gas Kinematics}

An advantage of IMACS-IFU is the broad wavelength coverage---approximately 3900 to 7100 \AA. As a result, we can investigate both the dynamical impact of AGN through the high-ionisation \oiii\ emission line, and the kinematics of the host galaxy disk through the star formation-related \ha\ emission line. 

The \ha\ velocity maps are dominated by rotation for 13 Seyferts and two control galaxies. Departures from axisymmetry can be seen in many of these galaxies, such as S-shaped zero-velocity lines (e.g., Mrk~1404, CGCG~005-043 and NGC~5806), and distortions in the very centre. Only SDSS~J033955.68--063237.5, SDSS~J015536.83--002329.4, CGCG 050-048 and NGC~5740 show regularly rotating \ha\ velocity fields.

Nine of the 11 Seyfert galaxies showing rotation in the low-ionisation gas also show rotation in the high-ionisation gas (\oiii). The \oiii\ velocity fields are more distorted, and the range of observed velocities with respect to the systemic velocity is generally lower than those of the \ha\ velocity fields. In seven Seyfert galaxies, however, there is evidence of an ionised outflow component, inferred by either purely blue-shifted \oiii\ velocities compared to a rotating \ha\ velocity component (SDSS~J033955.68--063237.5, Mrk~609 and SDSS~J034547.53--000047.3), by a spatial offset of the zero-velocity line with respect to the photometric centre (CGCG~005--043, CGCG~050--048, NGC~6500 and SDSS~J215259.07--000903.4), or by a strong warping of the zero-velocity line (NGC~5740). In the cases of the strongly blue-shifted \oiii\ velocity fields, the outflows show approaching velocities of over 200~\kms, extending over 2\as\ ($>1.5$~kpc). In one of these cases, SDSS~J033955.68--063237.5, the \ha\ velocity field is remarkably undisturbed suggesting that the ionised-gas outflow must be almost perpendicular to the plane of the rotating disk. 

All galaxies tend to show a maximum gas velocity dispersion (\ha\ and \oiii) in the centre, with the exception of Mrk~609, which shows a slight drop in \oiii\ dispersion in the centre---possibly reflecting the effect of the outflow component.

\section{Concluding Remarks}

We have presented the first two-dimensional maps of ionised gas and stars in an SDSS-selected sample of active and inactive galaxies obtained using the IMACS-IFU on the Magellan-I 6.5m telescope. This study required the development of a full reduction pipeline for IMACS-IFU, which was previously unavailable. In this paper we have provided a detailed description of the reduction process and analysis of IMACS-IFU data, from observing strategy to final extracted maps, with particular emphasis on aspects that affect IMACS-IFU such as telescope flexure. Following extensive testing of the IMACS-IFU pipeline, two-dimensional stellar and gas kinematics were derived for 26 of the 28 galaxies in the sample thus far.

In contrast to other IFUs, the unusually large wavelength coverage ($\sim 4000-7000$ \AA) provided by the IMACS-IFU, coupled with the fine pixel sampling across the \das415 $\times$ \das500 FOV has allowed the extraction of gaseous and stellar kinematics that probe both the AGN-related regions and the host galaxy. In particular, simultaneous observation of \oiii\ and \ha\ emission lines provide independent probes of galaxy rotation and AGN-driven outflow. 

Evidence of rotation was found in the stellar velocity fields of 11 out of the 26 galaxies analysed, while the S/N of the remaining galaxies was deemed insufficient to make any conclusions. Fifteen of twenty-six galaxies showed clear evidence of rotation in the \ha\ velocity fields. Seven Seyfert galaxies show possible evidence of an \oiii\ outflow component, the most extreme of which exceeds a blue-shifted line-of-sight velocity of 200\,\kms\ and extends over 2\as\ ($\sim$1.5~kpc). 

We have demonstrated the value of large simultaneous wavelength coverage, which allows for the derivation of the underlying host-galaxy dynamics from \ha\ kinematics, regardless of the perturbations revealed in \oiii\ kinematics. Full kinematics, modelling and its interpretation will be presented in Paper II (P. B. Westoby et al., in preparation).

\acknowledgments

PBW acknowledges financial support from STFC. CGM acknowledges financial support from the Royal Society, the Wolfson Foundation and Research Councils U.K. NN acknowledges funding from Conicyt-ALMA 31070013 and 3108022, the Fondap Center for Astrophysics, BASAL PFB-06/2007, and Fondecyt 1080324. We thank Pierre Ferruit for his role in the inception of the project, the referee for constructive comments that improved the paper and Alan Dressler, Gaelle Dumas, Davor Krajnovic, Dave Shone, Chris Simpson, Niranjan Thatte and Sylvain Veilleux for helpful discussions. We wish to acknowledge use of the high performance computing facilities at the ARI. We thank Michelle Cappellari and Marc Sarzi for the use of pPXF and GANDALF respectively. The Magellan-I telescope, operated at Las Campanas Observatory, Chile, is supported by the Magellan Project; a collaboration of the Carnegie Institution for Science, University of Arizona, Harvard University, University of Michigan, and Massachusetts Institute of Technology. This research was partially based on data from the MILES project. Data were also taken from the SDSS catalogue provided by the MPA Garching Group, and also from SDSS~DR5. Funding for the SDSS Archive has been provided by the Alfred P. Sloan Foundation, the Participating Institutions, the National Aeronautics and Space Administration (NASA), the National Science Foundation, the U.S. Department of Energy, the Japanese Monbukagakusho, and the Max Planck Society. This research has made use of the NASA Astrophysics Data System Abstract Service (ADS), and the NASA/IPAC Extragalactic Database (NED), which is operated by the Jet Propulsion Laboratory, California Institute of Technology, under contract with the National Aeronautics and Space Administration. We also acknowledge NASAs Astrophysics Data System Bibliographic Services.

%%	**************	References	***********************
%	\bibliographystyle{apj}
%	\bibliography{thesis}

%%%	****************************************************	Figures 		******************************************************	%%

%%	Figure showing parameter space covered in present suvey...

%%	Figure showing CMD locations of the IMACS-IFU sample...

%%	SDSS images and spectra figures...
%%		**	For ApJ style file use scale=0.060, for emulateapj use scale=0.07	**
\begin{figure*}[htp]
	\begin{centering}
		\subfigure{\includegraphics[scale=0.63]{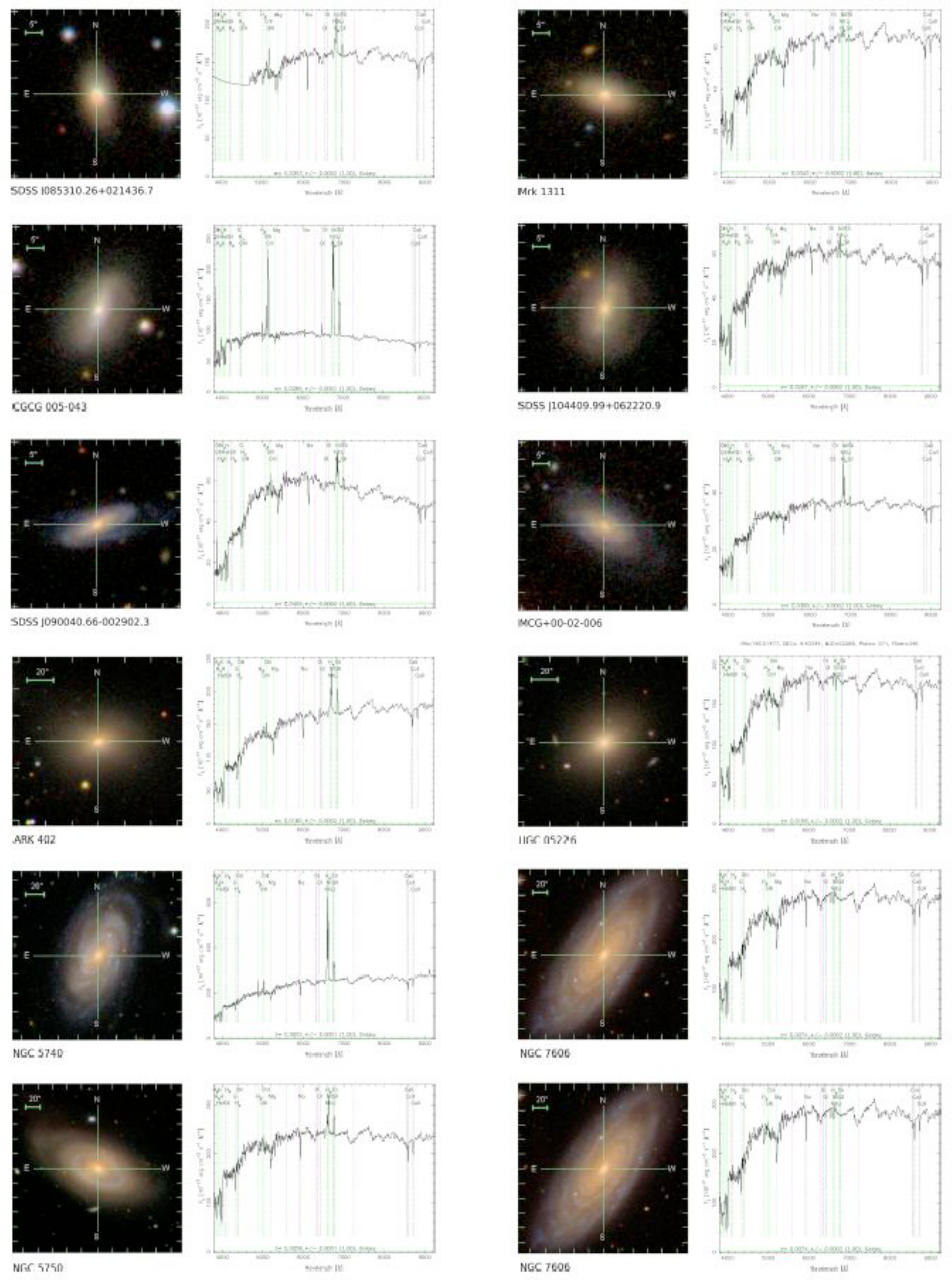}} \\
	\end{centering}
	\vspace{3mm}
		Fig.~\ref{fig:sdss} cont.
\end{figure*}
\begin{figure*}[htp]
	\begin{centering}
		\subfigure{\includegraphics[scale= 0.63]{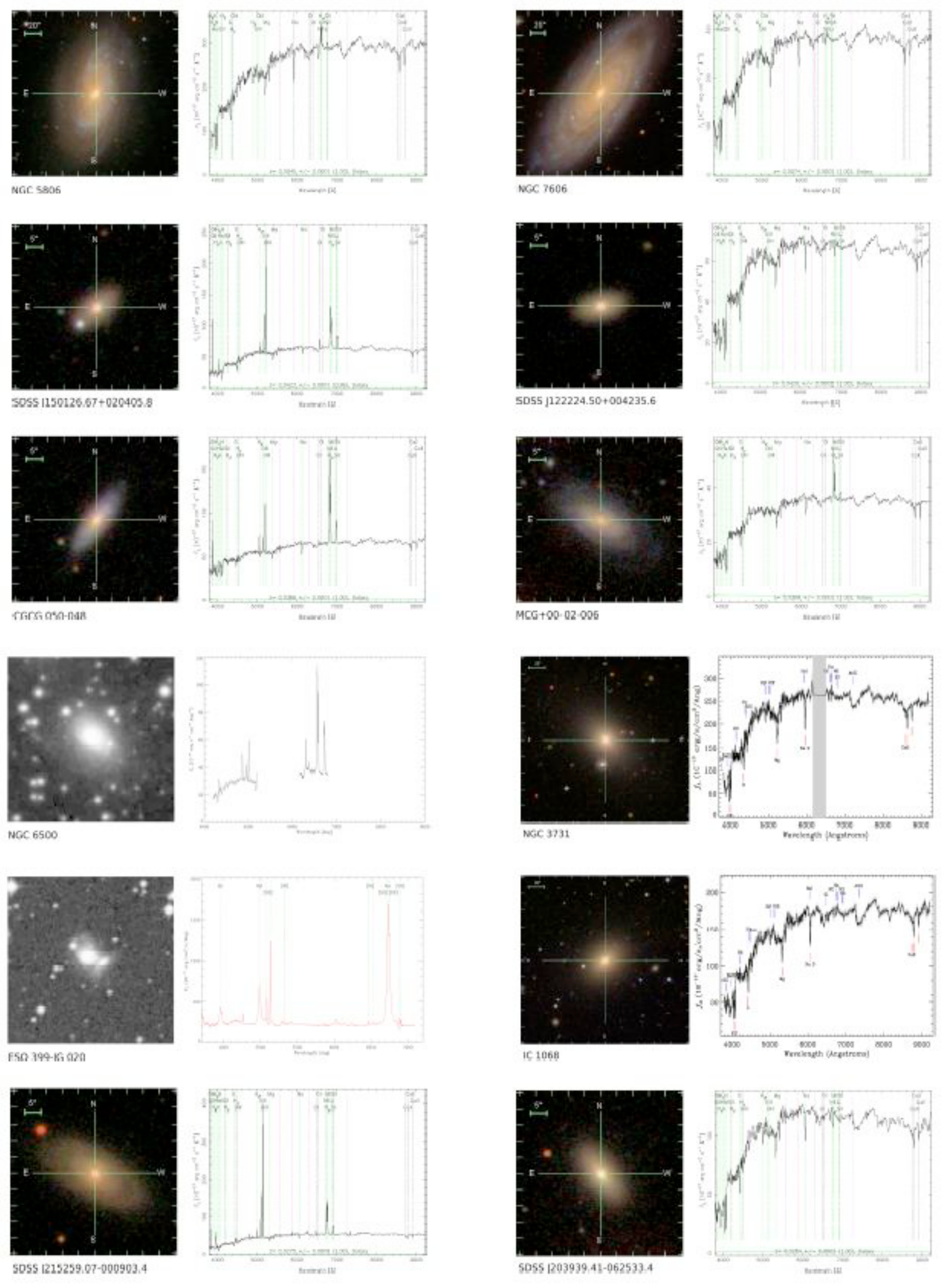}} \\
	\end{centering}
	\vspace{3mm}
		Fig.~\ref{fig:sdss} cont. \textit{Note:--} NGC~6500 and ESO\,399-IG\,020 are located beyond the SDSS footprint, and as yet no control galaxy observed or selected. The images displayed for these galaxies are therefore $R$-band Digitized Sky Survey (DSS) images. The spectrum for NGC~6500 was derived from long-slit spectroscopic observations \citep{ho95}, available on NED. The spectrum for ESO\,399-IG\,020 was created by integrating over the central 3\as\ of IMACS-IFU data, thus replicating an SDSS spectrum.
\end{figure*}

%%	Data reduction flow chart...

%%	IFU reduction figures...

	\clearpage

%%	Moment maps example figure... 

%%	All IMACS-IFU moment maps now...
\clearpage
\vspace{0.2in}

\vspace{0.3in}

\vspace{0.3in}

\vspace{0.3in}

\vspace{0.3in}

\vspace{0.3in}

\vspace{0.3in}

\vspace{0.3in}

\vspace{0.3in}

\vspace{0.3in}

\vspace{0.3in}

\vspace{0.3in}

\vspace{0.3in}

\vspace{0.3in}

\vspace{0.3in}

\vspace{0.3in}

\vspace{0.3in}

\vspace{0.3in}

\clearpage
\vspace{0.3in}

\vspace{0.3in}

\vspace{0.3in}

\vspace{0.3in}

\vspace{0.3in}

\vspace{0.3in}

\vspace{0.3in}

\vspace{0.3in}
%%	End of moment maps.

%%%	****************************************************	Tables 		******************************************************	%%

%%	IFUs...
	\begin{deluxetable}{lcccccc}
		\tabletypesize{\scriptsize}
		\tablecaption{Comparison with other selected IFU instrumental parameters\label{tab:IFUs}}	
		\tablewidth{0pt}
		\tablehead{	\colhead{Instrument} & \colhead{Telescope} &  \colhead{Telescope} & \colhead{FOV} & \colhead{Spaxel} & \colhead{Spaxel} & \colhead{Resolution} \\
					\colhead{} & \colhead{} & \colhead{Diameter} & \colhead{} & \colhead{Size} & \colhead{Number} & \colhead{} \\
					\colhead{} & \colhead{} & \colhead{(m)} & \colhead{(arcsec)} & \colhead{(arcsec)} & \colhead{} & \colhead{($\lambda / \Delta\lambda$)}}
		\startdata
			IMACS-IFU ($f/4$)		&	Magellan+1	&	6.5		&	$4.15 \times 5.0$	&	0.2	hexagonal		&	600 + 600		&	10\,000		\\
			FLAMES-ARGUS (1:1.67)	&	VLT			&	8.2		&	$6.6 \times 4.2$	&	0.3	circular		&	308 + 15		&	19\,000		\\
			IMACS-IFU ($f/2$)		&	Magellan+1	&	6.5		&	$6.92 \times 5.0$	&	0.2	hexagonal		&	1\,000 + 1\,000	&	1\,800		\\
			GMOS					&	Gemini		&	8.1		&	$7 \times 5$		&	0.2 hexagonal		&	1\,000 + 500	&	1\,700		\\
			FLAMES-ARGUS (1:1)		&	VLT			&	8.2		&	$11.5 \times 7.3$	&	0.52 circular		&	308 + 15		&	19\,000		\\
			INTEGRAL				&	WHT			&	4.2		&	$34 \times 29$		&	2.70 circular		&	115 + 20		&	4\,200		\\
			SAURON					&	WHT			&	4.2		&	$41 \times 33$		&	$0.94 \times 0.94$	&	1\,577			&	1\,250		\\
			DensePak				&	WIYN		&	3.5		&	$45 \times 30$		&	2.81 circular		&	91 + 4			&	20\,000		\\
			VIMOS					&	VLT			&	8.2		&	$54 \times 54$		&	$0.67 \times 0.67$	&	6\,400			&	220			\\
			SparsePak				&	WIYN		&	3.5		&	$72 \times 71$		&	4.69 circular		&	75 + 7			&	12\,000		\\
			PMAS PPak				&	Calar Alto	&	3.5		&	$74 \times 64$		&	2.68 circular		&	331 + 36		&	8\,000		\\
		\enddata
%		\tablecomments{}
	\end{deluxetable}

%%	Sample properties table...
\begin{deluxetable}{lllcccccc}
	\tabletypesize{\scriptsize}
	\tablecaption{Selection properties of the IMACS-IFU sample\label{tab:properties}}	
	\tablewidth{0pt}
	\tablehead{\colhead{Pairs$^{a}$} & \colhead{SDSS ID$^{b}$} & \colhead{Alternative} & \colhead{Hubble} & \colhead{Activity} &
					\colhead{$V_{syst}$$^{f}$} & \colhead{M$_{r}$$^{g}$} & \colhead{Inclination$^{h}$} & \colhead{$R_{90}$$^{i}$} \\
					\colhead{} & \colhead{ } & \colhead{Name$^{c}$} & \colhead{Type$^{d}$} & \colhead{Class$^{e}$} & 
					\colhead{(\kms)} & \colhead{(mag)} & \colhead{(deg)} & \colhead{(\as)}}
	\startdata
	1	&	SDSS\,J023311.04--074800.8			&					&	S0$^{4}$		&	S1		&  9289 & --19.89	&  43	& 7.0  \\
		&	SDSS\,J082323.42+042349.9$^{1}$	&					&				&			&  9163	& --20.08	&  40	& 8.0  \\
	\tableline
	2	&	SDSS\,J024440.23--090742.4			&					&	Sa$^{4}$		&	S2		&  7116	& --20.13	&  62	& 8.5  \\
		&	SDSS\,J015536.83--002329.4			&					&				&	HII		&  6735	& --19.74	&  49	& 7.0  \\
	\tableline
	3	&	SDSS\,J032525.35--060837.9			&	Mrk\,609			&	Im pec?		&	S1.8	& 10339 & --22.16	&  55	& 8.0  \\
		&	SDSS\,J031237.12--073422.2			&	Mrk\,1404			&	S0			&	HII		& 10798 & --21.98	&  73	& 9.5  \\
	\tableline
	4	&	SDSS\,J033955.68--063237.5			&					&	S0$^{4}$		&	S2		&  9384	& --19.75	&  47	& 4.0  \\
		&	SDSS\,J082323.42+042349.9$^{1}$	&					&				&			&       &           &       &      \\
	\tableline	
	5	&	SDSS\,J034547.53--000047.3			&					&	S0/a$^{4}$	&	S2		& 10850	& --20.50	&  64	& 7.0  \\
		&	SDSS\,J032519.40--003739.4			&					&	S0$^{4}$		&			& 10822	& --20.59	&  65 & 10.5 \\
	\tableline	
	6	&	SDSS\,J085310.26+021436.7			&					&	S0/a$^{4}$	&	S1		& 10581	& --20.96	&  35	& 5.0  \\
		&	SDSS\,J121032.17--011851.3			&	MRK\,1311		&	S0a$^{4}$	&			& 10338	& --20.79	&  33	& 8.0  \\
	\tableline	
	7	&	SDSS\,J085828.59+000124.4			&	CGCG\,005-043	&	S0-a$^{4}$	&	S1?		&  8564	& --20.99	&  41	& 11.5 \\
		&	SDSS\,J104409.99+062220.9			&					&				&			&  8921	& --20.77	&  56	& 14.0 \\
	\tableline
	8	&	SDSS\,J090040.66--002902.3			&					&	Sab			&	S2		& 12146	& --20.95	&  22	& 9.5  \\
		&	SDSS\,J002029.30--003317.9$^{2}$	&	MCG\,+00-02-006	&	Sab			&			& 11934	& --21.00	&  32	& 12.5 \\
	\tableline
	9	&	SDSS\,J130850.11--004902.3			&	ARK\,402			&	S0a$^{4}$	&	S2		&  5338	& --21.18   &  40& 26.5 \\
		&	SDSS\,J094603.54+042412.4			&	UGC\,05226		&	S0:			&			&  5044	& --21.13	&  49	& 20.5 \\
	\tableline
	10	&	SDSS\,J144424.44+014047.1			&	NGC\,5740		&	SAB(rs)b		&	S2		&  1572	& --19.68	&  22	& 37.0 \\
		&	SDSS\,J231904.77--082906.3$^{3}$	&	NGC\,7606		&	SA(s)b		&	HII		&  2231	& --20.98	&  11	& 58.5 \\
	\tableline
	11	&	SDSS\,J144611.12--001322.6			&	NGC\,5750		&	SB0/a(r)		&	S2		&  1687	& --20.12	&  30 & 34.5 \\
		&	SDSS\,J231904.77--082906.3$^{3}$	&					&				&			&       &           &       &      \\
	\tableline
	12	&	SDSS\,J150000.40+015328.7			&	NGC\,5806		&	SAB(s)b		&	S2		&  1359	& --19.77	&  23	& 43.5 \\
		&	SDSS\,J231904.77--082906.3$^{3}$	&					&				&			&       &           &       &      \\ 
	\tableline
	13	&	SDSS\,J150126.67+020405.8			&					&	S0$^{4}$		&	S1		& 12721	& --20.68	&  44	& 6.5  \\
		&	SDSS\,J122224.50+004235.6			&					&	S0$^{4}$		&			& 12609	& --20.61	&  41	& 6.0  \\
	\tableline
	14	&	SDSS\,J153747.90+030209.7			&	CGCG\,050-048	&				&	S2		& 11604	& --20.89	&  25	& 8.0  \\
		&	SDSS\,J002029.30--003317.9$^{2}$	&					&				&			&                &                &       &      \\
	\tableline
	15	&	N/A								&	NGC\,6500		&	SAab		&	S3		&  3003 	 &	--	&  --	&  --  \\
		&	SDSS\,J113411.67+123044.3			&	NGC\,3731		&	E			&			&  3124 	 &	--	& --	& --	 \\
	\tableline	
	16	&	N/A								&  ESO\,399-IG\,020 		&				&	S1		&  7480	 &	--	&  --	&  --  \\
		&	SDSS\,J145332.08+030456.5			&	IC\,1068			&				&			& 8437 	 &	--	& --  & --  \\
	\tableline
	17	&	SDSS\,J215259.07--000903.4			&					&				&	S2		&  8263	& --20.67	&  30	& 14.0 \\
		&	SDSS\,J203939.41--062533.4			&					&				&			&  8537	& --20.51	&  45	& 7.5  \\
	\enddata
	\tablecomments{Summary of the Magellan IMACS-IFU observations. \\
		$^{a}$Galaxy pair ID. \\
		$^{b}$SDSS identification number. \\
		$^{c}$Alternative galaxy name. \\
		$^{d}$ Galaxy morphological classification (NED). \\
		$^{e}$Activity classification: S1---Seyfert 1 galaxy; S2---Seyfert 2 galaxy; S3---LINER; HII---star-forming galaxy. \\
		$^{f}$Systemic velocity (NED). \\
		$^{g}$Absolute $r$-band magnitude (SDSS). \\
		$^{h}$Inclination (SDSS). \\
		$^{i}$Radius containing 90\% of the Petrosian flux, in arcseconds (SDSS). \\
		$^{1,2,3}$Common controls. \\
		$^{4}$Not all galaxies were classified on NED, so these galaxies have been classified by the authors. \\
		}
\end{deluxetable}

%%	Observations table...
\begin{deluxetable}{lccccc}
	\tabletypesize{\scriptsize}
	\tablecaption{Summary of IMACS-IFU Observations\label{tab:sample}}	
	\tablewidth{0pt}
	\tablehead{\colhead{Name} & \colhead{$\alpha$} &  \colhead{$\delta$} & 
				\colhead{Date} & \colhead{$T_{Exp}$} & \colhead{Seeing} \\
				\colhead{} & \colhead{(J2000)} & \colhead{(J2000)} & \colhead{(yyyy-mm-dd)} & \colhead{(sec)} & \colhead{(\as)}}
	\startdata
	SDSS\,J023311.04--074800.8			& 02:33:11.04 & --07:48:00.8 & 2005 12 10 &  $3\times1800$	& 0.6--1.2 \\
	SDSS\,J082323.42+042349.9$^{1}$		& 08:23:23.42 &  +04:23:49.9 & 2005 12 09 &  $4\times1800$	& 0.7--0.8 \\
	\tableline
	SDSS\,J024440.23--090742.4			& 02:44:40.64 & --09:07:35.6 & 2005 12 08 &  $4\times1800$	& 0.8--0.9 \\
	SDSS\,J015536.83--002329.4			& 01:55:35.89 & --00:23:28.2 & 2005 12 07 &  $4\times1800$	& 0.6--0.7 \\
	\tableline
	Mrk\,609							& 03:25:25.35 & --06:08:37.9 & 2005 12 07 &  $4\times1800$	& 0.7--1.0 \\
	Mrk\,1404							& 03:12:37.12 & --07:34:22.2 & 2005 12 08 &  $3\times1800$	& 0.6--0.8 \\
	\tableline
	SDSS\,J033955.68--063237.5			& 03:39:55.97 & --06:32:28.9 & 2005 12 10 &  $4\times1800$	& 1.0--1.1 \\
	SDSS\,J082323.42+042349.9$^{1}$		&			  &              &			  &			&		   \\
	\tableline
	SDSS\,J034547.53--000047.3			& 03:45:47.98 & --00:00:25.5 & 2005 12 09 &  $4\times1800$	& 0.4--0.6 \\
	SDSS\,J032519.40--003739.4			& 03:25:18.64 & --00:37:53.7 & 2005 12 09 &  $4\times1800$	& 0.7--1.1 \\
	\tableline
	SDSS\,J085310.26+021436.7			& 08:53:10.26 &  +02:14:36.7 & 2006 04 04 &  $4\times1800$	& 0.6--0.9 \\
	MRK\,1311							& 12:10:32.18 & --01:18:51.4 & 2006 04 04 &  $4\times1800$	& 0.6--0.7 \\
	\tableline
	CGCG\,005-043						& 08:58:28.59 &  +00:01:24.5 & 2006 04 05 &  $4\times1800$	& 0.6      \\
	SDSS\,J104409.99+062220.9			& 10:44:09.99 &  +06:22:20.9 & 2006 04 05 &  $1\times1800$	& 0.5      \\
										&             &              & 2006 04 30 &  $2\times1800$	& 0.5      \\
	\tableline
	SDSS\,J090040.66--002902.3			& 09:00:40.66 & --00:29:02.3 & 2005 12 07 &  $2\times1800$	& 0.7--1.4 \\
										&             &              & 2005 12 08 &  $2\times1800$	& 0.65     \\
	MCG\,+00-02-006$^{2}$				& 00:20:29.30 & --00:33:18.0 & 2007 08 21 &	 $1\times1000$  & 1.1      \\
	\tableline
	ARK\,402							& 13:08:50.11 & --00:49:02.4 & 2006 04 05 &  $4\times1800$	& 0.5      \\
	UGC\,05226						& 09:46:03.54 &  +04:24:12.4 & 2006 04 30 &  $3\times1800$	& 0.5      \\
	\tableline
	NGC\,5740						& 14:44:24.45 &  +01:40:47.2 & 2006 04 30 &  $3\times1800$	& 0.5--0.6 \\
										&             &              & 2007 08 20 &  $1\times1800$	& 1.0      \\
	NGC\,7606$^{3}$					& 23:19:04.78 & --08:29:06.3 & 2007 08 20 &	 $1\times1800$	& 1.0      \\
									&             &              & 2007 08 20 &   $1\times900$	& 1.1      \\
									&             &              & 2007 08 21 &  $1\times1800$	& 1.1      \\
	\tableline
	NGC\,5750						& 14:46:11.12 & --00:13:22.6 & 2006 04 04 &  $4\times1800$	& 0.6--0.7 \\
	NGC\,7606$^{3}$						&	  &              &			  &					&		   \\
	\tableline
	NGC\,5806						& 15:00:00.40 &  +01:53:28.7 & 2006 04 04 &  $3\times1800$	& 0.7      \\
	NGC\,7606$^{3}$					&			  &              &			  &				&		   \\
	\tableline
	SDSS\,J150126.67+020405.8			& 15:01:26.67 &  +02:04:05.8 & 2006 04 05 &  $3\times1800$	& 0.5      \\
	SDSS\,J122224.50+004235.6			& 12:22:24.50 &  +00:42:35.6 & 2006 04 30 &  $4\times1800$	& 0.5--0.6 \\
	\tableline
	CGCG\,050-048					& 15:37:47.90 &  +03:02:10.0 & 2006 04 30 &	 $2\times1800$  & 0.6      \\
									&             &              & 2007 08 21 &  $3\times1800$	& 1.1      \\
	MCG\,+00-02-006$^{2}$				&			  &              &			  &				&		   \\
	\tableline
	NGC\,6500						& 17:55:59.78 & +18:20:17.7	 & 2007 08 20 &	 $1\times1500$	& 1.0	   \\
	NGC\,3731						& 11:34:11.67 & +12:30:44.3	 &	-	  &	-	&	-	   \\
	\tableline
	ESO\,399-IG\,020					& 20:06:57.70 & --34:32:58.0 & 2007 08 20 &	 $3\times1800$	& 1.0	   \\
	IC\,1068							& 14:53:32.92 & +03:04:38.3  &	-	  &	-	&	-	   \\
	\tableline
	SDSS\,J215259.07--000903.4			& 21:52:59.08 & --00:09:03.5 & 2007 08 20 &  $2\times1800$	& 1.0      \\
	SDSS\,J203939.41--062533.4			& 20:39:39.41 & --06:25:33.4 & 2006 04 30 &  $2\times1800$	& 0.8      \\
										&             &              & 2007 08 20 &  $3\times1800$	& 0.9--1.0 \\
	\enddata
	\tablecomments{Summary of the Magellan IMACS-IFU observations. Listed are the date of observations, 
		the total integration time of each observation, and the range of seeing conditions at the time 
		of observation (in arcseconds). NGC~3731 and IC~1068 yet to be observed.}
\end{deluxetable}

%%	Instrument set-up table...
\begin{deluxetable}{lcc}
	\tabletypesize{\scriptsize}
	\tablecaption{Basic Specification and Instrument Set-Up\label{tab:basics}}	
	\tablewidth{0pt}
	\tablehead{\colhead{} & \colhead{f/2.5 Short mode} &  \colhead{f/4 Long mode}}
	\startdata
	Spatial coverage				&	\das692 $\times$ \das500	&	\das415 $\times$ \das500			\\
	Total apertures					&	1\,000 per field				&	600 per field				\\
	\vspace{3mm}
	Format						&	25 $\times$ 40 elements		&	25 $\times$ 24 elements		\\
	Shape						&	\multicolumn{2}{c}{Rectangular pattern}						\\
	Fibre size						&	\multicolumn{2}{c}{\das02}								\\
	Wavelength range				&	\multicolumn{2}{c}{400 -- 900 nm}							\\
	\vspace{3mm}
	Spectral resolution (FWHM)		&	-						&	1.6\AA\ at 5000\AA			\\
	Grating						&	-						&	600 lines/mm				\\
	Grating order					&	-						&	1st						\\
	Blaze angle					&	-						&	10.33\textdegree			\\
	Central wavelength				&	-						&	5520 \AA					\\
	Wavelength range				&	-						&	3975 \AA -- 7097 \AA		\\
	Dispersion					&	-						&	0.388 \AA \, / pixel			\\
	\enddata
\end{deluxetable}

%%	Observation strategy table...
\begin{deluxetable}{lcc}
	\tabletypesize{\scriptsize}
	\tablecaption{Typical observing procedure\label{tab:observed}}	
	\tablewidth{0pt}
	\tablehead{\colhead{} & \colhead{$N_{exp}$} &  \colhead{$T_{exp}$ (s)}}
	\startdata
	Bias						&				8			&			0		\\
	Domeflat (image)			&				2			&			3		\\
	Domeflats (spectra)			&				4			&			600		\\
	Skyflats (screen)			&				3			&			120		\\
	Skyflats (sky)				&				3			&			120		\\
	\vspace{3mm}
	Arcs						&				2			&			600		\\
	Acquisition					&				1			&			10		\\
	Arc							&				1			&			60		\\
	Science Observations		&				4			&			1800	\\
	\vspace{3mm}
	Arc							&				1			&			60		\\
	Acquisition					&				1			&			10		\\
	Arc							&				1			&			60		\\
	Standard Star Observations	&				2			&			10		\\
	Arc							&				1			&			60		\\
	\enddata
	\tablecomments{Overview of the observing strategy. The table lists the number of exposures  
		of each type of frame ($N_{exp}$), and the desired exposure time ($T_{exp}$).}
\end{deluxetable}

\end{document}